\documentclass[12pt,twoside]{report}
\usepackage{amsfonts,amsbsy,psfig}
\begin{document} 
\pagenumbering{roman}
\def\mycaption#1{\vskip 0.1truecm 
\rightskip=3truepc\leftskip=3truepc\baselineskip=13pt
	\noindent{\small#1}}

\newcommand{\del}{\partial}
\newcommand{\m}{\mathbf}
\newcommand{\K}{{\mathbf{k}}}
\newcommand{\x}{{\mathbf{x}}}
\newcommand{\y}{{\mathbf{y}}}
\newcommand{\z}{{\mathbf{z}}}
\newcommand{\n}{{\mathbf{n}}}
\newcommand{\M}{\mathbb}
\newcommand{\B}{\boldsymbol}
\newcommand{\bi}{{\mathbf{i}}}
\newcommand{\bj}{{\mathbf{j}}}
\newcommand{\bI}{{\mathbf{I}}}

\newcommand{\f}{\frac}
\newcommand{\T}{\tilde}
\newcommand{\N}{\nonumber}
\newcommand{\bb}{\bibitem}
\newcommand{\BF}{\begin{figure}}
\newcommand{\EF}{\end{figure}}
\newcommand{\BE}{\begin{equation}}
\newcommand{\EE}{\end{equation}}
\newcommand{\BEA}{\begin{eqnarray}}
\newcommand{\EEA}{\end{eqnarray}}
\newcommand{\ti}{\textit}
\newcommand{\bq}{\begin{quote}\small \noindent}
\newcommand{\eq}{\end{quote}}
\newcommand{\U}{\underline}
\newcommand{\tb}{\textbf}

\title{ Exploring Topology of the Universe \\
in the\\ 
    Cosmic Microwave Background}                      
                              
\author{Kaiki Taro Inoue \\
Yukawa Institute for Theoretical Physics\\
Kyoto University,~Kyoto 606-8502,~Japan\\
\\
Doctoral thesis
\\
\\
\\
submitted to
\\
\\Department of Physics
\\Kyoto University
\\
\\
}
\date{January 2001}
\maketitle
\begin{abstract}
We study the effect of global topology of the spatial geometry
on the cosmic microwave background (CMB) for closed flat and closed 
hyperbolic models in which the spatial hypersurface is
multiply connected. If the CMB temperature fluctuations 
were entirely produced at the last scattering, then
the large-angle fluctuations would be much suppressed in comparison with
the simply connected counterparts which is at variance with 
the observational data. However, as we shall show in
this thesis,  for low matter density models the observational 
constraints are less stringent since a large 
amount of large-angle fluctuations could 
be produced at late times. On the other hand, a slight suppression in 
large-angle temperature correlations in such models explains rather
naturally the observed anomalously low quadrupole which is
incompatible with the prediction of the ``standard''
Friedmann-Robertson-Walker-Lema$\hat{\textrm{i}}$tre 
models. Interestingly, moreover, 
the development in the astronomical observation technology
has made it possible to directly explore the imprint of the 
non-trivial topology by looking for identical 
objects so called ``ghosts'' in wide separated directions.
For the CMB temperature fluctuations identical patterns 
would appear on a pair of circles in the sky.
Another interesting feature 
is the non-Gaussianity in the temperature fluctuations. 
Inhomogeneous and anisotropic Gaussian fluctuations for a particular choice 
of position and orientation are regarded as non-Gaussian fluctuations
for a homogeneous and isotropic
ensemble.  If adiabatic non-Gaussian fluctuations with vanishing
skewness but non-vanishing kurtosis are found only on large angular
scales in the CMB then it will be a strong sign of 
the non-trivial topology.
Even if we failed to detect the identical patterns or objects, 
the imprint of the ``finiteness'' could be still observable
by measuring such statistical property.
\end{abstract}
\tableofcontents
\listoffigures
\listoftables
\clearpage
\mbox{}

\clearpage
\pagestyle{headings}
\pagenumbering{arabic}

\chapter{Introduction}
\thispagestyle{headings}
\bq
\ti{And so 
I'll follow on, and whereso'er thou set 
the extreme coasts, I'll query, "what becomes 
thereafter of thy spear?" 'Twill come to pass 
that nowhere can a world's-end be, and that 
the chance for further flight prolongs forever 
the flight itself. 
\vskip 0.1truecm
\rightline{-De Rerum Natura(Lucretius, 98?-55? BC)}
}
\eq

\section{Finite or infinite?} 
To ancient people
the sky had been appeared as an immense dome  
with stars placed just inside or on it.  
The daily movements of the stars were attributed to the 
rotation of the dome called 
the \ti{celestial sphere} around the Earth. 
Aristotle concluded that the size of the  
celestial sphere must be 
finite since the movement of an object of 
infinite size is not allowed by his philosophy. 
He denied the existence of any forms of
matter, space and time beyond the boundary.
On the other hand, Epicurus has thought that  
the universe has no beginning in time and the space is unlimited in 
size. If the universe were limited in size, he said,  
one could go to the end, throw a spear  
and where the spear was located
would be the new 'limit' of the universe. 
\\
\indent
In the 16th century Copernicus proposed a new cosmology 
where the Sun is placed at the center of the universe instead of
the Earth. Historically, the heliocentric cosmology had already been 
proposed by Aristarchus in the third B.C but soon the idea was 
rejected under the strong influence of Aristotle's philosophy.
The failure to detect the parallax at those
days must have provided a ground for the old 
geocentric (Earth-centered) cosmology.
Copernicus' new idea remained obscure for about 100 years after his
death.  However, in the 17th century, 
the works of Kepler, Galileo, and Newton built upon the 
heliocentric cosmology have completely swept away 
the old geocentric cosmology. The parallax effect owing to the 
movement of the Earth has been finally confirmed in the 19th century.
Thus the argument of Aristotle about the finiteness of the universe 
lost completely its ground.  
It seems that most of people in modern era 
are comfortable with the concept of an infinite space 
which extends forever as far as we could see.
\\
\indent 
However, one should be aware that 
the no-boundary condition 
does not require that the space is 
unlimited in size. Let us consider the surface of the Earth.
If the surface were flat, one might worry about the
``end''of it. Actually there is no end since the surface of the 
Earth is \ti{closed}. 
In the 19th century, Riemann proposed a 
new cosmological model whose spatial geometry is
described by a 3-sphere, a closed 3-space without boundary. 
If we were lived in such a space, 
an arrow which had been shot in the air
would be able to move back and hit the archer.
This could have surprised Epicurus who did not 
know the non-Euclidean geometry. 
\\
\indent
In the standard Friedmann-Robertson-Walker-Lema$\hat{\textrm{i}}$tre 
(FRWL) models, the spatial geometry 
is described by homogeneous and isotropic  
spaces, namely, a 3-sphere $S^3$, an 
Euclidean space $E^3$ and a hyperbolic space $H^3$ which have 
positive, zero and negative constant curvature, respectively.
The former one is spatially closed (finite) but the latter twos are
spatially open (infinite). 
The standard models have succeeded in explaining the following important  
observational facts:the expansion of
the universe, content of light elements, and the cosmic 
microwave background (CMB).
However, we are still not sure whether the spatial
hypersurface is closed (finite) or not (infinite). 
It seems that measuring the curvature of the spatial geometry 
is sufficient to answer the question.
\\
\indent
In fact, the above three spaces might not be the unique candidate
for the cosmological model that describes our universe. 
We can consider a various kinds of closed \ti
{multiply connected}\footnote{If there exists a closed curve 
which cannot be continuously contracted to a point then the space
is \ti{multiply connected} otherwise the space is 
\ti{simply connected}.} spaces with non-trivial global topology 
other than $S^3$ \footnote{It has been conjectured that 
$S^3$ is the only example of a closed connected and simply 
connected 3-space (Poincar\'e's conjecture).}.
For instance,
a flat 3-torus $T^3$ which is locally isometric to $E^3$ can be
obtained by identifying the opposite faces of a cube.
Identifying antipodal points in $S^3$ yields a projective space 
$\tb{R}P^3$. Furthermore, a plenty of examples of closed hyperbolic (CH) 
spaces have been known. It would be inappropriate 
to represent hyperbolic (=constantly negatively curved) 
spaces as ``open'' spaces, which have been
widely used in the literature of astrophysics.
The local dynamics of a closed multiply connected space 
is equivalent to that of the simply connected counterpart since 
Einstein's equations can only specify the local structure of 
spacetime and matter.
Therefore these multiply connected models do not confront with the 
above-mentioned three observational facts. One can 
also consider non-compact
finite-volume spaces which extends unlimitedly in particular
directions. However, in what follows we shall only consider
spatially ``closed'' (compact) models just for simplicity.
Regardless of each topology, the effect of the non-trivial topology 
appears only on scales of the order of the length of the 
closed curves that are not continuously contractable to a point.
\section{Imprint of topology}
Suppose that the spatial hypersurface is multiply connected
on scales of the order of the horizon or less. In this case,
the universe is called the ``small universe'' \cite{Ellis86}
for which we would be able to observe surprising periodical patterns  
in far distant place as if we were looking into a kaleidoscope.
Apparently it would look like the space extending to infinitely 
remote points, but the
observed images represent a series of ``snapshots'' of
one astronomical object at simultaneous or different epochs 
as the photons go around the space. 
Note that for a $S^3$ model without the cosmological constant, 
we would not be able to see such images
since the universe would be collapsed by the time the photons
go around the space. 
\\
\indent
Surprisingly, the recent advance in the observation technology has made it 
possible to determine the spatial global topology of the 
universe. The observation methods are divided broadly into 
two categories which are complementary: the detection of the 
periodical structure in the astronomical objects and the detection
of the peculiar property in the statistics of density or CMB 
temperature fluctuations.  The former method can directly prove the
presence of the non-trivial topology but it needs precise measurements
of relevant physical quantities. On the other hand, the latter methods
do not require high-precision measurements while the direct proof 
might be difficult since it relies on
a certain assumption of the initial conditions. 
\\
\indent
In multiply connected spaces, a number of geodesic segments that
connect arbitrary two points exist. We call the nearest image 
a ``real'' one for which the geodesic segment that connects the light source 
and the observer is the shortest, and others are called ``ghosts''.
If the comoving radius of the last scattering surface were much larger
than the injectivity radius (i.e. half of the shortest length of
a loop which cannot be continuously contracted to a point)
in the space theoretically we would be able to find
such ghost images. In general, the lengths of these geodesic segments
are different. The longer segments correspond to older images.
Therefore, the life time of the light source must be 
sufficently long and stable for recognizing the images as ``ghosts''.
To date, a number of authors have tried to search for ``ghosts''
using cluster catalogs (see \cite{LL95} and references therein)
but have failed to obtain any positive signals. 
\\
\indent
The appearance of the ``ghosts'' is relevant to the 
periodical structure in the distribution of observed objects
since each geodesic segment corresponds to a copy of 
one single domain called the 
\ti{fundamental domain} which tessellates the developed
space (the apparent space) called the 
\ti{universal covering space}. One way for detecting the 
periodical structure is to look for the spike
in the distribution of the physical distance between two 
arbitrary objects\cite{LLL96,LUL00}. Note that the method is 
applicable to only flat and some spherical geometry.  Using a 3-dimensional 
catalog of galaxy clusters by Bury a constraint $L\!>\!600 h^{-1}$Mpc
for closed flat 3-torus models in which 
the fundamental domain is a cube with
side L without cosmological constant has been obtained
\cite{LLL96}.
If the geodesics that connect the light source and the observer
have equivalent lengths, then we can see perfectly identical objects
\ti{without time delay} in wide separated directions in the sky.
For instance, one way is to look for any identical   
sets of neighboring QSOs\cite{Roukema}. If the relative 
position and the orientation of such QSOs were
equivalent to those of another set of QSOs that had been observed 
in a different direction then it would be a sign that we 
were looking at one object(neighboring QSOs) from different 
directions. In general such a special configuration in which one 
observes an identical object at a simultaneous time appears
on a pair of circles. Suppose that the observer 
sits at the center of a sphere. If the sphere is large enough 
then it wraps around the space and intersects with itself on a
circle. Then one would easily notice that the two geodesic segments that 
connect the center with a point on the circle have the same
length. All we have to do is to search for a pair of
identical circles in the sky. One might use QSOs or another high-z
objects such as $\gamma$-ray bursts for detecting the perfectly identical
``ghosts'' but ideal one is to use the CMB 
temperature fluctuations\cite{CSS98a}. The so-called ``circle in the
sky'' method is a powerful tool for detecting the non-trivial 
topology of the spatial geometry if the space is so small that 
the last scattering surface can wraps around the space. 
\\
\indent  
Another signature of the non-trivial topology appears in the 
statistical property of the CMB temperature fluctuations.
The break of the global homogeneity and isotropy in the spatial geometry
naturally leads to a non-Gaussian feature in the fluctuations 
regardless of the type of the primordial perturbation 
(adiabatic or isocurvature).
For a particular choice of orientation and position of the observer,
the fluctuations form an inhomogeneous and anisotropic
Gaussian random field. Marginalizing over the orientation and the 
position, then the ensemble of fluctuations can be regarded as 
a homogeneous and isotropic non-Gaussian random field.
The non-Gaussian signals might appear
in the topological quantities (total length and genus) of 
isotemperature contours in the sky map 
although the detection of the signal using the COBE data\cite{Smoot92} is 
unlikely owing to the significant instrumental noises. 
However, the observed non-Gaussian signals in the bispectrum 
\cite{Ferreira}
might be the positive sign of the non-trivial topology if
they are not related with any systematic errors.
\\
\indent
Using the COBE data, a lower bound $L\!\ge\! 4800~ h^{-1}
$Mpc for compact flat 3-torus toroidal models (for which the
fundamental domain is a cube with equal sides $L$)
without the cosmological constant (assuming that 
the initial perturbation is adiabatic and scale-invariant $(n\!=\!1)$ )
has been obtained\cite{Sokolov93,Staro93,Stevens93,Oliveira95a}.
The suppression of the fluctuations on scales beyond the 
size of the fundamental domain leads to a decrease in the 
large-angle power spectra $C_l$ which is at variance with the 
observed temperature correlations. 
\\
\indent
In contrast, for low matter density models, the constraint could be
considerably milder than the locally isotropic and homogeneous 
flat (Einstein-de-Sitter) models
since a bulk of large-angle CMB fluctuations 
can be produced at late epoch due to 
the so-called (late) integrated Sachs-Wolfe 
(ISW) effect \cite{HSS,CSS98b} which is the gravitational blueshift
effect of the free streaming photons by the decay of the gravitational 
potential. Because the angular sizes of the 
fluctuations produced at late time 
are large, the suppression of the fluctuations on scale larger than the 
topological identification scale may not lead to a
significant suppression of the large-angle power if the 
ISW effect is dominant.
\\
\indent
Thus the investigation of the CMB anisotropy in low matter density models
is quite important. If there is no cosmological constant, then the
geometry must be hyperbolic. the observational aspects of  
a limited number of closed(compact) hyperbolic (CH) models 
have been studied by Gott\cite{Gott80} and Fagundes
\cite{Fagundes85,Fagundes89,Fagundes93} but 
the property of the CMB in these models has not been investigated
until recently although they have a number of 
interesting features. For instance, we expect that 
the initial perturbations are smoothed out 
since the geodesic flows are strongly chaotic
\cite{LMP82, GK92, ET94}.  This may provide a 
solution to the pre-inflationary initial value
problem\cite{CSS96}.
Another interesting property is the existence of the lower bound for
the volume. It is known that the volume of CH manifolds
must be larger than 0.16668 times cube of the curvature
radius although no concrete examples of manifolds with such small volumes
are known \cite{GMT96}.
If the deep fundamental theory predicts higher probability 
of the creation of the universe with smaller volume then
we may answer the question why we see the periodical structure
at present.
\\
\indent
Unfortunately, the simulation of the CMB anisotropy in CH models
is not an easy task. Unlike closed flat spaces, the detailed 
property of the eigenmodes of the Laplacian has not yet been unveiled. 
One of the object of this thesis is to
establish a concrete method for analyzing the eigenmodes of CH 3-
spaces necessary for simulating the CMB anisotropy which enable us to
extract a generic property of the eigenmodes.  
For this purpose, we consider two numerical methods  
which had originally been developed
for the analysis of semiclassical behavior in
classically chaotic systems.
\\
\indent 
In chapter 2, we briefly review the necessary ingredients of 
mathematics of the three types of closed locally 
homogeneous and isotropic spaces.
In chapter 3 we describe two numerical methods, namely,
the direct boundary element method (DBEM) and
the periodic orbit sum method (POSM) 
for calculating eigenmodes of the Laplacian of CH manifolds.
Then we apply the methods to a number of known
examples of CH manifolds with small volume and
study the statistical property of eigenmodes. 
We also characterize the geometric property of manifolds in terms of the 
eigenvalue spectrum. 
Chapter 4 is devoted to the study of CMB anisotropy in
closed low mater density models with flat and hyperbolic geometry. 
First we estimate the degree of suppression in the large-angle   
power. Next, we carry out the Bayesian analysis
using the COBE-DMR 4-year data and obtain observational constraints
on the models. 
In chapter 5, we discuss methods for detecting
non-trivial topology in the CMB.
First we describe the direct method for searching identical
fluctuation patterns then we explore the expected 
non-Gaussian signatures.
In the last chapter, we summarize our results and
discuss future prospects.
\chapter{3-dimensional Topology}
\thispagestyle{headings}
\bq
\ti{\dots I think it is fair to say that until recently there was
little reason to expect any analogous theory for manifolds of
dimension 3 (or more)-except perhaps for the fact that so many 
3-manifolds are beautiful.  
\vskip 0.1truecm
\rightline{(W.P. Thurston)}
}
\eq
The classification of 2-dimensional
closed surfaces had already been completed in the 19th century.
A closed surface is homeomorphic to either a sphere or a 
2-torus with or without handles if orientable. 
A 2-torus without handles and that with handles 
can be endowed with flat geometry and hyperbolic geometry, respectively.
In other words,  2-dimensional closed (orientable) surfaces 
can be classified into three types of geometry of homogeneous and isotropic 
spaces, namely a 2-sphere $S^2$, a Euclidean plane $E^2$ and a 
hyperbolic plane $H^2$. 
\\
\indent
Similarly, it has been conjectured that the topology of closed 3-manifolds 
(complete and without any singularities)
can be classified into eight types of geometry of homogeneous
spaces\cite{Thurston82}. 
Suppose a closed 3-manifold $M$. First we 
repeatedly cut $M$ along 2-spheres embedded in $M$ and glue
3-balls to the resulting boundary components. Next we cut the
pieces along tori embedded nontrivially in $M$. The above-procedure is
called a \ti{canonical decomposition}. Thurston conjectured
that every closed complete 3-manifold $M$ has a canonical decomposition
into pieces which can be endowed with a complete, locally homogeneous 
Riemannian metric. Let us define a \ti{geometry} to be a pair $(X,G)$
where $X$ is a 3-manifold and $G$ acts transitively on $X$ 
such that the stabilizer of any point $x\in X$ is a compact
subgroup of $G$. 
A geometry is called \ti{maximal}
if $G$ is not contained in a larger group $G'$ for a given $X$. 
By considering $G$-invariant metrics on $X$, one can recover the 
ordinary viewpoint of differential geometry (the spaces are 
homogeneous Riemannian manifolds). $(X,G)$ and $(X',G')$
are \ti{equivalent} if there is a diffeomorphism 
$\phi:X \rightarrow X'$ with an isomorphism from $G$ to $G'$ 
$\T{\phi}:g \longmapsto \phi \circ g \circ \phi^{-1}$.  
Then we have the following theorem proved by Thurston 
\\
\ti{Any maximal 
simply connected geometry which admits a compact quotient\footnote
{A geometry is said to admit a compact quotient if $G$ has a subgroup 
$\Gamma$ which acts in $X$ as a covering group so that $X/\Gamma$
becomes compact} is
equivalent to one of the following eight geometries
$(X,\textrm{Isom}X)$ where $X$ is either $E^3, H^3, S^3,\\
 S^2\times \textbf{R}, H^2 \times \textbf{R},\widetilde{SL}(2,\textbf{R}),
\textrm{Nil}, or \textrm{Sol}$}
(see \cite{Scott83} for full details). 
Thus in addition to three geometries of constantly curved spaces
$E^3$, $S^3$, $H^3$ (flat, spherical, hyperbolic), there exist five
other geometries of homogeneous spaces. 
\\
\indent
Although it is not known whether
the canonical decomposition of a closed 
3-manifold $M$ into pieces of locally
homogeneous spaces always exist, we do know how to construct \ti{every}
closed 3-manifold. Suppose a \ti{link} $L$ to be a 1-dimensional 
compact submanifold. By removing a regular neighborhood of $L$ and
gluing it back by some new identification determined by a 
diffeomorphism of the torus for each component of $L$, one can obtain
a new manifold $N$. The procedure is called the {\it{Dehn surgery}}. 
It has been proved that every closed 3-manifold can be
obtained by a Dehn surgery along some link in a 3-sphere $S^3$
whose complement is hyperbolic. Suppose $L \subset M$
is a link whose complement $M-L$ can be endowed with hyperbolic
geometry. Then except for a finite number of choices,
all the remaining Dehn surgeries yield closed hyperbolic
manifolds\cite{Thurston82}.  This fact suggests that most closed 
3-manifolds are hyperbolic in some sense.
\\
\indent
In this chapter we describe 3-types of geometry of 
constant curvature $E^3,S^3$ and $H^3$ which are most relevant
to the cosmology.  

\section{Flat geometry}
First we consider the flat geometry 
$(\T{M},G)=(E^3,ISO(3)
=\tb{R}^3 \times SO(3))$. The actions of the 
isometry group on $E^3$ consist of an identity, 
a translation, a glide reflection, i.e. a reflection in a plane through
the origin followed by translation parallel to the plane and 
a helicoidal motion, i.e. a translation accompanied by a rotation, 
and their combination, which generate 
18 distinct types of locally Euclidean spaces\cite{Wolf67}.
Eight types of these spaces are open and other ten types are closed. The 
closed orientable ones are the following six types
$T^3,T^3/\tb{Z}_2, T^3/\tb{Z}_4,T^3/\tb{Z}_2\times \tb{Z}_2, 
T^3/\tb{Z}_3$ and $T^3/\tb{Z}_6$ where $T^3$ denotes a
3-torus and $\tb{Z}_m$ is a cyclic group of order $m$. 
The fundamental domains of the first four spaces can be 
a cube or a parallelepipe. Identifying the opposite faces by 
three translations one obtains a 3-torus $T^3$. Similarly,
identifying opposite faces by translations
but one pair being rotated by $\pi$ or $\pi/2$ yields
$T^3/\tb{Z}_2$ and $T^3/\tb{Z}_4$, respectively. Identifying the 
opposite faces by translations with a $\pi$ rotation, then we obtain 
$T^3/\tb{Z}_2\times \tb{Z}_2$. The fundamental domain of
the last two can be a hexagonal prism. Identifying the opposite
faces by translations but the hexagonal ones 
being rotated by $2 \pi/3$ or $\pi/3$ yield $T^3/\tb{Z}_3$ and 
$T^3/\tb{Z}_6$, respectively. 
It should be noted that there is no upper or lower
bound for the volume of the closed manifolds since there is no
particular scale in flat geometry. 
\\
\indent
$T^3$ is globally homogeneous but
the others are not globally homogeneous. If a Riemannian 
manifold $M$ is globally and locally homogeneous then every element 
$\gamma$ of the discrete isometry group $\Gamma \subset G$ is
equivalent to a Clifford translation of $\T{M}$, i.e. a transformation
$\gamma:x\rightarrow \gamma(x)$ such that the distance 
between $x$ and its image $\gamma(x)$ is the same 
for all $x\in \T{M}$. Thus the simplest model $T^3$ is 
belonging to a rather special class of multiply connected spaces.
\section{Spherical geometry}
Next we describe the topology of the spherical geometry 
$(S^3,SO(4))$. The most direct way to define $S^3$ is as the 
unit sphere $\{x_1^2+x_2^2+x_3^2+x_4^2=1\}$ in $\textbf{R}^4$ but here we
consider the unit sphere $|z_1|^2+|z_2|^2=1$ in $\textbf{C}^2$. Then
$S^3/Z_m, (m>1)$ is called the \ti{lens space} $L_m$ which is defined as
an orbit space of the 3-sphere $S^3$ under the action of $Z_m$, given
by \cite{DFN90},
\BE
(z_1,z_2)  \longmapsto (z_1 e^{2\pi i/m},z_2 e^{2\pi i/m}).
\EE
In the case of $m\!=\!2$ the space is called the \ti{real projective space}
denoted as $\textbf{R}P^3$. Each piece of a cell decomposition of 
of $L_m$ consists of points 
$(z_1,z_2)\in S^3$ such that $z_2=\rho e^{i \phi},\rho>0$
and
\BE
\f{2 \pi}{m}<\phi < \f{2 \pi(q+1)}{m},
\EE 
where $q=0,1,\dots , m-1$ specifies the cell. In stereographic
projection, each cell looks like a convex lens. 
\\
\indent
Identifying $\textbf{R}^4$ as the space of quaternions $\textbf{Q}$ we can 
obtain another representation of $S^3/\Gamma$. The basis of $\textbf{Q}$
is denoted as $\{1,i,j,k\}$ which satisfy
\BE
i^2=j^2=k^2=-1
\EE
and
\BE
ij=k=-ji,~~jk=i=-kj,~~ki=j=-ik.
\EE
The subspace spanned by the identity $1$ is identified with
$\textrm{R}$, and its elements are called \ti{real};the subspace 
$\textrm{R}i+\textrm{R}j+\textrm{R}k$ is called the space of 
\ti{pure} quaternions. The conjugate of a quaternion $q=a+bi+cj+dk$
is given by $\bar{q}=a-bi-cj-dk$ and its square root is the absolute
value of $q$ denoted by $|q|$. The set 
\BE
\textbf{Q}'=\{q\in \textbf{Q}:|q|=1\}
\EE
forms a multiplicative group, called the \ti{group of unit
quaternions} which can be identified as $S^3$.  The elements 
of $\textbf{Q}'$ are called the \ti{unit quaternions}.
It is known that the right or left multiplication by a unit quaternion 
gives a self-action of $S^3$ by orientation-preserving
isometries while conjugation gives a self-action which takes any
two-sphere onto itself (which fixes 1)\cite{Thurston97}. Conjugation can be
described as follows. Let us consider $q,q'\in \textbf{Q}'$. Then the
action by conjugation is given by $\rho(q)(q')=
qq'q^{-1}=\rho(-q)(q')$ which fixes
$1$. In other words, it leaves a 2-sphere invariant. Thus
$\rho(q)$ can be regarded as an element of $SO(3)$ but $\rho:S^3
\longmapsto SO(3)$ is a two to one 
($\rho(q)=\rho(-q)$) map (homomorphism). Therefore, $S^3$ can be 
regarded as the universal covering of $SO(3)$, i.e. $S^3/\{\pm1\}$
is isomorphic to $SO(3)$. 
\\
\indent
Thus the discrete isometry group of $SO(4)$ (without fixed points)is 
described by the discrete subgroups of the rotations $SO(3)$ which 
have fixed points, namely the cyclic group $Z_m$(order=$m$), 
the dihedral group  $D_m$(order=$2m$)
(i.e. the symmetry group of a regular $m$-gon lying in a
2-plane), three polyhedral groups, i.e., the symmetry groups
$T$(order=12) of a
regular tetrahedron, $O$(order=24) of a regular 
octahedron and $I$(order=60) of a regular
icosahedron in $\textbf{R}^3$. The \ti{binary dihedral} and \ti{binary
polyhedral} groups are defined by
\BE
D^*_m\equiv \rho^{-1}(D_m),~~T^*\equiv \rho^{-1}(T),~~
O^*\equiv \rho^{-1}(O),~~I^*\equiv \rho^{-1}(I).
\EE 
Globally homogeneous spherical spaces are  
classified as follows: 
(i) $S^3$, (ii) $RP^3$, (iii) $S^3/Z_m, (m>2)$,
(iv) $S^3/D^*_m, (m>2)$, (v)
$S^3/T^*$,(vi) $S^3/O^*$, (vii) $S^3/I^*$ \cite{Wolf67}. Note that 
all the groups are Clifford groups. 
On the other hand, globally inhomogeneous spaces are obtained by 
the quotient of $S^3$ by a group of the following:
(i) $Z_m\times H (m>2)$ where $H$ is either $D^*_n,T^*,O^*$ or $I^*$
and $m$ is relatively prime to the order of $H$, (ii) a subgroup of
index (i.e. a number of left(right) cosets with respect to the
subgroup) 3 in $Z_{3m}\times T^*$, where $m$ is odd (iii) a 
subgroup of index 3 in 
$Z_{2n}\times D^*_{2m}$, where $n$ is even and $m$ and $n$ are relatively
prime. Note that any isometry of $S^3$ that has no fixed point is
orientation-preserving. Therefore, even if we consider the maximal group 
$O(3)$ instead of $SO(3)$ the result does not change\cite{Thurston97}. 
\\
\indent
The volume of $M=S^3/\Gamma$ is simply given by
\BE
\textrm{vol}(M)=2 \pi^2 R^3/|\Gamma|
\EE
where $R$ is the curvature radius and $|\Gamma|$ is
the order of $\Gamma$. Thus, the largest manifold is $S^3$ itself.
On the other hand, there is no lower bound for the volume since one can
consider a group with arbitrary large number of elements (for
instance, consider $S^3/Z_m$ with arbitrary large $m$).
\section{Hyperbolic geometry} 
Finally, we consider the hyperbolic geometry $(H^3,PSL(2,\tb{C}))$.
The discrete
subgroup $\Gamma$ of $PSL(2,\tb{C})$ which is the orientation-preserving 
isometry group of the simply-connected hyperbolic 3-space $H^3$
is called the Kleinian group. Any CH 3-spaces (either manifold or
orbifold) can be described as compact quotients $M=H^3/ \Gamma$. 
If we represent $H^3$ as an upper half space
($x_1,x_2,x_3$), the metric is written as
\BE
ds^2=\f{R^2(dx_1^2+dx_2^2+dx_3^2)}{x_3^2},
\EE
where $R$ is the curvature radius. In what follows,
$R$ is set to unity without loss of generality.
If we represent a point $p$ on the upper-half space, as a quaternion whose
fourth component equals zero, then the actions of  $PSL(2,{\tb
C})$ on $H^3 \cup {\tb C} \cup\{ \infty\} $ take the form  
\begin{equation}
\gamma: p\rightarrow p'=\f{a p+ b}{c p+ d}, 
~~~~~ad-bc=1,~~~p\equiv z + x_3 {\bj}, ~~~z=x_1+x_2 {\bi},
\end{equation}
where a, b, c and d are complex numbers and $1$, $\bi$ and $\bj$ are
represented by matrices,
\BE
1=
\left(
\begin{array}{@{\,}cc@{\,}}
1 & 0\\ 
0 & 1 
\end{array}
\right),~~~
\bi=
\left(
\begin{array}{@{\,}cc@{\,}}
i & 0\\ 
0 & -i 
\end{array}
\right),~~~
\bj=
\left(
\begin{array}{@{\,}cc@{\,}}
0 & 1\\ 
-1& 0 
\end{array}
\right).
\EE
The action $\gamma$ is
explicitly written as   
\begin{eqnarray}
\gamma:H^3\cup{\tb C}\cup\{ \infty \} &\rightarrow& 
H^3\cup{\tb C}\cup\{ \infty \},
\nonumber
\\
\nonumber
\\
\gamma:(z(x_1,x_2),x_3)~~ &\rightarrow& 
\Biggl
( \f{(az+b)(\overline{cz+d})+a\bar{c}x_3^2}{|cz+d|^2+|c|^2x_3^2},
\f{x_3}{|cz+d|^2+|c|^2x_3^2}\Biggr), 
\end{eqnarray}
where a bar denotes a complex conjugate.
Elements of $\Gamma$ for orientable CH manifolds
are $SL(2,{\tb{C}})$ conjugate to 
\BE
\pm \left(
\begin{array}{@{\,}cc@{\,}}
\exp({l/2+i \phi/2}) & 0\\ 
0 & \exp({-l/2-i \phi/2})
\end{array}
\right )
\EE
which are called \ti{loxodromic} if $\phi\ne0$ and
\ti{hyperbolic} if $\phi=0$. An orientable CH manifold is obtained as
a compact quotient of $H^3$ by a discrete subgroup which consist
of loxodromic or hyperbolic elements. 
\\
\indent
Topological construction of CH manifolds starts with a cusped manifold
with finite volume $M_c$ obtained by gluing ideal tetrahedra. 
$M_c$ is topologically equivalent 
to the complement of a knot $K$ or link $L$(which consists of knots)
in 3-sphere $S^3$ or some other closed 3-spaces. A surgery in which 
one removes the tubular neighborhood $N$ of $K$
whose boundary is homeomorphic to a torus, and replace $N$
by a solid torus so that a meridian\footnote{Given 
a set of generators $a$ and $b$
for the fundamental group of a torus, a closed curve 
which connects a point $x$ in the torus with $ax$ is 
called a {\ti{meridian}} and
another curve which connects a point $x$ with $bx$ is called a
{\ti{longitude}}.}
in the solid torus
goes to $(p,q)$ curve\footnote{If $C$ connects 
a point $x$ with another point $(pa+qb)x$ where $p$ and $q$ 
are co-prime integer, $C$ is called a $(p,q)$ curve.}
on $N$ is called $(p,q)$ \ti{Dehn surgery}. Except for a finite number
of cases, Dehn surgeries on $K$ always yield CH 3-manifolds which
implies that most compact 3-manifolds are hyperbolic\cite{Thurston82}.
However, we are not still sure in what cases a Dehn surgery 
yields a hyperbolic manifold for any kinds of cusped manifolds. 
Furthermore there are a variety of ways to construct 
isometric CH manifolds by
performing various Dehn surgeries on various cusped hyperbolic 
manifolds. Thus the classification of CH manifolds has not
yet been completed. 
\\
\indent
However, with the help of the computer, we now know a number of 
samples of CH manifolds (and orbifolds). For instance a computer program
``SnapPea'' by Weeks\cite{SnapPea} can numerically perform 
Dehn surgeries on cusped manifolds which have made it possible to construct a 
large number of samples of CH 3-manifolds. SnapPea can also
analyze various property of CH manifolds such as volume,
fundamental group, homology, symmetry group, length spectra and so on. 
\\
\indent
In order to describe CH 3-manifolds, the volume 
plays a crucial role since only a finite number of CH 3-manifolds 
with the same volume exist\cite{Thurston82}. The key facts are:
the volumes of CH 3-manifolds obtained by Dehn surgeries
on a cusped manifold $M_c$ are always less than the volume 
of $M_c$; CH 3-manifolds converge
to $M_c$ in the limit $|p|,|q| \rightarrow \infty$. 
\BF
\centerline{\psfig{figure=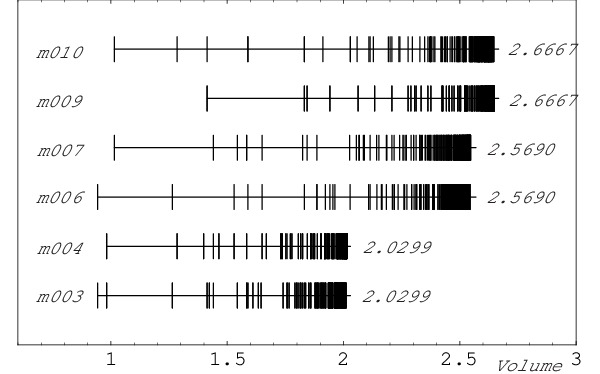,width=12cm}}
\caption{Volume Spectra of CH manifolds}
\mycaption{A prefix ``m'' in the labeling number represents a 
cusped manifold obtained by gluing five or 
fewer ideal tetrahedra. The numbers in the right side denote the volumes 
of the corresponding cusped manifold. The volumes have been computed
by using the SnapPea kernel.}
\label {fig:vol}
\EF
As shown in figure 
\ref{fig:vol}, the volume spectra are discrete but there are many 
accumulation points which correspond to the volumes of various cusped
manifolds. The smallest cusped manifolds in the known manifolds
have volume $2.0299$ which are labeled as 
``m003'' and ``m004''in SnapPea. m003 and m004 are
topologically equivalent  
to the complement of a certain knot in the lens space
$L_{5,1}$ and the complement of a
``figure eight knot'' (figure \ref{fig:fig8}), respectively\cite{Matveev}.
\BF[b]
\centerline{\psfig{figure=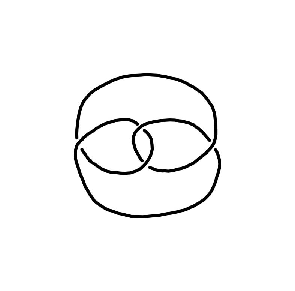,width=6cm}
\psfig{figure=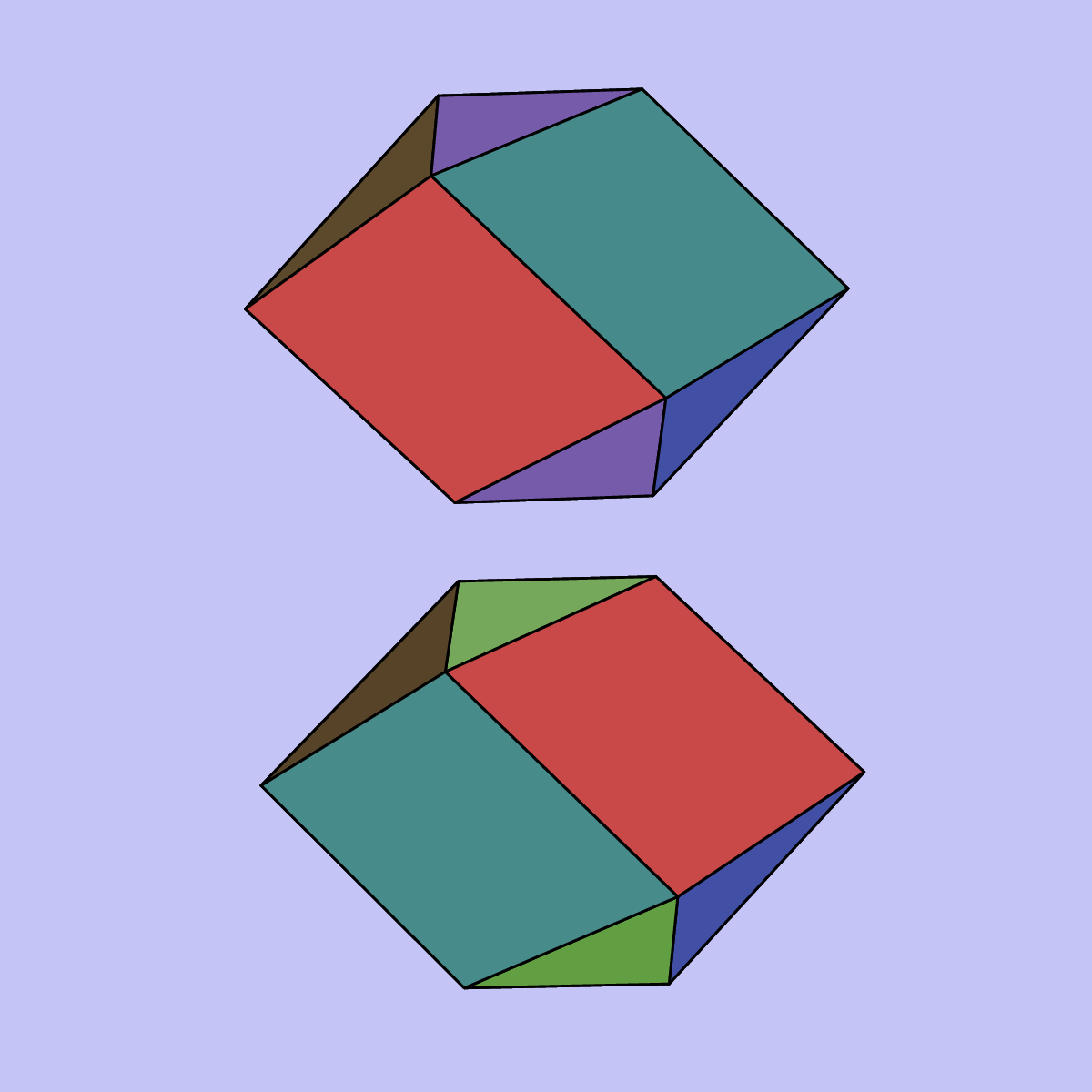,width=6cm}
}
\caption{Cusped Hyperbolic Manifold}
\mycaption{The figure in the left side represents a 
``figure eight knot'' and the figure in the right shows
the Dirichlet domain of a cusped manifold m004
viewed from two opposite directions
in the Klein(projective) coordinates where geodesics and planes are
mapped into their Euclidean counterparts. 
The two vertices on the left 
and right edges of the polyhedron which are 
identified by an element of the discrete isometry group 
correspond to a cusped point. Colors on the faces correspond to the
identification maps. One obtains the Dirichlet domain of m003 
by interchanging colors on quadrilateral faces in
the lower(or upper) right figure.}
\label{fig:fig8}
\EF
\BF[b]
\centerline{\psfig{figure=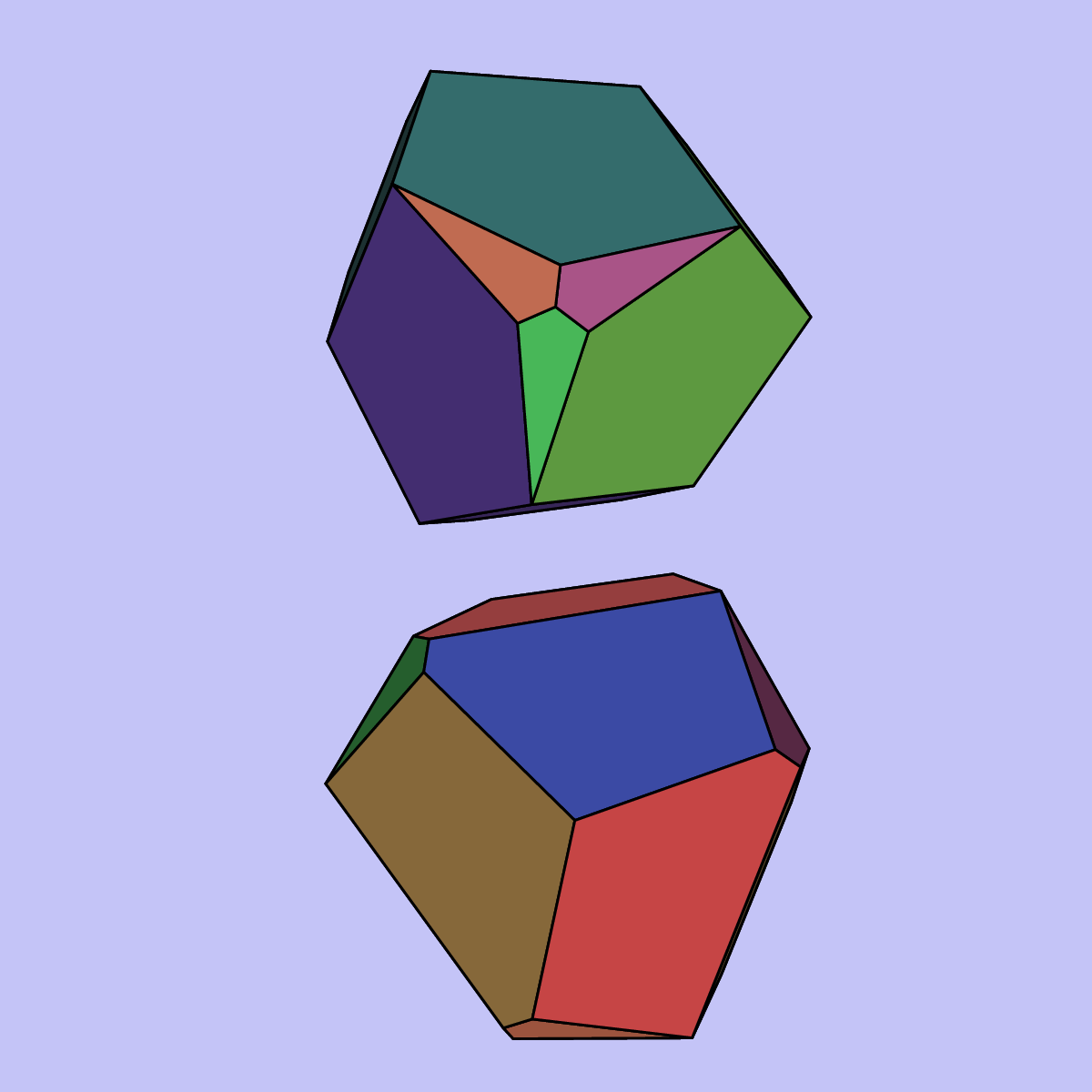,width=6.5cm}
\psfig{figure=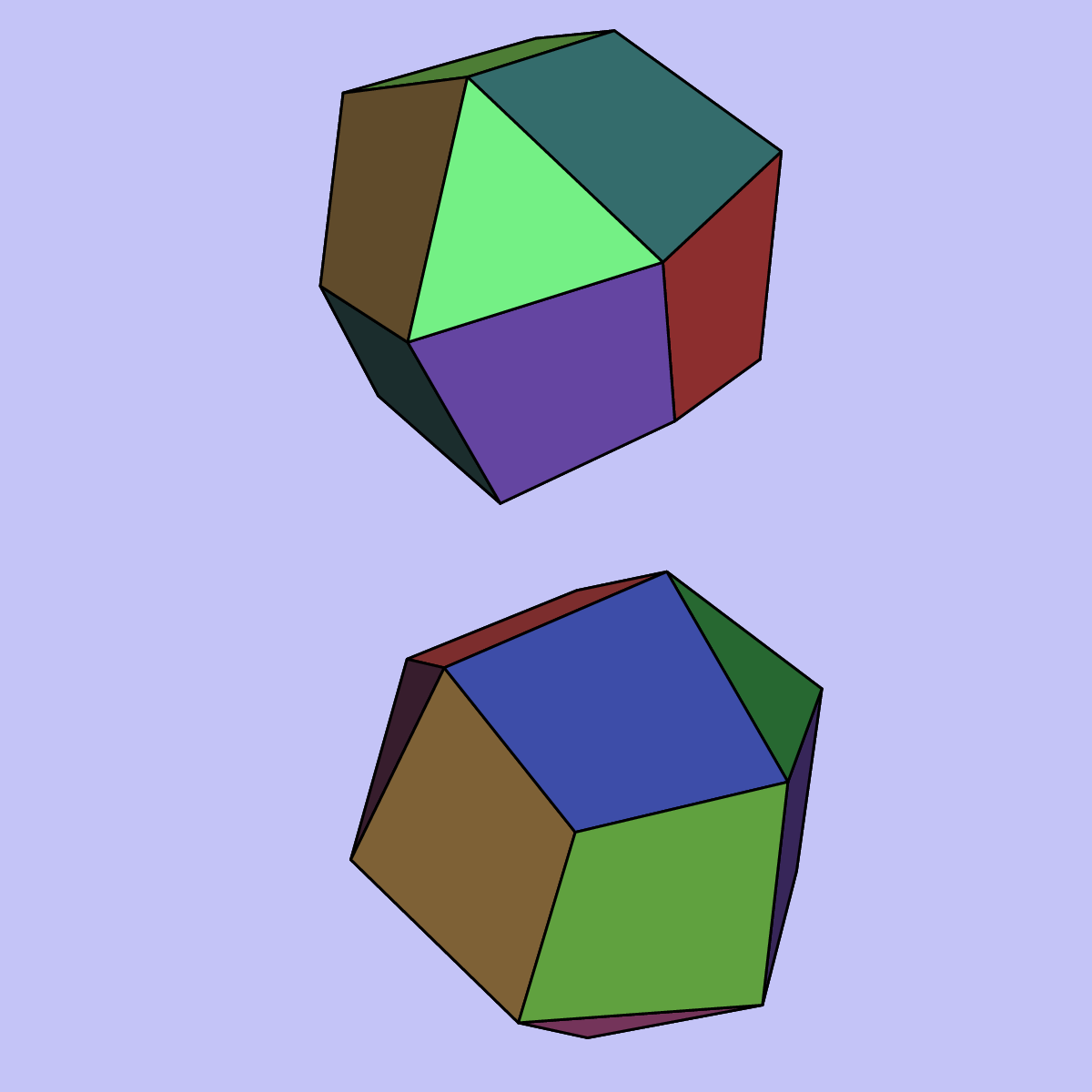,width=6.5cm}
}
\caption{Smallest CH Manifolds}
\mycaption{Dirichlet domains 
of the Weeks manifold (left) and the Thurston manifold(right)   
viewed from opposite directions in the
Klein(projective) coordinates. Colors on the faces correspond to the
identification maps. The manifolds can be obtained by performing 
(3,-1) and (-2,3) Dehn surgeries on m003, respectively.}
\EF
(3,-1) and (-2, 3) Dehn surgeries on m003 yield
the smallest and the second smallest known manifolds, which are
called the Weeks manifold (volume=0.9427) and the Thurston
manifold (volume=0.9814), respectively. 
It should be noted that the volume of CH manifolds
must be larger than 0.16668... times cube of the curvature
radius although no concrete examples of manifolds with such small volumes
are known\cite{GMT96}.  
As $|p|$ and $|q|$ becomes large, the volumes 
converge to that of $M_c$. Similarly, one can do Dehn surgeries on
m004 or other cusped manifolds to 
obtain a different series of CH manifolds.


\chapter{Mode Functions}
\thispagestyle{headings}
\bq
\ti{Science grows slowly and gently; reaching the truth by a variety
of errors. One must prepare the introduction of a new idea through
long and diligent labour; then, at a given moment, it emerges as if
compelled by a divine necessity ...
\vskip 0.1truecm
\rightline{(Karl Gustav Jacob Jacobi, 1804-1851)}
}

\eq
In locally isotropic and homogeneous background spaces
each type (scalar, vector and tensor) of
first-order perturbations can be decomposed into a decoupled set of
equations. In order to solve the decomposed linearly perturbed
Einstein equations, it is useful to expand the  
perturbations in terms of eigenmodes ($\!=\!$
eigenfunctions that are square-integrable)
of the Laplace-Beltrami operator which satisfies the 
Helmholtz equation with certain boundary conditions
\BE
(\nabla^2+k^2)u_{k}(x)=0,
\label{eq:Helmholtz}
\EE
since each eigenmode evolves independently.
Then one can easily see that the time evolution of the 
perturbations in multiply-connected locally isotropic and
homogeneous spaces coincide with those in the locally and globally
homogeneous and isotropic FRWL spaces whereas 
the global structure of the background space is determined by these 
eigenmodes. In the simplest closed flat toroidal spaces obtained by gluing
the opposite faces of a cube by three translations, the mode functions
are given by plane waves with discrete wave numbers. In general, 
the mode functions in locally isotropic and
homogeneous spaces can be written as a linear combination of
eigenfunctions on the universal covering space.
\\
\indent
The property of eigenmodes of CH spaces
have not been unveiled until recently although a number of theorems
concerning with lower bounds
and upper bounds of eigenvalues in terms of
diffeomorphism-invariant quantitities have been known in mathematical 
literature. However, despite the lack of knowledge about the property,
eigenmodes of CH spaces have been numerically investigated by 
a number of authors in the field of number theory and harmonic
analysis and of quantum chaology which are connected each other 
at a deep level. For the last 15 years, various numerical 
techniques have been applied
to the eigenvalue problems for solving the Helmholtz 
equation (\ref{eq:Helmholtz})
with periodic boundary conditions (manifold
case), Neumann and Dirichlet boundary conditions (orbifold case).
Eigenvalues of CH 2-spaces has been numerically obtained
by many authors\cite{Balazs,AS1,AS2,BSS,AS4,Hejhal}. Eigenvalues of
cusped arithmetic and cusped non-arithmetic 3-manifolds 
with finite volume have been obtained by Grunewald
and Huntebrinker using a
finite element method\cite{Grunewald}.  Aurich and Marklof have computed
eigenvalues of a non-arithmetic 3-orbifold using the direct boundary
element method(DBEM)\cite{AM}. The author
has succeeded in computing low-lying eigenmodes of the 
Thurston manifold, the second smallest one using the 
DBEM\cite{Inoue1}, and later the Weeks
manifold, the smallest one in the known CH manifolds using 
the same method\cite{Inoue3}. Cornish and Spergel 
have also succeeded in calculating eigenmodes of these manifolds and
10 other CH manifolds based on the Trefftz method\cite{CS99}.
\\
\indent
In quantum mechanics, an eigenmode can be interpreted as a wave function 
of a free particle at a stationary state with energy $E\!=\!k^2$.
The statistical property of the energy eigenvalues $E$ and 
the eigenmodes $u_E$ of the classically chaotic systems
has been investigated for exploring the imprints of 
classical chaos in the corresponding quantum system
(e.g. see\cite{Boh} and other articles therein).
Because any classical dynamical systems of a free particle 
in CH spaces (known as the Hadamard-Gutzwiller models)
are strongly chaotic (or more precisely they are K-systems 
with ergodicity, mixing and Bernoulli properties \cite{Balazs}), it is
natural to expect a high degree of complexity for each eigenstate. 
In fact, in many classically chaotic systems, it has been found that 
the short-range correlations observed
in the energy states agree 
with the universal prediction of random matrix theory (RMT) for 
three universality classes:the Gaussian 
orthogonal ensemble(GOE), the Gaussian unitary ensemble(GUE) and the
Gaussian symplectic ensemble (GSE)\cite{Meh,Boh}. 
For the Hadamard-Gutzwiller models 
the statistical property of the eigenvalues and 
eigenmodes is described by GOE (which
consist of real symmetric $N \!\times\!N$ matrices $H$ which obey the
Gaussian distribution $\propto \exp{(-\textrm{Tr}H^2/(4a^2))}$ 
(where $a$ is a constant) as the systems possess a time-reversal
symmetry. RMT also predicts that
the squared expansion coefficients of an eigenstate with respect to a 
generic basis are distributed as Gaussian random numbers \cite{Brody}
which had been numerically confirmed for some classically 
chaotic systems (\cite{AS4,Haake}).
Although these results are for highly excited modes in the 
semiclassical region, low-lying modes may also retain the 
random property.
For the analysis of CMB temperature fluctuations,
the statistical property of the 
expansion coefficients of low-lying eigenmodes
is of crucial importance since it
is relevant to the ensemble averaged temperature correlations  
on the present horizon scales. 
\\
\indent
For classically chaotic systems, the semiclassical correspondence
is given by the Gutzwiller trace
formula\cite{Gutzwiller} which relates a set of periodic
orbits in the phase space
to a set of energy eigenstates. Note that the correspondence is no longer
one-to-one as in classically integrable systems.
Interestingly, for CH spaces,
the Gutzwiller trace formula gives an \textit{exact} relation which had 
been known as the Selberg trace formula in mathematical
literature\cite{Selberg}.  The trace formula gives an 
alternative method to compute the
eigenvalues and eigenmodes in terms of periodic orbits. The poles of energy
Green's function are generated as a result of interference of 
waves each one of which corresponds to a periodic orbit. 
Roughly speaking, periodic 
orbits with shorter length contribute to the deviation from the
asymptotic eigenvalue distribution on larger energy scales. In fact,
a set of zero-length orbits produce Weyl's
asymptotic formula. 
Because periodic orbits can be obtained algebraically, the periodic orbit
sum method enables
one to compute low-lying eigenvalues for a large sample of manifolds
or orbifolds systematically if each fundamental group is known beforehand.
The method has been used
to obtain eigenvalues of the Laplace-Beltrami operator on 
CH 2-spaces and a non-arithmetic 3-orbifold\cite{AS1,AS3,AM}. 
However, it has not been applied to any CH
3-manifolds so far.
\\
\indent
In section 1, we first formulate the DBEM
for computing eigenmodes of the Laplace-Beltrami operator 
of CH 3-manifolds. In section 2, we study the statistical property of
low-lying eigenmodes on the Weeks and the Thurston maniolds, 
which are the smallest examples in the known CH 3-manifolds.
In section 3, we describe the method for computing the length spectra 
and derive an explicit form for computing the spectral staircase 
in terms of length spectrum which is applied for CH manifolds
with volume less than 3. 
In section 4, we analyze the relation between the low-
lying eigenvalues and several diffeomorphism-invariant geometric 
quantities, namely, volume, diameter and length of the 
shortest periodic geodesics.
In the last section, the deviation of low-lying eigenvalue spectrum 
from the asymptotic distribution is measured by $\zeta-$ function 
and the spectral distance.

\section{Boundary element method}
\subsection{Formulation}
The boundary element methods (BEM) use a free Green's function as
the weighted function, and the Helmholtz equation is
rewritten as an integral equation defined on the
boundary using Green's theorem. Discretization of the 
boundary integral equation yields a
system of linear equations. Since one needs the discretization on only
the boundary, BEM reduces the dimensionality of the
problem by one, which leads to economy in the numerical task.   
To locate an eigenvalue, the DBEM \footnote{The DBEM uses only
boundary points in evaluating the integrand in 
Eq.(\ref{eq:re2}). The indirect
methods use internal points in evaluating the integrand in 
Eq.(\ref{eq:re2}) as well as the boundary points.} requires
one to compute many determinants of the corresponding boundary
matrices which are dependent on the wavenumber $k$. 
\\
\indent
First, let us consider the Helmholtz equation with certain boundary 
conditions,
\begin{equation}
(\nabla^2+k^2)u(\x)=0,\label{eq:helmholtz}
\end{equation}
which is defined on a bounded
M-dimensional connected and simply-connected domain 
$\Omega$ which is a subspace of a 
M-dimensional Riemannian manifold ${\cal M}$ and the boundary 
$\del\Omega$ is piecewise smooth. $\nabla^2\equiv\nabla^i \nabla_i,~
(i=1,2,\cdot\cdot\cdot,M)$, and $\nabla_i$ is the covariant derivative 
operator defined on $\cal M$. A function $u$ in Sobolev space 
$H^2(\Omega)$ is the
solution of the Helmholtz equation if and only if 
\begin{equation}
{\cal R}[u(\x),v(\x)]\equiv \Bigl\langle(\nabla^2+k^2)\
u(\x),v(\x)\Bigr\rangle=0,\label{eq:residue}
\end{equation}
where $v$ is an arbitrary function in Sobolev space $H^1(\Omega)$ called 
\textit{weighted function} and $\langle\,\rangle$ is defined as 
\begin{equation}
\langle a,b \rangle\equiv \int_{\Omega}  ab\sqrt{g}\, dV. 
\end{equation}    
Next, we put $u(\x)$ into the form
\begin{equation}
u=\sum_{j=1}^M u_j \phi_j,
\end{equation}
where $\phi_j$'s are linearly independent
square-integrable functions.  Numerical
methods such as the finite element methods try 
to minimize the residue function $\cal R$ for a fixed
weighted function $v(\x)$ by changing the coefficients $u_j$. 
In these methods, one must 
resort to the variational principle to find the $u_j$'s which 
minimize ${\cal R}$.
\\
\indent
Now, we formulate the DBEM
which is a version of BEMs. Here we search $u(\x)$'s for the space 
$C^1(\bar{\Omega})\cap C^2(\Omega)\cap L^2(\Omega)$. 
First, we slightly 
modify Eq.(\ref{eq:residue}) using the Green's theorem
\begin{equation}
\int_\Omega (\nabla^2 u) v \sqrt{g} dV
-\int_\Omega (\nabla^2 v) u \sqrt{g} dV
=\int_{\del\Omega} (\nabla_i u) v \sqrt{g} dS^i
-\int_{\del\Omega} (\nabla_i v) u \sqrt{g} dS^i,
\label{eq:re2}
\end{equation}
where $g \equiv \mathrm{det}\{ g_{ij} \}$ and $dV\equiv dx_1 \ldots dx_M$;
the surface element $dS^i$ is given by
\begin{eqnarray}
dS_i &\equiv& \f{1}{M !} \epsilon_{i j_1 \cdot \cdot \cdot j_M} 
dS^{j_1 \cdot \cdot \cdot j_M},
\nonumber
\\
\nonumber
\\
dS^{j_1 \ldots j_M}&\equiv&
\left|
\begin{array}{@{\,}cccc@{\,}}
dx^{(1)j_1} & dx^{(2)j_1} & \ldots & dx^{(M)j_1} \\
dx^{(1)j_2} & dx^{(2)j_2} & \ldots & dx^{(M)j_2} \\
\vdots      &   \vdots    & \ddots & \vdots      \\
dx^{(1)j_M} & dx^{(2)j_M} & \ldots & dx^{(M)j_M}
\end{array}
\right|,
\end{eqnarray}
\\
where $\epsilon_{j_1 \cdot \cdot \cdot j_{M+1}}$ denotes the M$
\!+1$-dimensional Levi-Civita tensor. 
Then Eq.(\ref{eq:residue}) becomes
\begin{equation}
\int_\Omega (\nabla^2 v+k^2 v)u  \sqrt{g}\, dV
+\int_{\del\Omega} (\nabla_i u) v \sqrt{g}\, dS^i
-\int_{\del\Omega} (\nabla_i v) u \sqrt{g}\, dS^i=0.
\label{eq:IN}
\end{equation} 
As the weighted function $v$, we choose the fundamental solution
$G_E(\x,\y)$ which satisfies  
\begin{equation}
(\nabla^2+E)G_E(\x,\y)=\delta_D(\x-\y),
\label{eq:DE}
\end{equation}
where $E\equiv k^2$, and $\delta_D(\x-\y)$ 
is Dirac's delta function.
$G_E(\x,\y)$ is also known as the free Green's function 
whose boundary 
condition is given by 
\BE
\lim _{d(\x,\y) \rightarrow \infty} G_E(\x,\y) =0, 
\EE
where $d(\x,\y)$ is the geodesic distance between $\x$ and $\y$.
 Let $\y$ be an internal point of $\Omega$. Then
we obtain from Eq.(\ref{eq:IN}) and Eq.(\ref{eq:DE}),
\begin{equation}
u(\y)
+\int_{\del\Omega} G_E(\x,\y) \nabla_i u \,\sqrt{g}\, dS^i
-\int_{\del\Omega} (\nabla_i G_E(\x,\y)) u \,\sqrt{g}\, dS^i=0.\label{eq:inter}
\end{equation}
Thus the values of eigenfunctions at internal points can be 
computed using only the boundary integral. If $\y \in
\del\Omega$, we have to evaluate the limit of the
boundary integral terms as  $G_E(\x,\y)$ becomes divergent at $\x=\y$
(see appendix A). The boundary integral
equation is finally written as 
\begin{equation}
\f{1}{2}u(\y)
+\int_{\del\Omega} G_E(\x,\y) \nabla_i u \,\sqrt{g}\, dS^i
-\int_{\del\Omega} (\nabla_i G_E(\x,\y)) u \,\sqrt{g}\, dS^i=0,
\label{eq:bem0}
\end{equation}
or in another form, 
\begin{equation}
\f{1}{2}u(\y)
+\int_{\del\Omega} G_E(\x,\y) \f{\del u}{\del x^i} n^i \,\sqrt{g}\, dS
-\int_{\del\Omega} \f{\del G_E(\x,\y)}{\del x^i}n^i\, u \,\sqrt{g}\, dS=0,
\label{eq:bem}
\end{equation} 
where $n^i\equiv dS^i/dS$ and $dS\equiv \sqrt{dS^i\,dS_i}$. Note that
we assumed that the boundary surface at $\y$ is sufficiently smooth.
If the boundary is not smooth, one must calculate the internal solid
angle at $\y$ (see appendix A). Another approach is to rewrite 
Eq.(\ref{eq:inter}) in a regularized form \cite{Tan}. We see from 
Eq.(\ref{eq:bem0}) or Eq.(\ref{eq:bem}) that the approximated
solutions can be obtained without resorting to the variational principle.
Since it is virtually impossible to solve Eq.(\ref{eq:bem})
analytically, we discretize it using boundary elements. Let
the number of the elements be N. We approximate $u$  by some
low-order polynomials (shape function) on each element as
$u=c_1+c_2\, \eta+c_3\, \xi$ where $\eta$ and $\xi$
denote the coordinates on the corresponding standard
element \footnote{It can be proved that the approximated 
polynomial solutions converge to $u(\x)$ as the number of boundary 
elements becomes large \cite{Tabata,Johnson}.}.
   
Then we have the following equation:
\begin{equation}
[H]\{u\}=[G]\{q\},~~~~q\equiv \f{\del u}{\del n},\label{eq:BEM}
\end{equation}
where \{u\} and \{q\} are $N$-dimensional vectors which consist of 
the boundary values of 
an eigenfunction and its normal derivatives, respectively. 
[H] and [G] are 
$N\times N$-dimensional coefficient matrices which are
obtained from integration of 
the fundamental solution $G_E(\x,\y)$ and its normal derivatives multiplied 
by $u_i$ and $q_i$, respectively. Note that the elements in 
[H] and [G] include $k$ implicitly.
Because Eq.(\ref{eq:BEM}) includes both $u$ and $q$, the boundary 
element method can naturally incorporate the periodic boundary conditions: 
\begin{equation}
u(\x)=u(g_i(\x)),~~~q(\x)=-q(g_i(\x)),~~\textrm{on}~~ \del\Omega,
\label{eq:po}
\end{equation}
where $g_i$'s are the face-to-face identification maps 
defined on the boundary. The boundary
conditions constrain the number of unknown constants to $N$. 
Application of the boundary condition (\ref{eq:po}) to Eq.(\ref{eq:BEM})
and permutation of the columns of the 
components yields
\begin{equation}
[A]\{x\}=0, \label{eq:Ax}
\end{equation}
where  $N\times N$-dimensional matrix $A$ is constructed from
$G_{ij}$ and $H_{ij}$ and $N$-dimensional vector 
$x$ is constructed from $u_i$'s and $q_i$'s. For the presence of 
the non-trivial 
solution, the following relation must hold,
\begin{equation}
\mathrm{det}[A]=0. \label{eq:dA}
\end{equation}
Thus, the eigenvalues of the Laplace-Beltrami operator acting on the
space $C^1(\bar{\Omega})\cap C^2(\Omega)\cap L^2(\Omega)$ are obtained by
searching for $k$'s which satisfy Eq.(\ref{eq:dA}). 
\subsection{Computation of low-lying eigenmodes}

In this section, we apply the DBEM for computing
the low-lying eigenmodes of CH 3-manifolds.    
The Helmholtz equation in the Poincar$\acute{\textrm{e}}$ coordinates 
is written as
\begin{equation}
\f{1}{4}\biggl (1-|\x|^2\biggr )^2 \Biggl
[\Delta_E+ \f{2}{1-|\x|^2}~~ \x \cdot \nabla_E  \Biggr ] u + k^2 u=0,
\label {eq:Hel}  
\end{equation}
where $\Delta_E$ and $\nabla_E$ are the Laplacian and the gradient 
on the corresponding three-dimensional Euclidean space, respectively.
Note that we have set the curvature radius $R=1$ without loss of 
generality. By using the DBEM, the Helmholtz equation (\ref{eq:Hel}) 
is converted to an integral representation on the boundary. 
Here Eq.(\ref{eq:bem})  can
be written in terms of Euclidean quantities as
\begin{equation}
\f{1}{2}u(\y)
+\int_{\del\Omega} G_k(\x,\y) \f{\del u}{\del x^i}\, n_E^i\,
  dS
-\int_{\del\Omega} \f{\del G_k(\x,\y)}{\del x^i}\,u \,n_E^i\,
 dS=0,
\label{eq:bem2}
\end{equation}
where $dS=2(1-|\x|^2)^{-1}\, dS_E$.
The fundamental solution is given as \cite{Els,Tom}
\begin{equation}
G_k\,(\x,\y)=-\f{1}{4 \pi}\,\,
\f{\Bigl( \sigma+\sqrt{\sigma^2-1}\Bigr)^{-s}}{\sqrt{\sigma^2-1}},~~~
-\f{\pi}{2} < \mathrm{arg}\,\, s \leq \f{\pi}{2},
\end{equation}  
where $s=\sqrt{1-k^2}$ and ~$\sigma=\cosh d(\x,\y)$. 
Then Eq.(\ref{eq:bem2}) is discretized on the boundary elements
$\Gamma_J$ as
\begin{equation}
\f{1}{2}u(\x_I)
+\sum_{J=1}^N \Biggl[ \int_{\Gamma_J} G_k(\x_I,\y_J) 
\f{\del u(\y_J)}{\del n}\, dS
-\int_{\Gamma_J} \f{\del G_k(\x_I,\y_J)}{\del n}u(\y_J)\, dS\Biggr]=0,
\label{eq:bem2d}
\end{equation}
where $N$ denotes the number of the boundary elements.  An example of 
$N\!=\!1168$ elements on
the boundary of the fundamental
domain in the Poincar$\acute{\textrm{e}}$ coordinates is shown in
figure \ref{fig:1168BE}. These elements are firstly generated in Klein
coordinates in which the mesh-generation is convenient.  
\BF
\centerline{\psfig{figure=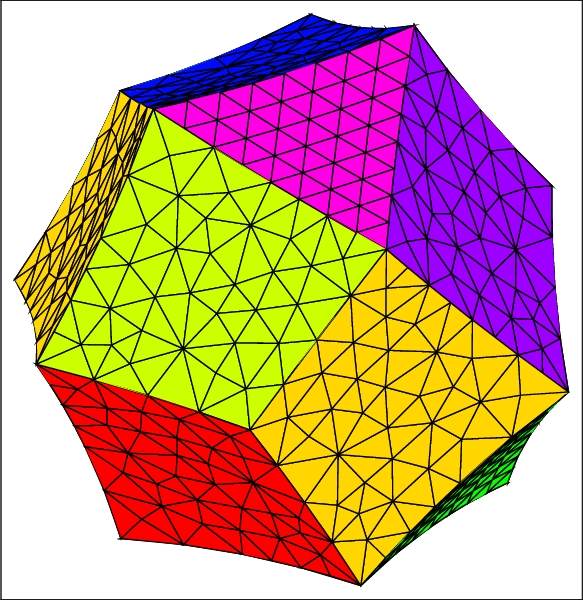,width=15cm}}
\caption{1168 Boundary Elements}
\label{fig:1168BE}
\EF
The maximum length of the edge $\Delta l$ in 
these elements is 0.14. The condition that
the corresponding de Broglie wavelength $2 \pi/k$ is longer than 
the four times of
the interval of the boundary elements yields a rough estimate of
the validity condition of the calculation as $k\!<\!11$. On each
$\Gamma_J$, u and q $\equiv \del u/\del n$ are approximated by
low order polynomials. For simplicity, we use constant elements:
\begin{equation}
u(\x_J)=u^J=Const.\,,~~~~q(\x_J)=q^J=Const.\,,~~~ \mathrm{on}~~ \Gamma_J.\label{eq:qu}
\end{equation}  
Substituting Eq.(\ref{eq:qu}) into Eq.(\ref{eq:bem2d}), we obtain
\begin{eqnarray}
\sum_{J=1}^N H_{IJ} u^J &=& \sum_{J=1}^N G_{IJ} q^J,
\nonumber
\\
H_{IJ}&=&\left\{
\begin{array}{@{\,}ll}
\T{H}_{IJ} & \mbox{$I \neq J$}\\
\T{H}_{IJ}-\f{1}{2} & \mbox{$I=J$}, 
\nonumber
\end{array}
\right.
\end{eqnarray}
where 
\begin{equation}
\T{H}_{IJ}\equiv \int_{\Gamma_{J}}\f{\del G_k}{\del n}
(\x_I,\y_J) \,dS(\y_J),~~G_{IJ}\equiv  \int_{\Gamma_{J}}
G_k(\x_I,\y_J) \,dS(\y_J). \label{eq:bemIJ}
\end{equation}
The singular integration must be carried out
for I-I components as the fundamental solution diverges at
($\x_{I}=\y_{I}$). This is not an intractable problem. Several numerical 
techniques have already been proposed by some authors \cite{Tel,Hay}. 
We have applied Hayami's method 
to the  evaluation of  the singular integrals \cite{Hay}. Introducing
coordinates similar to spherical coordinates 
centered at $\x_I$, the singularity is 
canceled out by the Jacobian which makes the integral regular.    
\\
\indent
Let $g_i\,{(i\!=\!1,2,\ldots,8)}$ be the generators of the discrete 
group $\Gamma$ which identify a boundary face $F_i$ 
with another boundary face $g_i(F_i)$: 
\begin{equation}
g_i(\x_i)=\x_i,~~~~~~~~~~~ \x_i \in F_i.
\end{equation}
The boundary of the fundamental domain can be divided into two
regions $\del \Omega_A$ and $\del \Omega_B$ and each of them consists
of $N/2$ boundary elements,
\begin{equation}
 \del \Omega_A={\cup F_i},~~~~~~~~ \del \Omega_B=
{\cup g_i(F_i)},~~i=1,2, \ldots,8.
\end{equation}
The periodic
boundary conditions 
\begin{equation}
u(g_i(\x_i))=u(\x_i),~~~~q(g_i(\x_i))=-q(\x_i),~~~~~~i=1,2,\ldots ,8
\end{equation}   
reduce the number of the independent variables to N, \textit{i.e.}
for all $\x_B \in \del \Omega_B$, there exist $g_i \in \Gamma$ and
$\x_A \in \del \Omega_A$ such that 
\begin{equation}
u(\x_B)=u(g_i(\x_A))=u(\x_A),~~~~q(\x_B)=
-q(g_i(\x_A))=-q(\x_A).
\end{equation}  
Substituting the above relation into Eq.(\ref{eq:bemIJ}), we obtain
\begin{equation}
\left[
\begin{array}{@{\,}cc@{\,}}
H_{AA}+H_{AB} & -G_{AA}+G_{AB}\\ 
H_{BA}+H_{BB} & -G_{BA}+G_{BB } 
\end{array}
\right]
\left\{
\begin{array}{@{\,}c@{\,}}
u_A\\ 
q_A
\end{array}
\right\}
=0, \label{eq:uq}
\end{equation}
where $u_A=(u^1,u^2, \ldots u^{N/2})$ and $q_A=(q^1,q^2, \ldots
q^{N/2})$
and
matrices $H=\{H_{I J}\}$ and $G=\{G_{I J}\}$ are written as 
\begin{equation}
H=
\left[
\begin{array}{@{\,}cc@{\,}}
H_{AA}& H_{AB}\\ 
H_{BA}& H_{BB} 
\end{array}
\right],~~~
G=
\left[
\begin{array}{@{\,}cc@{\,}}
G_{AA}& G_{AB}\\ 
G_{BA}& G_{BB} 
\end{array}
\right].
\end{equation}
Eq. (\ref{eq:uq}) takes the form
\begin{equation}
[A(k)]\{x\}=0, \label{eq:Ax2}
\end{equation}
where $N \times N$-dimensional matrix $A$ is constructed from  $G$ and
$H$ and $N$-dimensional vector $x$ is constructed from $u_A$ and $q_A$.
For the presence of the non-trivial solution, the following relation 
must hold,
\begin{equation}
\mathrm{det}[A(k)]=0. \label{eq:detA}
\end{equation}
Thus the eigenvalues of the Laplace-Beltrami operator in a
CH space are obtained by
searching for $k$'s which satisfy Eq.(\ref{eq:detA}). 
In practice, Eq.(\ref{eq:detA}) cannot be exactly satisfied as the
test function which has a locally polynomial behavior is slightly
deviated from the exact eigenfunction. 
Instead, one must search for 
the local minima of \textrm{det}[A(k)]. This process needs long
computation time as $A(k)$ depends on k implicitly. Our numerical
result ($k\!<\!13$) is shown in table \ref{tab:eg}.
\begin{table}
\begin{center}
\begin{tabular}{cccc|cccc}  
\hline \hline
\multicolumn{4}{c}{Weeks}&\multicolumn{4}{|c}{Thurston} 
\\ \hline
\multicolumn{1}{c}{$k$} &
\multicolumn{1}{c}{$m(k)$} &
\multicolumn{1}{c}{$k$} &
\multicolumn{1}{c|}{$m(k)$} &
\multicolumn{1}{c}{$k$} &
\multicolumn{1}{c}{$m(k)$} &
\multicolumn{1}{c}{$k$} &
\multicolumn{1}{c}{$m(k)$} 
\\ \hline
5.268&1&  11.283&1& 5.404&1& 10.686&2\\ \hline
5.737&2&  11.515&1& 5.783&1& 10.737&1 \\ \hline
6.563&1&  11.726&4& 6.807&2&  10.830&1\\ \hline
7.717&1&  12.031&2& 6.880&1&  11.103&2\\ \hline
8.162&1&  12.222&2& 7.118&1&  11.402&1\\ \hline
8.207&2&  12.648&1& 7.686&2&  11.710&1\\ \hline
8.335&2&  12.789&1& 8.294&1&  11.728&1\\ \hline
9.187&1& & &         8.591&1&  11.824&1\\ \hline  
9.514&1& & &         8.726&1&  12.012&2\\ \hline  
9.687&1& & &         9.246&1&  12.230&1\\ \hline 
9.881&2& & &         9.262&1&  12.500&1\\ \hline  
10.335&2& & &        9.754&1&  12.654&1\\ \hline  
10.452&2& & &        9.904&1&  12.795&1\\ \hline  
10.804&1& & &        9.984&1&  12.806&1\\ \hline 
10.857&1& & &        10.358&1& 12.897&2\\ \hline 
\hline
\end{tabular}
\caption{ Eigenvalue k and Multiplicity $m(k)$} 
\label{tab:eg}
\end{center} 
\end{table}
\\
\indent
The first ''excited state'' that corresponds to $k\!=\!k_1$ is 
important for the
understanding of CMB anisotropy. Our numerical result $k_1\!=\!5.41$
is consistent with the value $5.04$ obtained from Weyl's
asymptotic formula 
\begin{equation}
N[\nu]=\f{V\!o\,l ({\cal M})\nu^3}{6 \pi^2},~~\nu\equiv \sqrt{k^2-1},
~~~~\nu\!>\!>\!1,
\end{equation} 
assuming that no degeneracy occurs. 
One can interpret the first excited state as the mode that has 
the maximum de Broglie wavelength $2 \pi/k_1$. Because of the 
periodic boundary conditions, the de Broglie wavelength  
can be approximated by the 
''average diameter'' of the fundamental domain defined as a sum of
the inradius $r_-$ and the outradius $r_+$\footnote{The inradius 
$r_-$ is the radius of the largest
simply-connected sphere in the fundamental domain, and the outradius $r_+$
is the radius of the smallest sphere that can enclose the
fundamental domain. $r_-\!=\!0.535$, $r_+\!=\!0.7485$ for
the Thurston manifold.},
which yields $k_1=4.9$ just 
$10\!\%$ less than the numerical value. From these estimates,
supercurvature modes in small CH spaces ($Vol({\cal M})\sim 1$) 
are unlikely to be observed.
\\
\indent
To compute the value of eigenfunctions inside the fundamental domain,
one needs to solve Eq.(\ref{eq:Ax2}). The singular decomposition
method is the most suitable numerical method for solving any linear
equation with a singular matrix $A$, which can be 
decomposed as 
\begin{equation}
A=U^{\dagger}D V,
\end{equation}
where $U$ and $V$ are unitary matrices and D is a diagonal matrix. If 
$D_{ii}$ in $D$ is almost zero then the complex conjugate of the i-th
row in V is an approximated solution of Eq.(\ref{eq:Ax2}).    
The number of the ''almost zero'' diagonal elements in D is equal to 
the multiplicity number.  
\BF[t]
\centerline{\psfig{figure=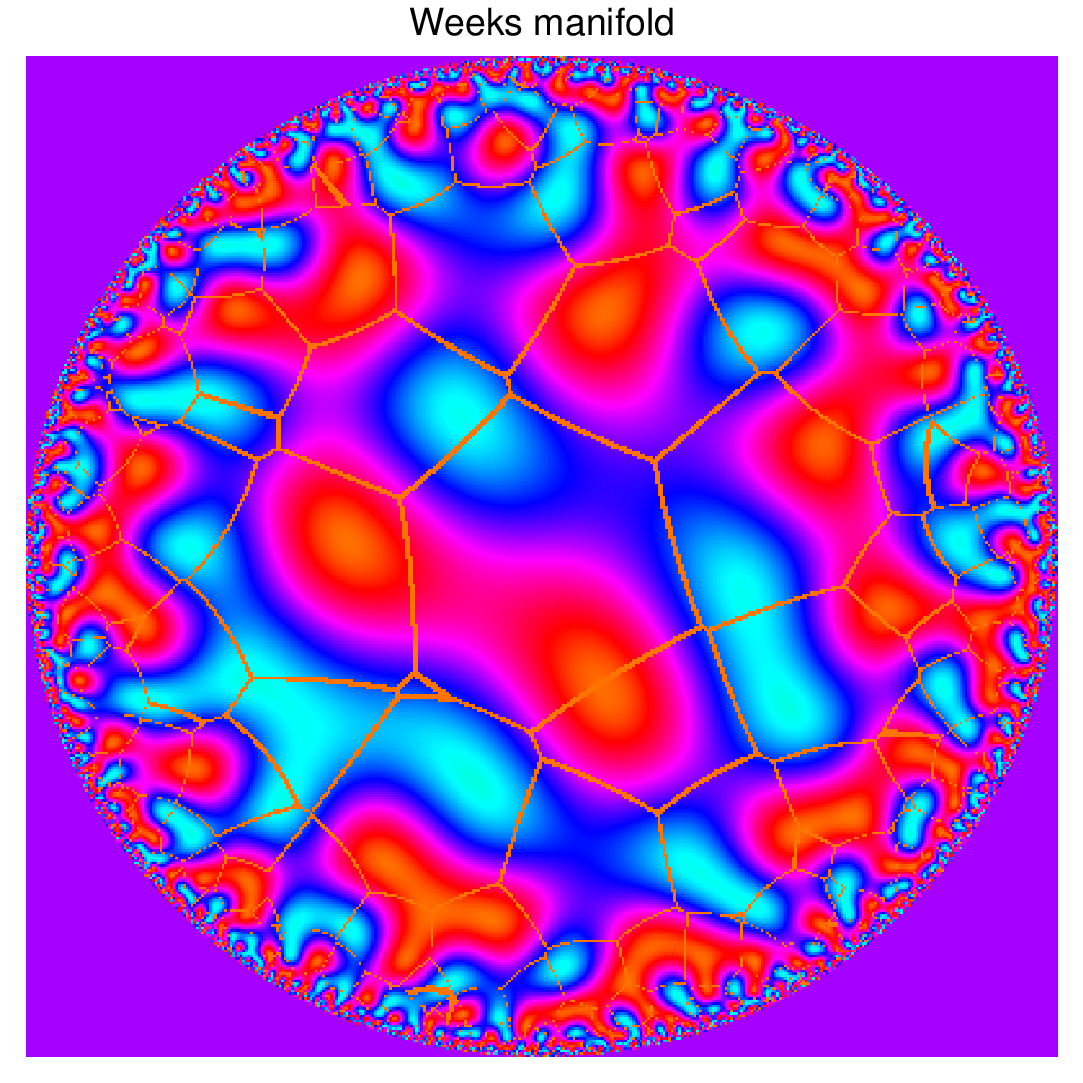,width=8cm}
\psfig{figure=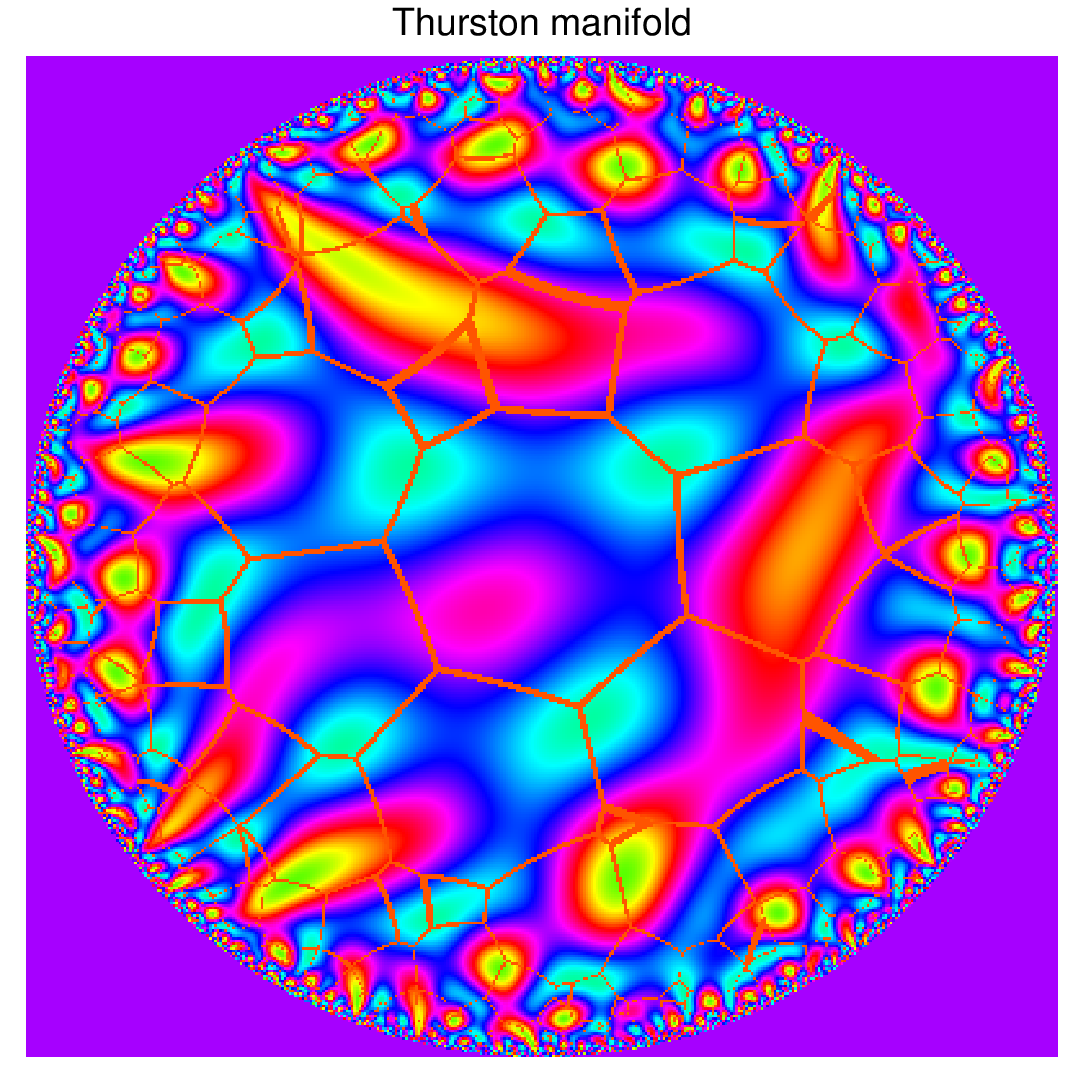,width=8cm}
}
\caption{Lowest Eigenmodes}
\mycaption{The lowest eigenmodes of the 
Weeks manifold $(k\!=\!5.268)$ 
and the Thurston manifold $(k\!=\!5.404)$ 
on a slice $z\!=\!0$ in the Poincar$\acute{\textrm{e}}$ coordinates. 
The boundaries of the copied Dirichlet domains (\textit{solid
curves}) are plotted in full curve.}
\label{fig:EF}
\EF
Substituting the values of the eigenfunctions and their normal 
derivatives on the
boundary into Eq.(\ref{eq:inter}), the values of the 
eigenfunctions inside the fundamental domain can be computed. 
Adjusting the normalization factors, one would obtain real-valued 
eigenfunctions. Note that 
non-degenerated eigenfunctions must be always real-valued.  
\\
\indent
The numerical accuracy of the obtained eigenvalues is roughly estimated as
follows. First, let us write the obtained numerical solution in terms
of the exact solution as $k=k_0+\delta k$ 
and $u_k(\x)=u_{k_0}(\x)+\delta u_k(\x)$,
where $k_0$ and $u_{k_0}(\x)$ are the exact eigenvalue and
eigenfunction, respectively. The singular decomposition method
enables us to find the best approximated solution which satisfies
\BE
[A]\{x\}=\epsilon, ~~~|\epsilon | <<1, 
\label{eq:epsilon}
\EE
where $\epsilon$ is a N-dimensional vector and $|~ |$ denotes
the Euclidean norm. It is expected that 
the better approximation gives the smaller $|\epsilon |$.
Then Eq. (\ref{eq:epsilon}) can be written as, 
\BE
\int_{\Omega} G_{k_0+\delta k}(\x,\y_J) (\Delta+(k_0+\delta k)^2)
(u_{k_0}(\x)+\delta u_k(\x))\,\sqrt{g}\, dV_\x=\epsilon(\y_J).
\label{eq:eps}
\EE
Ignoring the terms in second order, Eq.(\ref{eq:eps}) is reduced to 
\BE
\int_{\Omega} G_{k}(\x,\y_J)((\Delta+k_0^2)\delta u_k(\x)
+2 k \delta k u_{k}(\x))\,\sqrt{g}\, dV_\x=\epsilon(\y_J).
\EE
Since it is not unlikely that $(\Delta+k_0^2)\delta u_k(\x)$
is anticorrelated to $2 k \delta k u_{k}(\x)$, 
we obtain the following relation by averaging over $\y_J$,
\BE
2 k |\delta k| \Biggl <\biggl|\int_{\Omega} G_{k}(\x,\y_J)u_{k}(\x)\,
\sqrt{g}\, dV_\x \biggr|\Biggr> \sim\,\,<|\epsilon|>,
\EE
where $< >$ denotes the averaging over $\y_J$  
Thus one can estimate the expected deviation of 
the calculated eigenvalue $|\delta k|$ from $u_k(\x)$ and $\epsilon(\y_J)$.  
We numerically find that $|\delta k|=0.005$ for $k=5.41$ 
and $|\delta k|=0.01$ for
$k=9.91$(Thurston manifolds). 
The other deviation values lie in between $0.005$ and $0.01$. 
\\
\indent
By computing the second derivatives, one
can also estimate the accuracy of the computed eigenfunctions. 
The accuracy parameter $err$ is defined as
\BE
err(k,\x)\equiv (\Delta + k^2) u_k(\x),
\EE
\BF
\centerline{\psfig{figure=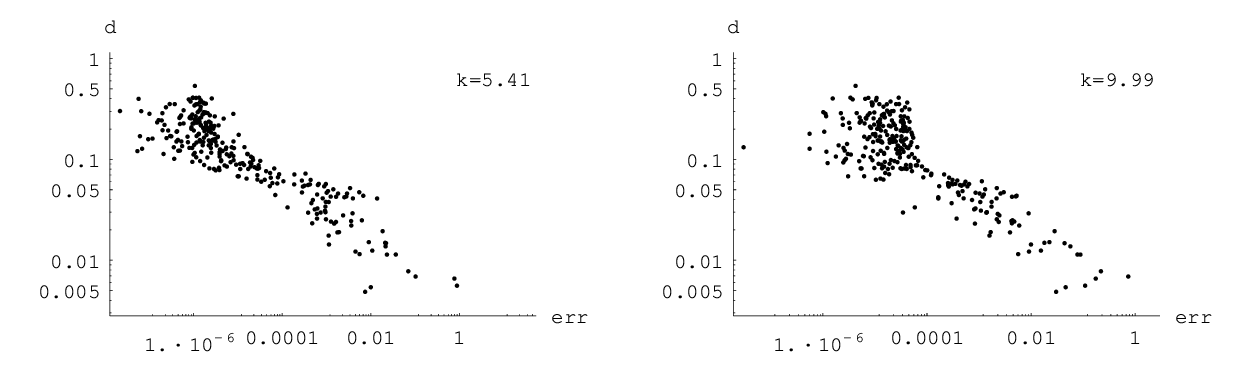,width=17cm}}
\caption{Error Estimation}
\mycaption{Errors ($err$) in terms of hyperbolic distance 
$d$ to the boundary from 291 points in the fundamental domain of 
the Thurston manifold.}
\label{fig:err}
\EF
where $u_k(\x)$ is normalized (${\cal O}(u_k(\x))\sim 1$). 
We see from figure 
\ref{fig:err} that the accuracy becomes worse
as the evaluation point approaches the boundary. However, for points with 
hyperbolic distance $d>0.1$ between the evaluating point and the
nearest boundary, the errors are very small 
indeed: $err=
10^{-4\sim -5}$. This result is considered to be 
natural because the characteristic 
scale $L$ of the boundary elements is $\sim 0.07$ for our 1168 elements. 
If $d<L$, the integrands in Eq.(\ref{eq:inter}) become appreciable on
the neighborhood of the nearest boundary
point because the free Green's
function approximately diverges on the point. 
In this case, the effect of the deviation from the exact 
eigenfunction is significant. If $d>>L$, the integrand on 
all the boundary points contributes almost equally to the 
integration so that the local deviations are cancelled out.  
\\
\indent
As we shall see in the next section, expansion coefficients
are calculated using the values of eigenfunctions on a sphere. 
Since the number of evaluating points which
are very close to the boundary is negligible on the sphere, 
expansion coefficients can be computed with
relatively high accuracy.
\subsection{Statistical property of eigenmodes}
In pseudospherical coordinates ($R,\chi,\theta,\phi$), 
the eigenmodes are written in terms of complex expansion 
coefficients $\xi_{\nu l m}$
and eigenmodes on the universal covering space, 
\BE
u_\nu=\sum_{l m} \xi_{\nu l m}\,X_{\nu l}(\chi) Y_{l m}(\theta,\phi),
\label{eq:u}
\EE
where $\nu=\sqrt{k^2-1}$ and $Y_{l m}$ is a (complex) 
spherical harmonic. The radial eigenfunction $X_{\nu l}$ 
is written in terms of the associated Legendre function $P$, 
and the $\Gamma$ function as
\BE
X_{\nu l}
=\f{\Gamma  (l+1+\nu i)}{\Gamma (\nu i)} \sqrt{\f{1}{\sinh \chi}}
P^{-l-1/2}_{\nu i-1/2}(\cosh
\chi),~~~~\nu^2=k^2-1.
\EE
Then the real expansion 
coefficients $a_{\nu l m}$ are given by
\begin{eqnarray}
\nonumber
a_{\nu 0  0}&=&-\textrm{Im}(\xi_{\nu 0  0}),~~~~
a_{\nu l  0}=\sqrt{c_{\nu l}}\textrm{Re}(\xi_{\nu l 0}),
\\
\nonumber
a_{\nu l  m}&=&\sqrt{2}\textrm{Re}(\xi_{\nu l m}),~~m>0,
\\
a_{\nu l  m}&=&-\sqrt{2}\textrm{Im}(\xi_{\nu l -m}),~~m<0,
\end{eqnarray}
where
\begin{eqnarray}
\nonumber
c_{\nu l}&=&\f{2}{(1+\textrm{Re}(F(\nu,l)))},
\\
F(\nu, l)&=&\f{\Gamma(l+\nu i+1)}{\Gamma(\nu i)}
\f{\Gamma(-\nu i)}{\Gamma(l-\nu i+1)}.
\end{eqnarray}
\\
\indent
$a_{\nu l m}$ 's can be promptly obtained after the normalization and
orthogonalization of these eigenmodes. The orthogonalization is
achieved at the level of $10^{-3}$ to $10^{-4}$ (for the inner product
of the normalized eigenmodes) which implies that each eigenmode is 
computed with relatively high accuracy.
\BF[tpb]
\centerline{\psfig{figure=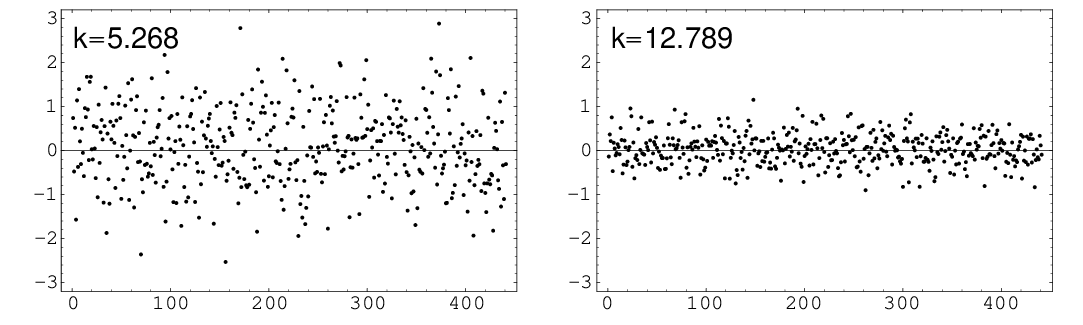,width=16cm}}
\caption{Random Nature of Expansion Coefficients} 
\mycaption{Expansion coefficients $a_{\nu l m}$ are plotted 
in ascending order as $l(l+1)+m+1, 0\le l \le 20$ for eigenmodes 
$k\!=\!5.268$(left) and $k\!=\!12.789$(right) on the Weeks manifold
at a point which is randomly chosen.}
\label{fig:anulmW}
\EF
\\
\indent
In figure \ref{fig:anulmW} one can see that the distribution of 
$a_{\nu l m}$'s, which are ordered as $l(l+1)+m+1$ 
are qualitatively random.
In order to estimate the randomness 
quantitatively, we consider a cumulative distribution of 
\begin{equation}
b_{\nu l m}=\f{|a_{\nu l m}-\bar{a}_{\nu}|^2}{\sigma_{\nu}^2}
\end{equation}
where $\bar{a}_{\nu}$ is the mean of $a_{\nu
l m}$'s and $\sigma_{\nu}^2$ is the variance.
If $a_{\nu l m}$'s are Gaussian then $b_{\nu l m}$ 's obey
a $\chi^2$ distribution $P(x)=(1/2)^{1/2}\Gamma(1/2)x^{-1/2}e^{-x/2}$
with 1 degree of freedom. To test the goodness of fit between the
the theoretical cumulative distribution $I(x)$ and the empirical 
cumulative distribution function $I_N(x)$, we use 
the Kolmogorov-Smirnov statistic $D_N$ which is defined as 
the least upper bound of all pointwise differences $|I_N(x)-I(x)|$
\cite{Hog},
\begin{equation}
D_N\equiv \sup_{x} |I_N(x)-I(x)|,
\end{equation}
where $I_N(x)$ is defined as
\begin{eqnarray}
 I_N(x)&=&\left\{ \begin{array}{@{\,}ll}
0, & x<y_1,
\\
j/N,~~& y_j \leq x < y_{j+1},~~~~j=1,2,\ldots,N\!-\!1,
\\
1, & y_N \leq x, 
\end{array}
\right. 
\end{eqnarray}
and $y_1<y_2< \ldots <y_N$ are the computed values of a random
sample which consists of $N$ elements. 
For random variables $D_N$ for any $z>0$, it can be shown that 
the probability of $D_N\!<\!d$ is given by \cite{Bir}
\begin{equation}
\lim_{N \rightarrow \infty} ~P(D_N<d=z N^{-1/2})=L(z),\label{eq:P} 
\end{equation}  
where
\begin{equation}
L(z)=1-2 \sum_{j=1}^{\infty} (-1)^{j-1} e^{-2j^2 z^2}.
\end{equation}
From the observed maximum difference $D_N\!=\!d$, we obtain 
the significance level $\alpha_D\!=\!1-P$, which is equal to the probability 
of $D_N\!>\!d$. If $\alpha_D$ is found to be large enough, 
the hypothesis $I_N(x)\!\neq\! I(x)$ is not verified. 
The significance levels $\alpha_D$ for $0 \!\leq\! l \!\leq\! 20$ and
for eigenmodes $k\!<\!13$ on the Thurston manifold are shown in table 
\ref{tab:KScenter}. 
\begin{table}
\begin{center}
\begin{tabular}{cccc}
\hline \hline
\multicolumn{4}{c}{Thurston} 
\\ \hline 
\multicolumn{1}{c}{k} &
\multicolumn{1}{c}{$\alpha_D$} &
\multicolumn{1}{c}{k} &
\multicolumn{1}{c}{$\alpha_D$} 
\\ \hline
5.404&0.98          &10.686b& $7.9\times 10^{-4}$ \\ \hline
5.783&0.68          &10.737 & 0.96 \\ \hline
6.807a &0.52       &10.830 & 0.67 \\ \hline
6.807b & $7.1\times 10^{-4}$  &11.103a & 0.041 \\ \hline
6.880&1.00          &11.103b & $8.8\times 10^{-15}$ \\ \hline
7.118&0.79          &11.402 & 0.98 \\ \hline
7.686a&0.26         &11.710 & 0.92 \\ \hline
7.686b& $2.3\times 10^{-8}$ &11.728 & 0.93 \\ \hline
8.294&0.45          &11.824 & 0.31 \\ \hline
8.591&0.91          &12.012a &0.52 \\ \hline
8.726&1.00          &12.012b &0.73 \\ \hline  
9.246&0.28          &12.230 &0.032 \\ \hline  
9.262&0.85          &12.500 &0.27 \\   \hline  
9.754&0.39          &12.654 & 0.88 \\ \hline  
9.904&0.99          &12.795 &0.76 \\ \hline  
9.984&0.20          &12.806 &0.42 \\ \hline  
10.358&0.40         &12.897a &0.87  \\ \hline        
10.686a&0.76        &12.897b &$6.9 \times 10^{-4}$
\\ \hline 
\hline
\end{tabular}
\caption{Kolmogorov-Smirnov Test I}
\mycaption{Eigenvalues $k$ and the corresponding 
significance levels $\alpha_D$ for the test of the hypothesis
 $I_N(x) \neq I(x)$ for the Thurston manifold are listed. 
The injectivity radius 
is maximal at the base point.} 
\label{tab:KScenter}
\end{center} 
\end{table}
The agreement with the RMT prediction is
fairly good for most of eigenmodes which is consistent with the
previous computation in \cite{Inoue1}. However, for five degenerated modes, 
the non-Gaussian signatures are prominent (in \cite{Inoue1}, two modes 
in ($k\!<\!10$) have been missed). Where does this non-Gaussianity come
from? 
\\
\indent
First of all, we must pay attention to the fact that the
expansion coefficients $a_{\nu l m}$ depend on the observing point.
In mathematical literature, the point is called the \textit{base point}. 
For a given base point, it is possible to construct a particular
class of fundamental domain called the \textit{Dirichlet}
(\textit{fundamental}) \textit{domain} which is a convex polyhedron.
A Dirichlet domain $\Omega(x)$ centered at a base point $x$ is defined as
\BE
\Omega(x)=\bigcap_{g} H(g,x)~~~,H(g,x)=\{z|d(z,x)<d(g(z),x)\},
\EE
where $g$ is an element of a Kleinian group $\Gamma$(a discrete isometry
 group of $PSL(2,{\mathbb C})$) 
and $d(z,x)$ is the proper distance between $z$ and $x$. 
\\
\indent 
The shape of the Dirichlet domain depends on the base point but the
volume is invariant. Although the base point can be chosen arbitrarily, 
it is a standard to choose a point $Q$ 
where the injectivity radius\footnote{The injectivity radius $r_{inj}$ 
at a point $p$ is equal to the 
radius of the largest connected and simply connected 
ball centered at $p$ which does not cross
itself.} is locally maximal. 
More intuitively, $Q$ is a center where one
can put a largest connected ball on the manifold.  
If one chooses other point as the base point, the nearest copy of the base
point can be much nearer. The reason to choose $Q$ as a base point is 
that one can expect the corresponding Dirichlet domain to have 
many symmetries at $Q$ \cite{Weeks2}.
\\
\indent
\begin{figure}[tpb]
\centerline{\psfig{figure=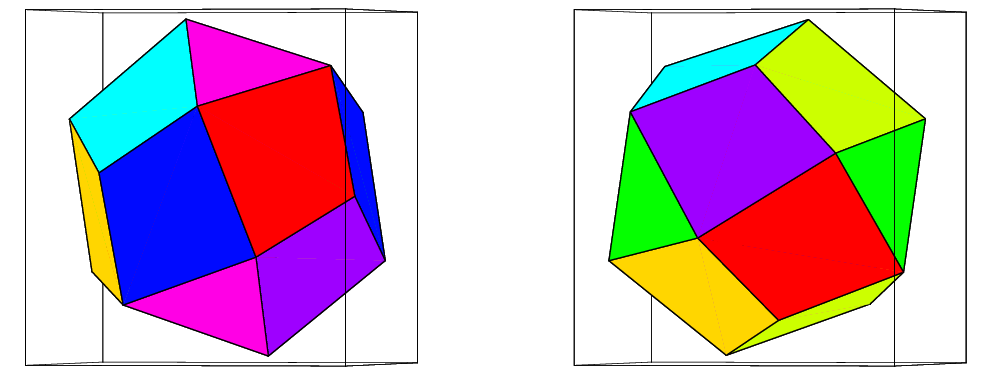,width=16cm}}
\caption{$Z^2$ Symmetry}
\mycaption{A Dirichlet domain of the Thurston manifold
in the Klein coordinates viewed from opposite directions at $Q$
where the injectivity radius is locally maximal. The Dirichlet domain
has a $Z2$ symmetry(invariant by $\pi$-rotation)at $Q$.}
\label{fig:SYMFDT}
\end{figure}
As shown in figure \ref{fig:SYMFDT} the Dirichlet domain at $Q$ has 
a $Z2$ symmetry (invariant by $\pi$-rotation) 
if all the congruent faces are identified. Generally, 
congruent faces 
are distinguished but it is found that these five modes have exactly the same
values of eigenmodes on these congruent faces. Then one can no longer
consider $a_{\nu l m}$'s as ''independent'' random numbers. Choosing    
the invariant axis by the $\pi$-rotation as the $z$-axis,
$a_{\nu l m}$'s are zero for odd $m$'s, leading 
to a non-Gaussian behavior.
It should be noted that the observed $Z2$ symmetry is not the subgroup
of the isometry group (or \textit{symmetry group} 
in mathematical literature)
$D2$ (dihedral group with order $2$) of the Thurston manifold 
since the congruent faces must be actually distinguished in the 
manifold\footnote{The observed $Z2$
symmetry might be a ``hidden symmetry'' 
which is a symmetry of the finite sheeted cover of the manifold (which
can be tessellated by the manifold).}. 
\\
\indent 
Thus the observed non-Gaussianity is caused by a particular choice of the
base point. However, in general, the chance that we actually observe any
symmetries (elements of the isometry group of the manifold or the finite
sheeted cover of the manifold) is expected to be 
very low. Because a fixed point by an element of the isometric group
is either a part of 1-dimensional line (for instance, an axis of a
rotation) or an isolated point (for instance, a center of an antipodal 
map). 
\\
\indent
In order to confirm that the chance is actually low, the KS statistics
$\alpha_D$ of $a_{\nu l m}$'s are computed at 300 base points which are 
randomly chosen. \begin{table}
\begin{center}
\begin{tabular}{cccp{22mm}|cccc}
\hline \hline   
\multicolumn{4}{c|}{Weeks}&\multicolumn{4}{c}{Thurston} 
\\ \hline
\multicolumn{1}{c}{k} &
\multicolumn{1}{c}{$<\alpha_D>$} &
\multicolumn{1}{c}{k} &
\multicolumn{1}{p{22mm}|}{$<\alpha_D>$} &
\multicolumn{1}{c}{k} &
\multicolumn{1}{c}{$<\alpha_D>$} &
\multicolumn{1}{c}{k} &
\multicolumn{1}{c}{$<\alpha_D>$} 
\\ \hline
5.268&0.58& 10.452b &~~0.62&  5.404&0.63 &10.686b& 0.62 \\ \hline
5.737a&0.61& 10.804&~~0.63&5.783&0.61 &10.737 & 0.62 \\ \hline
5.737b&0.61& 10.857&~~0.62&6.807a &0.62 &10.830 &  0.63\\ \hline
6.563&0.62 & 11.283&~~0.57&6.807b &0.62 &11.103a &0.59 \\ \hline
7.717&0.59&  11.515&~~0.61&6.880&0.63 &11.103b &0.60\\ \hline
8.162&0.61&  11.726a&~~0.63&7.118&0.61 &11.402 & 0.61 \\ \hline
8.207a&0.65& 11.726b&~~0.59&7.686a&0.61 &11.710 & 0.62 \\ \hline
8.207b&0.61& 11.726c&~~0.61&7.686b&0.63 &11.728 &0.64 \\ \hline
8.335a&0.59& 11.726d&~~0.61&8.294&0.60  &11.824 &0.62  \\ \hline
8,335b&0.62& 12.031a&~~0.60&8.591&0.60  &12.012a &0.63 \\ \hline
9.187&0.59&  12.031b&~~0.60&8.726&0.60  &12.012b &0.61 \\ \hline  
9.514&0.56&  12.222a&~~0.61&9.246&0.60  &12.230 &0.60 \\ \hline  
9.687&0.61&  12.222b&~~0.62&9.262&0.63  &12.500 &0.63 \\   \hline  
9.881a&0.61& 12.648&~~0.59&9.754&0.62  &12.654 & 0.62 \\ \hline  
9,881b&0.62& 12.789&~~0.59&9.904&0.60  &12.795 &0.62\\ \hline  
10.335a&0.63&  & &          9.984&0.60  &12.806 &0.62 \\ \hline  
10.335b&0.60&  & &          10.358&0.62 &12.897a &0.62  \\ \hline        
10.452a&0.63&  & &           10.686a&0.60 &12.897b &0.56
\\ \hline
\hline
\end{tabular}
\caption{Kolmogorov-Smirnov Test II}
\mycaption{Eigenvalues $k$ and corresponding 
averaged significance levels $<\alpha_D>$ based on 300 realizations
of the base points for the test of the hypothesis
 $I_N(x) \neq I(x)$ for the Weeks and the Thurston manifolds.} 
\label{tab:KSII}
\end{center} 
\end{table}
As shown in table (\ref{tab:KSII}), 
the averaged significance levels $<\!\alpha_D\!>$
are remarkably consistent with the 
Gaussian prediction. $1\sigma$ 
of $\alpha_D$ are found to be 0.26 to 0.30. 
\\
\indent
Next, we apply the run test for testing the randomness of $a_{\nu l
m}$'s where each set of $a_{\nu l m}$ 's are ordered as $l(l+1)+m+1$ 
(see \cite{Hog}). Suppose that we have $n$ observations
of the random variable $U$ which falls above the median and n 
observations of the random variable
$L$ which falls below the median. 
The combination of those variables into $2 n$ observations
placed in ascending order of magnitude yields
\begin{center}
\large\textit{
\U{UUU} \U{LL} \U{UU} \U{LLL} \U{U} \U{L} \U{UU} \U{LL}},
\end{center}
Each underlined group which consists of successive 
values of $U$ or $L$ is called \textit{run}. The total number of 
run is called the \textit{run number}.
The run test is useful because the run number 
always obeys the Gaussian statistics in the limit
$n\!\rightarrow\!\infty$ regardless of the type of the
distribution function of the random variables.
\begin{table}
\begin{center}
\begin{tabular}{cccp{22mm}|cccc}
\hline \hline   
\multicolumn{4}{c|}{Weeks}&\multicolumn{4}{c}{Thurston} 
\\ \hline 
\multicolumn{1}{c}{k} &
\multicolumn{1}{c}{$<\alpha_r>$} &
\multicolumn{1}{c}{k} &
\multicolumn{1}{p{22mm}|}{$<\alpha_r>$} &
\multicolumn{1}{c}{k} &
\multicolumn{1}{c}{$<\alpha_r>$} &
\multicolumn{1}{c}{k} &
\multicolumn{1}{c}{$<\alpha_r>$} 
\\ \hline
5.268&0.51& 10.452b &~~0.52&  5.404&0.48 &10.686b& 0.51 \\ \hline
5.737a&0.48& 10.804&~~0.52&5.783&0.45 &10.737 & 0.49 \\ \hline
5.737b&0.45& 10.857&~~0.53&6.807a &0.53 &10.830 &0.53\\ \hline
6.563&0.54 & 11.283&~~0.49&6.807b &0.50 &11.103a &0.52 \\ \hline
7.717&0.50&  11.515&~~0.51&6.880&0.47 &11.103b &0.53\\ \hline
8.162&0.54&  11.726a&~~0.51&7.118&0.50 &11.402 & 0.51 \\ \hline
8.207a&0.52& 11.726b&~~0.48&7.686a&0.49 &11.710 & 0.51 \\ \hline
8.207b&0.49& 11.726c&~~0.49&7.686b&0.52 &11.728 &0.49 \\ \hline
8.335a&0.53& 11.726d&~~0.48&8.294&0.50 &11.824 &0.54  \\ \hline
8,335b&0.50& 12.031a&~~0.54&8.591&0.50  &12.012a &0.51 \\ \hline
9.187&0.53&  12.031b&~~0.51&8.726&0.51  &12.012b &0.49 \\ \hline  
9.514&0.55&  12.222a&~~0.54&9.246&0.43  &12.230 &0.51 \\ \hline  
9.687&0.53&  12.222b&~~0.50&9.262&0.50  &12.500 &0.48 \\   \hline  
9.881a&0.51& 12.648&~~0.54&9.754&0.54 &12.654 & 0.48 \\ \hline  
9,881b&0.51& 12.789&~~0.48&9.904&0.52  &12.795 &0.50\\ \hline  
10.335a&0.54&  & &          9.984&0.49  &12.806 &0.51 \\ \hline  
10.335b&0.51&  & &          10.358&0.53 &12.897a &0.57  \\ \hline        
10.452a&0.53&  & &           10.686a&0.51 &12.897b &0.55 \\ \hline
\hline
\end{tabular}
\caption{Run Test}
\mycaption{Eigenvalues $k$ and corresponding 
averaged significance levels $<\!\alpha_r\!>$ for the test of the hypothesis
that the $a_{\nu l m}$'s are not random numbers for the Weeks and
Thurston manifolds. $\alpha_r$'s at 300 points which are randomly
chosen are used for the computation.} 
\label{tab:KSrun}
\end{center} 
\end{table}
As shown in table \ref{tab:KSrun}, 
averaged significance levels  $<\!\alpha_r\!>$ are
very high (1$\sigma$ is 0.25 to 0.31). 
Thus each set of $a_{\nu l m}$'s ordered as $l(l+1)+m+1$ can be
interpreted as a set of Gaussian pseudo-random numbers except for limited
choices of the base point where one can observe symmetries of 
eigenmodes. 
\\
\indent
Up to now, we have considered $l$ and $m$ as the index numbers of $a_{\nu
l m}$ at a fixed base point. However, for a fixed $(l,m)$, the
statistical property of a set of $a_{\nu l m}$'s at a number of 
different base points is also important since the temperature 
fluctuations must be averaged all over the places 
for spatially inhomogeneous models. 
From figure \ref{fig:mKS}, one can see the 
behavior of m-averaged significance levels  
\BE
\alpha_D(\nu, l)\equiv \sum_{m=-l}^{l} \f{\alpha_D(a_{\nu l m})}{2l+1}
\EE
which are calculated based on 300 realizations of the base points.  
It should be noted that each $a_{\nu l m}$ at a particular
base point is now considered to be ''one realization'' whereas
a choice of $l$ and $m$ is considered to be ''one realization'' in
the previous analysis.
The good agreement with the RMT prediction for 
components $l\!>\!1$ has been found. 
For components $l\!=\!1$, the disagreement
occurs for only several modes. However, the non-Gaussian 
behavior is distinct in $l\!=\!0$ components. 
What is the reason of the non-Gaussian 
behavior for $l\!=\!0$? 
\begin{figure}[tpb]
\centerline{\psfig{figure=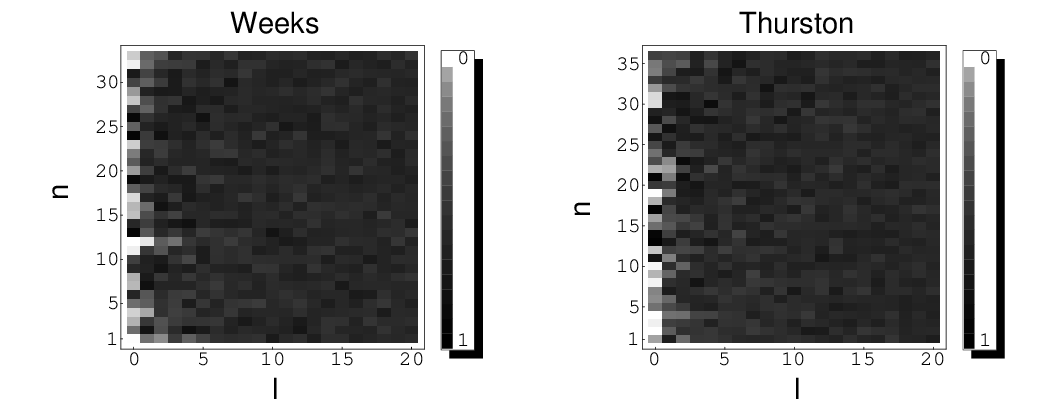,width=16cm}}
\caption{Kolomogorov-Smirnov Test III}
\mycaption{Plots of $m$-averaged significance 
levels $\alpha_D(\nu, l)$ 
based on 300 realizations for the Weeks and the Thurston manifolds
($0 \! \le\! l \!\le\! 20$ and $k<13$). $n$ denotes the index number
which corresponds to an eigenmode $u_k$ where the number of eigenmodes 
less than $k$ is equal to $n$ ($k(n\!=\!1)$ is the lowest non-zero
eigenvalue). The accompanying palettes show the 
correspondence between the level of the grey and the value.}
\label{fig:mKS}
\end{figure}
Let us estimate the values of the expansion coefficients for
$l\!=\!0$. 
In general, the complex expansion coefficients $\xi_{\nu l m}$ 
can be written as,
\BE
\xi_{\nu l m}(\chi_0)=\f{1}{X_{\nu l}(\chi_0)} 
\int u_\nu(\chi_0,\theta,\phi)\, 
Y^*_{l m}(\theta,\phi) d \Omega. \label{eq:xi}
\EE 
For $l\!=\!0$, the equation becomes
\BE
\xi_{\nu 0 0}(\chi_0)=-\f{i}{2\sqrt{2}} \f{\sinh{\chi_0}}{\sin{\nu
 \chi_0}} \int u_\nu(\chi_0,\theta,\phi)\, d \Omega. \label{eq:xi0}
\EE 
Taking the limit $\chi_0 \rightarrow 0$, one obtains,
\BE
\xi_{\nu 0 0}=-\f{2 \pi u_\nu(0) i}{\nu}.
\EE
Thus $a_{\nu 0 0}$ can be written in terms of the value of the
eigenmode at the base point. 
As shown in figure \ref{fig:EF} the lowest eigenmodes have only 
one ''wave'' on scale of the topological identification scale $L$ 
(which will be defined later on) 
inside a single Dirichlet domain which implies that the random
behavior within the domain may be not present. 
Therefore, for low-lying eigenmodes,
one would generally expect non-Gaussianity in a set of 
$a_{\nu 0 0}$ 's. 
However, for high-lying eigenmodes, this may not be the case 
since these modes have a number of ''waves'' on scale of $L$ 
and they may change their values locally in an almost random fashion.
\\
\indent
The above argument cannot be applicable to $a_{\nu l m}$ 's for 
$l \!\neq\! 0$ where ${X_{\nu l}}$ approaches zero in the limit 
$\chi_0\!\rightarrow\! 0$ while the integral term
\BE
 \int u_\nu(\chi_0,\theta,\phi)\, 
Y^*_{l m}(\theta,\phi) d \Omega 
\EE 
also goes to zero because of the symmetric property of the spherical
harmonics. Therefore $a_{\nu l m}$ 's  cannot be written in terms of
the local value of the eigenmode for $l \!\neq\! 0$. 
For these modes, it is better to consider the opposite
limit $\chi_0 \rightarrow \infty$. It is numerically found that 
a sphere with a very large radius $\chi_0$ intersects each copy of 
the Dirichlet domain almost randomly (the 
pulled back surface into a single Dirichlet domain chaotically
fills up the domain). Then the values of the eigenmodes
on the sphere with a very large radius vary in an almost random
fashion. For large $\chi_0$, we have 
\BE
X_{\nu l}(\chi_0)\!\propto\!e^{-2
\chi_0+\phi(\nu, l)i},
\EE
where $\phi(\nu, l)$ describes the phase factor. Therefore, the order
of the integrand in Eq. (\ref{eq:xi}) is approximately $e^{-2
\chi_0}$ since Eq. (\ref{eq:xi}) does not depend on the choice of
$\chi_0$. As the spherical harmonics do not have correlation with 
the eigenmode $u_{\nu}(\chi_0,\theta,\phi)$, the integrand varies
almost randomly for different choices of $(l,m)$ or base points. 
Thus, we conjecture that Gaussianity of $a_{\nu l m}$'s have their origins 
in the chaotic property of the sphere with large radius in CH spaces. 
The property may be related to the 
classical chaos in geodesic flows\footnote{If one considers a great
circle on a sphere with large radius, the length of the circle is
very long except for rare cases in which the circle ``comes back'' 
before it wraps around in the universal covering space. Because the 
long geodesics in CH spaces chaotically (with no particular direction 
and position) wrap through the manifold, it is natural to assume that 
the great circles also have this chaotic property.}  
. 
\\
\indent
Let us now consider the average and variance 
of the expansion coefficients.
As the eigenmodes have oscillatory features, it is natural to expect
that the averages are equal to zero. In fact, the averages of
$<\!a_{\nu l m}\!>$ 's over $0\!\le\! l\!\le\! 20$ and $-l\!\le\!
m\!\le\! -l$ and 300 realizations of base points for each $\nu$-mode
are numerically found to be $0.006\pm 0.04-0.02$(1$\sigma$) 
for the Weeks manifold, 
and $0.003\pm 0.04-0.02$(1$\sigma$) for the Thurston
manifold. Let us next consider the $\nu$-dependence ($k$-dependence) 
of the variances $Var(a_{\nu l m})$. In order to crudely 
estimate the $\nu$-dependence, 
we need the angular size $\delta \theta $
of the characteristic length of the eigenmode $u_{\nu}$
at $\chi_0$\cite{Inoue1}
\BE
\delta \theta^2\approx
\f{16 \pi^2~V\!o\,l(M)}  
{k^2 (\sinh(2(\chi_o+r_{ave}))-\sinh(2(\chi_o-r_{ave}))-4 r_{ave})},
\label{eq:deltheta}
\EE
where $V\!o\,l(M)$ denotes the volume of a manifold $M$ and $r_{ave}$
is the averaged radius of the Dirichlet domain. There is an arbitrariness
in the definition of $r_{ave}$. Here we define $r_{ave}$ as the radius 
of a sphere with volume equivalent to
the volume of the manifold, 
\BE
V\!o\,l(M)=\pi(\sinh(2 r_{ave})-2 r_{ave}),
\EE
which does not depend on the choice of a base point.
Here we define the topological identification length $L$ as 
$L\!=\!2r_{ave}$.
For the Weeks and the Thurston manifolds, $L\!=\!1.19$ and 
$L\!=\!1.20$ respectively. From
Eq. (\ref{eq:deltheta}), for large $\chi_0$, one can approximate
$u_{\nu}(\chi_o)\sim u_\nu'(\chi'_o)$ by choosing an appropriate
radius $\chi'_o$ that satisfies 
$\nu^{-2}\exp(-2\chi_o)=\nu'^{-2}\exp(-2\chi'_o)$. 
Averaging Eq. (\ref{eq:xi}) over $l$'s and $m$'s or the base
points, for large $\chi_0$, one obtains,
\BE
\bigl <|\xi_{\nu' l m} |^2 \bigr >
\sim 
\f{\exp(-2\chi_o)}{\exp(-2\chi'_o)}
\bigl<|\xi_{\nu l m} |^2 \bigr >,
\EE
which gives $\bigl<|\xi_{\nu l m}|^2 \bigr> \sim \nu^{-2}$.
Thus, the variance of $a_{\nu l m}$'s is proportional to $\nu^{-2}$.
The numerical results for the two CH manifolds shown in figure 
\ref{fig:SEC} clearly support the $\nu^{-2}$ dependence of the variance. 
\BF[tpb]
\centerline{\psfig{figure=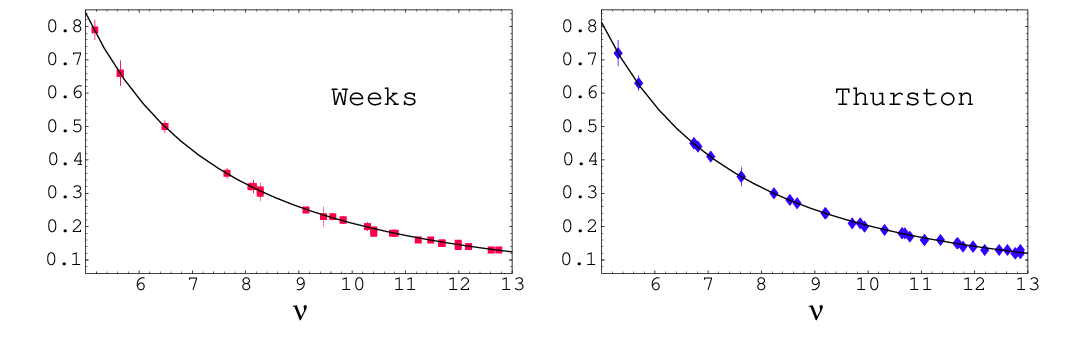,width=16cm}}
\caption{Variance of Expansion Coefficients} 
\mycaption{Averaged squared $a_{\nu}$'s ($k\!<\!13$)
based on 300 realizations of the base points for the Weeks and the Thurston
manifold with $\pm 1\sigma$ run-to-run variations.  
 $a_{\nu}$ is defined to be $Var(a_{\nu l m})$ averaged over $0\! \le\! l\!
\le\!20$ and $-l\! \le\! m\!\le\!l$. 
The best-fit curves for the Weeks and the Thurston manifolds are 
$21.0\nu^{-2}$ and $20.3\nu^{-2}$, respectively.} 
\label{fig:SEC}
\EF
\\
\indent
As we have seen, the property of eigenmodes on general CH manifolds is
summarized in the following conjecture:
\\
\\
\textit{Conjecture: Except for the base points which are too close to any
fixed points by symmetries, for a fixed $\nu$, a set of the expansion 
coefficients $a_{\nu l m}$ over $(l,m)$'s can be regarded as 
Gaussian pseudo-random numbers. For a fixed $(\nu l m)~(l>0)$, 
the expansion coefficients at different base points that are 
randomly chosen 
can be also regarded as Gaussian pseudo-random numbers. In either
case, the variance is proportional to $\nu^{-2}$ and the average is
zero.}
\section{Periodic orbit sum method}
\subsection{Length spectra}
Computation of periodic orbits (geodesics) is of crucial
importance for the semiclassical quantization of 
classically chaotic systems.
However, in general, solving a large number of
periodic orbits often becomes an intractable problem since
the number of periodic orbits grows exponentially with an 
increase in length.
For CH manifolds periodic orbits can be calculated  
algebraically since 
each periodic orbit corresponds to a conjugacy class
of hyperbolic or loxodromic elements of the discrete isometry group 
$\Gamma$. 
The conjugacy classes can be directly computed from generators
which define the Dirichlet domain of the CH manifold
\\
\indent
Let $g_i,(i=1,...,N)$ be the generators and ${\cal{I}}$ be the
identity. In general these generators are 
not independent. 
They obey a set of relations 
\BE
\prod g_{i_1}g_{i_2}\ldots g_{i_n}={\cal{I}}, \label{eq:relation}
\EE
which describe the fundamental group of $M$.
Since all the elements of $\Gamma$ can be represented by certain
products of generators, an element $g\in \Gamma$ can be written
\BE
g=g_{i_1}g_{i_2}\ldots g_{i_m},
\EE
which may be called a ``\textit{word}'' .
Using relations, each word can be shorten to a
word with minimum length. Furthermore, all cyclic permutations
of a product of generators belonging to the same conjugacy class
can be eliminated.
Thus conjugacy classes of $\Gamma$ can be computed by generating words with 
lowest possible length to some threshold length which are reduced  
by using either relations  (\ref{eq:relation}) among the generators
or cyclic permutations of the product. 
\\
\indent
In practice, we introduce a cutoff length $l_{cut}$ 
depending on the CPU power
because the number of periodic orbits grows exponentially 
in $l$ which is a direct consequence of the exponential proliferation
of tiles (copies of the fundamental domain) in tessellation.
Although it is natural to expect a long
length for a conjugacy class described in a word with many letters,
there is no guarantee that all the periodic orbits 
with length less than $l_{cut}$ are actually computed or not for
a certain threshold of length of the word.
\\
\indent
Suppose that each word as a transformation
that acts on the Dirichlet fundamental domain $D$. For example,
$D$ is transformed to $D'\!=\!gD$ by an element $g$.
 We can consider $gD'$s as tessellating tiles in the 
universal covering space.
If the geodesic distance $d$ between the center (basepoint) of  
$D$ and that of $gD$ is large, we can expect a long periodic orbit 
that corresponds to the conjugacy class of $g$. Tessellating tiles
to sufficiently long distance $d>l_{cut}$ makes it 
possible to compute the complex primitive
length spectra \{$L_j=l_j \exp (i \phi_j),|l_j<l_{cut}$\} where
$l_j$ is the real length of the periodic orbit of a conjugacy class
with one winding number and $\phi_j$ is the phase of the 
corresponding transformation. We also compute multiplicity number
$m(l_j)$ which counts the number of orbits having the same $l_j$ 
and $\phi_j$.  
\\
\indent
In general, the lower limit of the
distance $d$ for computing a complete set of length spectrum 
for a fixed $l_{cut}$ is not known 
but the following fact has been proved by
Hodgson and Weeks\cite{HW}.
In order to compute a length spectra of a CH 
3-manifold (or 3-orbifold) with length less than $l$,
it suffices to compute elements $\{g\}$ satisfying
\BE
d(x,gx)< 2 \cosh^{-1} (\cosh R \cosh l/2),
\label{eq:sufficient}
\EE
where $x$ is a basepoint
and $R$ is the spine radius\footnote{Spine radius $R$ is equal to the
maximum over all the Dirichlet fundamental domain's edges of the 
minimum distance from the edge to the basepoint. Note that R is finite 
even for a cusped manifold with finite volume.}. Note that there is a unique
geodesic which lies on an invariant axis for each hyperbolic
or loxodromic element.
SnapPea can compute
length spectra of CH spaces either by the ``rigorous method''
based on the inequality (\ref{eq:sufficient}) 
or the ``quick and dirty'' method by setting the tessellating radius 
$d$ by hand. 
The former method has been used for manifolds 
with small volume ($<1.42$), but  
the latter method ($d=l_{cut}+0.5$) has been also used for some 
manifolds with large volume ($>1.42$) since the 
tessellating radius given by the former method
is sometimes so large that the computation time becomes too long. 
The detailed algorithm is summarized in appendix B.
\\
\indent
The asymptotic behavior of the classical staircase $N(l)$ which counts 
the number of primitive 
periodic orbits with length equal to or less than $l$
for CH 3-spaces can be written in terms of $l$ and the topological entropy 
$\tau$ \cite{Margulis}
\BE
N(l)\sim \textrm{Ei}(\tau l) \sim \f{\exp({\tau l})}{\tau l}, 
~~l \rightarrow \infty.
\label{eq:asympl}
\EE
\BF
\centerline{\psfig{figure=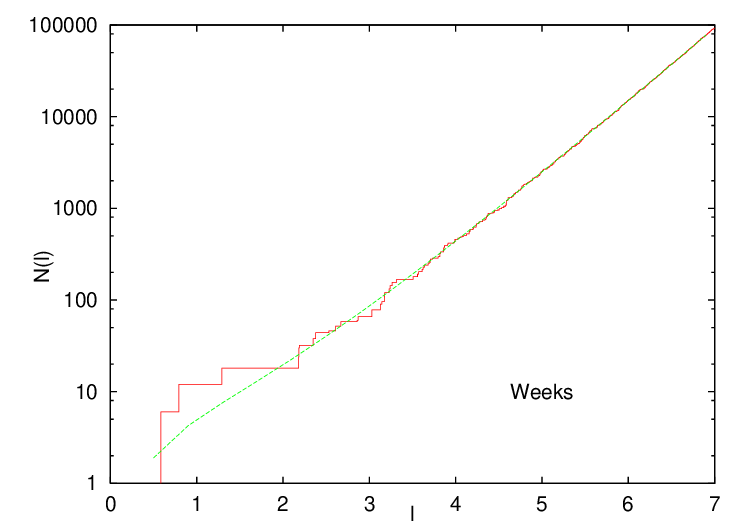,width=9cm}
\psfig{figure=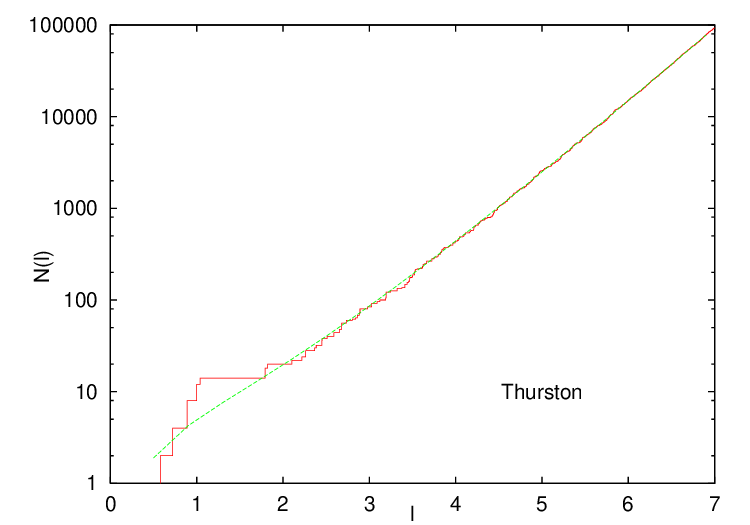,width=9cm}}
\caption{Classical Staircases}
\mycaption{The classical staircases $N(l)$ for the 
Weeks manifold and the Thurston manifold are well consistent with 
the asymptotic distribution (\ref{eq:asympl}) for large $l$.}
\label{fig:length}
\EF
The topological entropy for $D$-dimensional CH spaces is given by 
$\tau=D-1$. A larger topological entropy implies that the efficiency
in computation of periodic orbit is much less for higher dimensional
cases\cite{AM}.
\\
\indent
In figure \ref{fig:length}, 
the computed classical staircases ($l_{cut}=7.0$) 
are compared with the asymptotic formula for the smallest (Weeks) 
manifold and the second smallest (Thurston) manifold. For both cases, an
asymptotic behavior is already observed at $l\sim3.5$. 
\\
\indent
Although the asymptotic behavior of the 
classical staircase $N(l)$ does not depend on the topology
of the manifold, the multiplicity number $m(l)$ does. 
In fact, it was Aurich and Steiner who firstly noticed that 
the locally averaged multiplicity number 
\BE
<\!m(l)\!>=\f{1}{N} \sum_{l-\Delta l/2 < l_i < l+\Delta l/2} g(l_i), 
~~~(N\!=\!\textrm{total number of terms} )
\EE 
grows exponentially $<\!m(l)\!>\sim e^{l/2}/l$
as $l\!\rightarrow\!\infty$ for
arithmetic 2-spaces (manifolds and orbifolds)\cite{BSS,AS0,ASS}. 
since the length $l$ of the periodic orbits
are determined by algebraic integers in the form
$2 \cosh(l/2)=\ti{algebraic integer}$\footnote{For a two-dimensional
space, the classical staircase has an asymptotic form
$N(l)\sim \exp(l)/l $. On the other hand, the  classical staircase for 
distinct periodic orbits has a form $\hat{N}(l)\sim \exp(l/2)$
for arithmetic systems. Because $m(l)d\hat{N}=dN$ we have $m(l)\sim
\exp(l/2)/l$ as $l \rightarrow \infty$.}.
The failure of 
application of the random matrix theory 
to some CH spaces may be attributed to the 
arithmetic property. For non-arithmetic
spaces, one expects that the multiplicities are determined
by the symmetries (elements of the isometry group) of the space.  
However, in the case of a non-arithmetic 3-orbifold,
it has been found that $<\!m(l)\!>$ grows exponentially in the form
$e^{b l}/(c l)$ where $b$ and $c$ are fitting
parameters\cite{AM}. This fact might implies that the symmetries of long 
periodic orbits are much larger than that of the space
even in the case of non-arithmetic systems.
\\
\indent
For 3-manifolds, one can expect that the property of  $<\!m(l)\!>$ 
for 3-orbifold also holds.
The locally averaged multiplicities $(\Delta l=0.2)$ for
the smallest twelve examples which include seven
arithmetic and five non-arithmetic 3-manifolds \cite{CGHN}
have been numerically computed using SnapPea.
\BF
\centerline{\psfig{figure=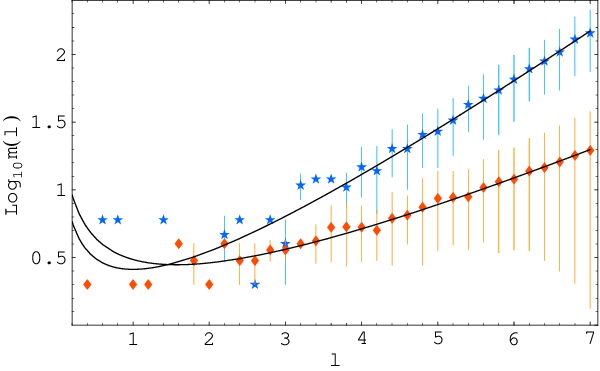,width=12cm}}
\caption{Locally Averaged Multiplicities}
\mycaption{Plots of locally averaged multiplicities ($\Delta l=0.2$)
with one-sigma errors and the fitting curves
for the Weeks manifold m003(-3,1) which is
arithmetic (star) and a non-arithmetic manifold m004(1,2)
(diamond). The fitting curves ($a \exp(l)/l $ for the former and
$\exp(b l)/(c l)$ for the latter) are obtained by the least square method
using data $3.0<l<7.0$. }
\label{fig:aveg}
\EF
\begin{table}
\begin{center}
\setlength{\tabcolsep}{3pt}
\begin{tabular}{rrccrrr} 
\hline \hline
\multicolumn{1}{c}{manifold} &
\multicolumn{1}{c}{volume} &
\multicolumn{1}{c}{A/N} &
\multicolumn{1}{c}{G} &
\multicolumn{1}{c}{$a$} &
\multicolumn{1}{c}{$b$} &
\multicolumn{1}{c}{$c$} 
\\ \hline
m003(-3,1)&0.9427   &A  &D6  &0.9514 &- &- \\ \hline  
m003(-2,3)&0.9814   &A  &D2  &0.5667 &- &- \\   \hline  
m007(3,1)&1.0149    &A  &D2$\dagger$ & 0.7108 &- &- \\ \hline  
m003(-4,3)&1.2637   &A  &D4  &0.8364  &-  &- \\ \hline  
m004(6,1)&1.2845    &A  &D2  &0.6066  &- &-  \\ \hline  
m004(1,2)&1.3985    &N  &D2  &-       &0.6360 & 0.6180\\ \hline        
m009(4,1)&1.4141    &A  &D2  &0.5362  &- &- \\ \hline
m003(-3,4)&1.4141   &A  &D2  &0.5655  &- &- \\ \hline  
m003(-4,1)&1.4236   &N  &D2  &-       &0.5933 &0.5912\\   \hline  
m003(3,2)&1.4407    &N  &D2  &-       &0.6018 &0.5532 \\ \hline  
m004(7,1)&1.4638    &N  &D2  &-       &0.5693 &0.4829\\ \hline  
m004(5,2)&1.5295    &N  &D2  &-       &0.5780 &0.5590\\ \hline  
\hline
\end{tabular}
\end{center} 
\label{tab:coefficients}
\caption{Coefficients of Fitting Curves for Averaged Multiplicities}
\mycaption{Coefficients of the fitting curves which describe the average
behavior of locally averaged multiplicities $<\!m(l)\!>$ for
arithmetic (A) and non-arithmetic (N) 3-manifolds. G denotes the
isometry group (symmetry group). Fitting parameters
$a,b$ and $c$ are obtained by the least square method
using data $3.0<l<7.0$. $\dagger$ For m007(3,1), the isometry group may 
be larger than D2.} 
\end{table}
From figure \ref{fig:aveg}, one can see that the  
difference in the behavior of $<\!m(l)\!>$ between the
non-arithmetic manifold and the arithmetic one is manifest. 
As observed in 3-orbifolds, averaged multiplicities behave as 
\BEA
<\!m(l)\!>&=&a \f{\exp{l}}{l},~~~~~~\textrm{(arithmetic)},
\\
&=&\f{\exp{b l}}{c l},~~~~~\textrm{(non-arithmetic)},
\EEA
where $a$ depends on the discrete isometry group while $b$ and
$c$ are fitting parameters. From table I, one observes that 
arithmetic manifolds having a larger
symmetry group have a larger value of $a$.
The growth rates for
non-arithmetic manifolds ($b\sim 0.56$) 
are always less than that for arithmetic
manifolds ($b=1$) 
but nevertheless exponential. 
\subsection{Trace formula}
Gutzwiller's periodic orbit theory provides a semiclassical
quantization rule for classically chaotic systems.
The theory is expressed in form of a semiclassical approximation 
($\hbar \rightarrow 0$) of the trace of the energy Green's operator
(resolvent operator) $\hat{G_E}\!=\!(\Delta+E)^{-1}$ 
in terms of the length of periodic 
orbits (geodesics) $\{ L_i \}$ which is known as the
\ti{Gutzwiller trace formula}\cite{Gutzwiller}.
For the dynamical system
of a free massive particle on a CH space known as the 
Hadamard-Gutzwiller model, the periodic orbits give 
the exact eigenvalues, and the relation is no longer 
semi-classical approximation. 
In mathematical literature, the trace formula is known as the 
{\it{Selberg trace formula}} \cite{Selberg}. 
In what follows we consider
only orientable CH 3-manifolds(denoted as CH manifolds)
(for general cases including
orbifolds, see \cite{AM}). 
The Selberg trace formula for a CH manifold
$M\!=\!H^3/ \Gamma$ ($\Gamma$ is a
discrete isometry group containing
only hyperbolic or loxodromic elements) can be written as 
\BEA
Tr(\hat{G}_E-\hat{G}_{E'})
&=&-\f{v(M)}{4 \pi i}(p-p')
\nonumber
\\
&&
-\sum_{\{g_\tau\}}
\f{l(g_{\tau_0})}{4(\cosh l(g_\tau)-\cos \phi(g_\tau))} \Biggl ( \f{\exp(-i 
p l(g_\tau))}{ip}-\f{\exp(-i p' l(g_\tau))}{ip'}\Biggr ), \label{eq:ST}
\EEA
where $p^2\!=\!E-1$, $v(M)$ 
denotes the volume of $M$, $l(g_\tau)$
is the (real) length of the periodic orbit of transformation $g_\tau
\in \Gamma$. $g_{\tau_0}$ is a transformation 
that gives the shortest length of the periodic orbit $l(g_{\tau_0})$
which commutes with $g_\tau$. $\phi(g_\tau)$ is the phase of the
transformation $g_\tau$. The sum in (\ref{eq:ST}) 
extends over $\Gamma$ conjugacy classes 
\BE
\{g_\tau\}:=\{g'_\tau|g_\tau'=h g_\tau h^{-1}, h \in \Gamma \}
\EE
of hyperbolic ($\phi\!=\!0$) or loxodromic elements ($\phi\!\neq\!0$).
However the periodic orbit sum in (\ref{eq:ST}) which is known as 
Maa\ss-Selberg series converges at only
complex energy such that $\textrm{Im}~ p\!<\!-1$ and $\textrm{Im}~p'\!<\!-1$. In order to obtain real eigenvalues, one needs to multiply
the trace by some suitable analytic ``smoothing'' function  $h(q)$
that satisfies:
\\
(i):$h(q)=h(-q)$;
\\
(ii):$h(q)={\cal{O}}(|q|^{-3-\delta})$ for $\delta>0$ as $|q|\rightarrow \infty$;
\\
(iii):$h(q)$ is analytic in the strip ${|\textrm{Im}~ q|<1+\epsilon}$ for 
$\epsilon>0$.
\\ 
Multiplying (\ref{eq:ST}) by $q~ h(q)/(\pi~ i)$ and integrating it
over $q$ from $-\infty$ to $\infty$, one obtains the \textit{general Selberg trace formula},
\BEA
\sum_{n=0}^\infty h(p_n) = &-& \f{v(M)}{2 \pi} \tilde{h}''(0)
\nonumber
\\
&+&\sum_{\{g_\tau\}} \f{l(g_{\tau_0})}{2(\cosh~l(g_\tau)-\cos~
\phi(g_\tau))}~\tilde{h}(l(g_\tau)), \label{eq:GTF}
\EEA  
where 
\BE
\tilde{h}(l)=\f{1}{2 \pi} \int_{\infty}^{\infty} dq~ h(q)~\exp(-iql),
\EE
and $\tilde{h}''(0)$ is the second derivative of $\tilde{h}$ and
$p_n$ denotes a wavenumber of the corresponding eigenmode.
The sum in (\ref{eq:GTF}) is absolutely convergent for any real
eigenvalues $E_n=p_n^2+1$. One can obtain various functions of eigenvalues
such as heat kernels and energy level densities from periodic orbits by choosing an appropriate ``smoothing''
function $h(q)$. 
\\
\indent
In order to obtain eigenvalues, a simple approach is to compute the
spectral staircase
\BE
N(E)=\sum_{n=0}^\infty \theta(E-E_n),
\EE
where $E_0\!=\!0, E_1,E_2,\cdots $ are the eigenvalues of the
Laplacian and $\theta$ is the Heaviside function\cite{AS1}. 
To explore supercurvature modes $u_E$,
 $0\leq E<1$, we choose the ``smoothing'' function of
$N(E)$ as 
\BEA
h_{p\!,\epsilon }(p')
&=&
\f{1}{2}\Biggl(1-\textrm{Erf}
\biggl( \f{E'-E}{\epsilon^2} \biggr ) \Biggr),
\nonumber
\\
&=&
\f{1}{2}\biggl(1-\textrm{Erf}
\biggl( \f{p'^2-p^2}{\epsilon^2} \biggr ) \biggr),
\EEA
which is real for $E>0$. Note that $h_{p}(p')$ satisfies all the 
conditions (i) to (iii).
By taking the limit $\epsilon \rightarrow 0$, one obtains
the spectral staircase
\BE
N(E)=\lim_{\epsilon \rightarrow 0} \sum_{n=0}^\infty h_{p\!,\epsilon }(p_n). 
\EE
\\
\indent
Let us first estimate the behavior of the trace in (\ref{eq:GTF}). 
From a straightforward calculation, the zero-length
contribution can be written as
\BEA
 \tilde{h}''_{p\!,\epsilon}(0)=&-& \f{\epsilon^3}{6~ \pi^{\f{3}{2}}}
 \exp(-p^4/\epsilon^4)
\nonumber
\\
&\times& \biggl ( \f{\sqrt{2}\pi}{4} (\Gamma(3/4))^{-1}
 F(5/4,~1/2,~p^4/\epsilon^4)
+\f{3~p^2}{2~\epsilon^2}\Gamma(3/4)F(7/4,~3/2,~p^4/\epsilon^4)
\biggr ), \label{eq:zerolength}
\EEA
where $\Gamma(x)$ is the Gamma function and $F(a,b,z)$ is the 
confluent hypergeometric function. 
For $x\!=\!p^4/\epsilon^4
\!\rightarrow\!\infty $, it is asymptotically expanded as  
\BEA
F(a,b,x)&=&\f{\Gamma(b)}{\Gamma(b-a)}e^{i \pi a}x^{-a} 
\Biggl \{ \sum_{n=0}^{R-1} 
\f{(a)_n (1+a-b)_n}{n !
}(-x)^{-n}+{\cal{O}}(x^{-R})
\Biggr \}
\nonumber
\\
&+&\f{\Gamma(b)}{\Gamma(a)}e^{x}x^{a-b} \Biggl \{
\sum_{n=0}^{S-1} \f{(b-a)_n (1-a)_n}{n !
}(x)^{-n}+{\cal{O}}(x^{-S}) \Biggr\}, \label{eq:as}
\EEA
where $(a)_m\equiv \Gamma(a+m)/\Gamma(a)$.
From (\ref{eq:zerolength}) and (\ref{eq:as}), in the lowest order,
we have the average part, 
\BE
\hat{N}(p)=\lim_{\epsilon \rightarrow 0} -\f{v(M)}{2 \pi}
\tilde{h}''_{p\!,\epsilon}(0)=\f{v(M)}{6\pi^2}|p^2|^{\f{3}{2}}, 
\label{eq:Weyl}
\EE
that gives the dominant term in the Weyl asymptotic formula 
for $p\!>>\!1$.
\\
\indent
Next, we estimate the oscillating term in (\ref{eq:GTF})
\BE
\tilde{h}_{p,\epsilon}(l)=\alpha \int_{-\infty}^{\infty} dq~ \textrm{exp}
(-i q l)~q~  \textrm{exp} \Bigl(-\f{(q^2-p^2)^2}{\epsilon^4}
\Bigr),~~~~\alpha 
\equiv \f{i}{l \epsilon^2 \pi^{3/2}}. \label{eq:longh}
\EE
In the long orbit-length limit $l\!>\!>\!1$ with $p\!>\!0$,
the integrand in Eq.(\ref{eq:longh}) oscillates so rapidly that 
the dominant contribution comes from $q\!\sim\!p,$ or $-p$ 
where $(q^2-p^2)^2/\epsilon^4$ can be approximately given as 
$-4(q-p)^2p^2/\epsilon^4$. Then (\ref{eq:longh}) can be written as
\BE
\tilde{h}_{p,\epsilon}(l)\sim \f{1}{\pi l}\sin(p
l)\textrm{exp}\biggl(\f{-\epsilon^4 l^2}{16 p^2}\biggr ). \label{eq:tilh}
\EE
Thus each periodic orbit corresponds
to a wave with wavelength $2 \pi/l$ and an amplitude
which is exponentially suppressed with an increase in $l$ or a decrease
in $p$.
For a finite subset of length spectra $l_j<l_{cut}$, 
the appropriate choice for the
smoothing scale is given by   
$\epsilon\! =\!\alpha p^{1/2}$ for $p^2\!>\!0$ where 
$\alpha$ depends on $\l_{cut}$ since for a reasonable value 
of the proportional factor $\alpha$ all the contributions
from periodic orbits with large length $l>l_{cut}$ can be negligible.
For  $p^2\!<\!0$, 
an optimal choice can be obtained from a numerical
computation of Eq.(\ref{eq:longh}) directly. Comparing the obtained 
smoothed spectral staircase with the one based on the computed ``true''
eigenvalues using the direct boundary element method (DBEM), 
it is numerically found
that for  $l_{cut}\!=\!7.0$, an appropriate smoothing scale is given by 
\BE
\epsilon(k)=\cases{
0.116k^2+0.184~k+1.2 ~~(k<1) \cr
0.832k^{1/2}+0.668 ~~(k \ge 1) \cr},
\EE
where $E=k^2=p^2+1$. The eigenvalues can be computed by searching
$E$ at which $N(E)-0.5$ becomes positive integer.
It should be emphasized that the precision of computation depends on the 
value of $l_{cut}\!$ which determines the resolution scale in the
eigenvalue spectra.
\\
\indent  
In order to get eigenvalues from the smoothed spectral staircase,
one must take into account the effect of the multiplicity 
number (degeneracy number) for each eigenvalue since the 
spectral staircase is 
smoothed on larger scales for degenerate modes. 
Therefore, the numerical accuracy becomes worse if
the eigenmode has a large multiplicity number.
Fortunately, the order of the symmetry group is not so large for a
small manifold (volume$<3$). For instance, of the twelve 
smallest examples, nine manifolds have a symmetry group with order 4.    
If one assumes
that the multiplicity number is either 1 or 2 then the 
deviation $\Delta k$ from a precise value is approximately 
given by
\BE
\f{1}{2}\Biggl(1-\textrm{Erf}
\biggl( \f{k^2-(k+\Delta k)^2}{\epsilon^2} \biggr ) \Biggr)=\f{1 
\pm 0.5}{2}.
\label{eq:deltak}
\EE
For instance, if one uses a length spectrum $l<7.0$ then  
(\ref{eq:deltak}) gives 
$\Delta k\!=\!0.30,0.33$ for $k\!=\!5.0,3.0$, respectively. 
If one permits the multiplicity number as much as $6$,
then the expected precision becomes 
$\Delta k\!=\!0.49,0.54$ for $k\!=\!5.0,3.0$, respectively.
\\
\BF
\centerline{\psfig{figure=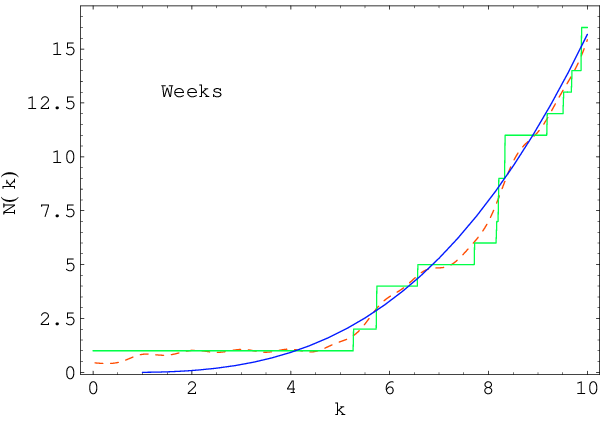,width=8cm}
\psfig{figure=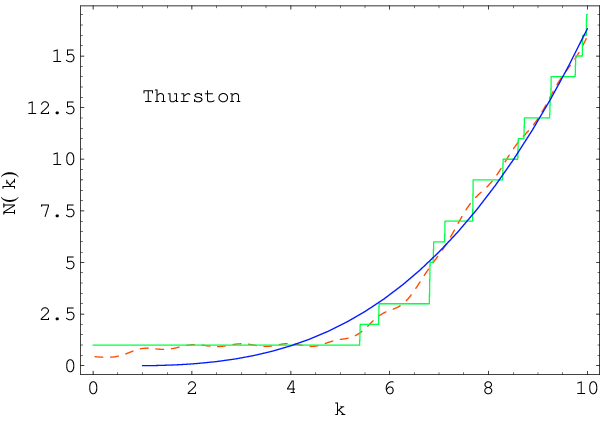,width=8cm}}
\caption{Spectral Staircases}
\mycaption{The spectral staircases $N(k)$ for the Weeks manifold 
and the Thurston manifold 
obtained by the DBEM are
compared with the average parts $\hat{N}(k)$ (solid curves), namely, 
Weyl's asymptotic formula (\ref{eq:Weyl}) 
and smoothed spectral staircase (dotted curves) obtained by the 
periodic orbit sum method(POSM)
using all periodic orbits $l<7.0$.} 
\label{fig:ssT}
\EF
\begin{table}
\begin{center}
\setlength{\tabcolsep}{3pt}
\begin{tabular}{cccccc}   
\hline \hline
\multicolumn{1}{c}{manifold} &
\multicolumn{1}{c}{volume} &
\multicolumn{1}{c}{$k_1$(DBEM)} &
\multicolumn{1}{c}{$k_1$(POSM)} &
\multicolumn{1}{c}{m($k_1$)} &
\multicolumn{1}{c}{$\Delta k_1/k_1$}  
\\ \hline
m003(-3,1)&0.9427   &5.27  &5.10  &1 &0.03  \\ \hline  
m003(-2,3)&0.9814   &5.40  &5.34  &1 &0.01  \\   \hline  
m007(3,1)&1.0149    &5.29  &5.37  &1 &0.02   \\ \hline  
m003(-4,3)&1.2637   &4.58  &4.31  &2 &0.06     \\ \hline  
m004(6,1)&1.2845    &4.53  &4.35  &1 &0.04   \\ \hline  
m004(1,2)&1.3985    &4.03  &3.93  &1 &0.02 \\ \hline        
m009(4,1)&1.4141    &5.26  &4.84  &2 &0.08 \\ \hline
 \hline
\end{tabular}
\end{center} 
\label{tab:accuracy}
\caption{First Eigenvalues}
\mycaption{The first (non-zero) wavenumbers $k_1=E_1^{1/2}$ are 
calculated using the DBEM and
using the POSM for seven smallest manifolds.
$k_1$'s agree with 
relative accuracy $\Delta k_1 /k_1=0.01-0.08$.
The multiplicity number of the first eigenmode is 
calculated using the DBEM. }
\end{table}
We can see from table II that the first eigenvalues 
calculated by using length spectra
$l<7.0$ for some smallest known 
CH manifolds lie within several per-cent
of those obtained by the DBEM. Note that 
the eigenvalues are also consistent with 
those obtained by the Trefftz method\cite{CS99}.
For two examples in which the first non-zero mode is
degenerated, the eigenvalues are much shifted to lower values owing
to the smoothing effect.
From figure \ref{fig:ssT}, one can see that the curves of the 
obtained smoothed spectral stairs cross the ``true'' stairs 
at almost half height.  A slight deviation from the 
average part of the spectral staircase
is caused by the interference of waves each one of which corresponds to
a periodic orbit.
\subsection{First eigenvalue and geometrical quantity}
The estimate of the first (non-zero) eigenvalue 
$E_1=k_1^2$ of the Laplace-Beltrami operator 
plays a critical role in describing the global topology and 
geometry of manifolds. A number of estimates of $E_1$ 
for  $n$-dimensional compact 
Riemannian manifolds $M$ 
using diffeomorphism-invariant quantities have been 
proved in mathematical literature.
\\
\indent
First of all, we consider the relation between the first eigenvalue 
$E_1$ and the \ti{diameter} $d$
which is defined as the maximum of the minimum geodesic distance
between two arbitrary points on $M$. Various analytic upper and 
lower bounds of $E_1$ in terms of $d$ have been known. 
 Suppose $M$ with Ricci curvature bounded
below by $-L(L>0)$. Cheng and Zhou proved that $E_1$ satisfies 
\BE
E_1 \ge \max \biggl [ \f{1}{2} \f{\pi^2}{d^2}-\f{1}{4}L,
\sqrt{\f{\pi^4}{d^2}+\f{L^2}{16}}-\f{3}{4}L,
\f{\pi^2}{d^2}\exp \bigl(-C_n \sqrt{Ld^2}/2\bigr) 
\biggr ],
\label{eq:Cheng}
\EE
where $C_n=\max[\sqrt{n-1},\sqrt{2}]$\cite{CZ}. Another lower bound 
has been obtained by Lu\cite{Lu}. Suppose that the Ricci
curvature of $M$ is bounded below as $R_{ab}\ge -K g_{ab},(K\ge 0)$
for a some real number $K$ 
where $g_{ab}$ is the metric tensors of $M$. 
Then $E_1$ satisfies
\BE
E_1 \ge \max \biggl [ \f{\pi^2}{d^2}-K,
\f{8}{d^2}-\f{K}{3},\f{8}{d^2} \exp \biggl ( - \f{d^2 K}{8}
\biggr ), 
\f{8}{d^2}\biggl (1+\f{d}{3}\sqrt{K(n-1)} \biggr) 
\exp\biggl ( - \f{d}{2} \sqrt{K(n-1)} \biggr ) \biggr ].
\label{eq:Lu}
\EE   
As for upper bounds, the following theorem has been proved by 
Cheng\cite{Cheng}. Suppose M with Ricci curvature 
larger than $(n-1)c$, then we have
\BE
E_1\le \tilde{E}_1\bigl ( V_n (c,d/2) \bigr )
\label{eq:Chen}
\EE
where $V_n(c,r)$ denotes a geodesic ball 
with radius $r$ in the $n$-dimensional simply-connected space
with sectional curvature $c$ and $\tilde{E}_1$ is the first Dirichlet
eigenvalue.  
Setting $L\!=\!K\!=\!2,n\!=\!3$, and $c\!=\!-1$, we obtain the upper and 
lower bounds of $E_1$ for CH 3-manifolds. For the upper bound,
(\ref{eq:Chen}) gives a simple relation, $d\le 2 \pi/\nu_1$,
where $\nu_1^2=k_1^2-1$. The physical interpretation is 
clear: the wavelength $\lambda_1\!\equiv \!2 \pi/\nu_1$
of the lowest non-zero mode 
must be larger than the diameter.   
\BF
\centerline{\psfig{figure=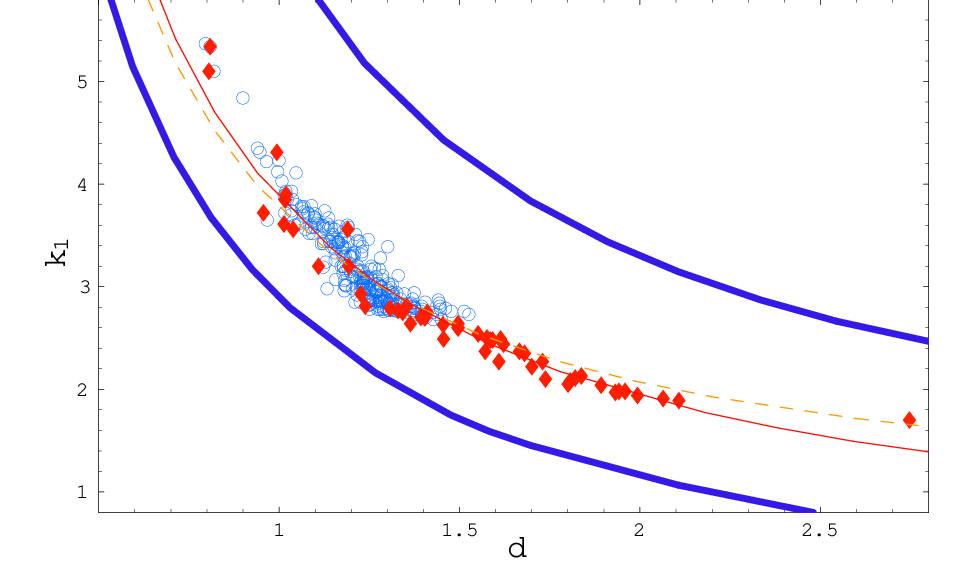,width=13cm}}
\caption{Diameter vs First Eigenvalue}
\mycaption{Diameter $d$ versus $k_1$ for 263 examples of CH 3-manifolds
with $l_{min}\!>\!0.3$ and $v\!<\!3$ (group A, circle) and 52 examples
that are obtained by performing Dehn surgeries (51 examples for 
$|p|,|q|<10$ plus 1 example $(p,q)=(16,13)$ group B, diamond) 
on a cusped manifold m003 with 
the best-fit curves for group B corresponding to (\ref{eq:fitA}) 
(dashed curve) assuming $k_1(\textrm{cusp})=1$ 
and (\ref{eq:fitB}) (solid curve) assuming  
$k_1(\textrm{cusp})=0.1$. The fitting curves also agree with 
the computed values for group A.
The upper and lower thick curves denote the
analytic bounds (\ref{eq:Cheng}), (\ref{eq:Lu}) and (\ref{eq:Chen}).  }
\label{fig:dk1}
\EF
\\
\indent
Now, we compare the first eigenvalues  
of 263 examples of CH 3-manifolds with volume less than 3
which have the length of the shortest periodic orbit 
$l_{min}>0.3$
(the Hodgson-Weeks census\cite{HW}) 
and of 45 other examples obtained
by Dehn surgeries ($|p|<17,|q|<14$) 
on a cusped manifold m003\footnote{The Hodgson-Weeks 
census with volume less than 3
also includes 8 manifolds
obtained by Dehn surgeries on m003.} with the analytic
bounds. The diameter of a CH 3-manifold is
given by the supremum of the outradius\footnote{The outradius at 
a basepoint $x$ is equal 
to the minimum radius of the simply-connected 
ball which encloses the 
Dirichlet domain at $x$.}
over all the basepoints,  which has been numerically computed 
using the SnapPea kernel\footnote{I would like to thank J. Weeks
for providing me a code to compute the diameter using the SnapPea kernel.}.
The numerical accuracy is typically $\Delta
d=0.03-0.09$ depending on the topology of the manifold.
\\
\indent
As shown in figure \ref{fig:dk1}, the eigenvalues are well described
by an empirical fitting formula $\lambda_1=\beta d$, or
\BE
k_1=\sqrt{1+\f{4 \pi^2}{\beta^2 d^2}}.
\label{eq:fitA}
\EE
Applying the least square method for the 263 manifolds in the 
Hodgson-Weeks census (group A), and 
53 manifolds obtained by Dehn surgeries on
m003 (group B), the best-fit values $\beta=1.70,1.73$ have been
obtained for each group, respectively.
Note that $\beta\sim1.7$ is slightly larger than the 
values 1.3-1.6 for 12 examples in the previous result 
by Cornish and Spergel\cite{CS99}. The deviation from the
fitting formula (\ref{eq:fitA}) is found to be remarkably small
(with one sigma error $\Delta k_1=0.18,0.19$ for each group,
respectively), which implies the existence of much  
sharper bounds. 
\\
\indent
The empirical formula (\ref{eq:fitA}) asserts that no supercurvature 
modes ({\it{i.e.}} $k_1\!<\!1$) exist in the limit 
$d \rightarrow \infty$ where the manifold converges
to the original cusped manifold. However, cusped manifolds may
have some supercurvature modes even for those with small volume. 
Instead of (\ref{eq:fitA}), we consider a generalized empirical formula 
\BE
k_1=\sqrt{(k_1 (\textrm{cusp}))^2+\f{4 \pi^2}{\beta^2 d^2}},
\label{eq:fitB}
\EE
where $k_1 (\textrm{cusp})\le 1$ is the smallest wavenumber for the 
original cusped manifold. Although no supercurvature modes
were observed in this analysis, the non-existence of 
such modes in the limit $d\rightarrow \infty$ was not confirmed
since the
numerical accuracy becomes worse for manifolds with 
small $k_1$ and large $d$.
\\
\indent
Next, we consider the relation between diameter $d$ and volume $v$
of the manifold. Since  diameter is given by the supreme
of the outradius (minimum radius of a sphere which circumscribes the
Dirichlet domain) all over the basepoints, one expect that 
$v$ is estimated as the volume of a sphere (in a hyperbolic space) 
with radius $r$ somewhat
smaller than $d$ if there is no region which resembles the
neighborhood of a cusp (``thin part''
\footnote{A ``thin'' part is defined as
a region where the injectivity radius is short.}) 
or equivalently $l_{min}$ is 
sufficiently large. 
Suppose that $r=\alpha d$ with $\alpha <1$
then the volume of a sphere  
\BE
v=\pi (\sinh (2 \alpha d)-2 \alpha d)
\label{eq:dvola}
\EE
gives the approximate value of a CH manifold
with diameter $d$. The best fit value for a sample of 79
manifolds with $l_{min}\!>\!0.5$ is $\alpha=0.69$.
If a manifold has a ``thin'' part then 
the relation (\ref{eq:dvola}) is no longer valid
since the diameter becomes too long.  Let $v_c$ be the volume of a
cusped manifold $M_c$ (with only one cusp) and $v(d)$ be the 
volume of a CH manifold $M$ 
obtained by a Dehn surgery on $M_c$. 
In the limit $d \rightarrow \infty$, one can show that 
the following approximation holds (see appendix C):
\BE
v(d)=v_c\biggl (1-\f{\exp(-2(d-d_0))}{\delta+1} \biggr ),
\label{eq:volapp}
\EE
where $\delta$ denotes a ratio of the volume of the complementary part
to that of the ``thin'' part and $d_0$ is the diameter of
the complementary part. For a sample of 41 manifolds with diameter
longer than 1.65 in group B, the best fits are
$\delta=0.0$ and $d_0=0.25$. As shown in figure \ref{fig:dvol},
the volume-diameter relation for CH manifolds with 
large $l_{min}$ is well described
by the fitting formula (\ref{eq:dvola}). As $l_{min}$ 
becomes smaller, or equivalently, $d$ becomes larger, 
a CH manifold $M$ converges to the
original cusped manifold $M_c$ in which (\ref{eq:volapp})
holds.
\BF
\centerline{\psfig{figure=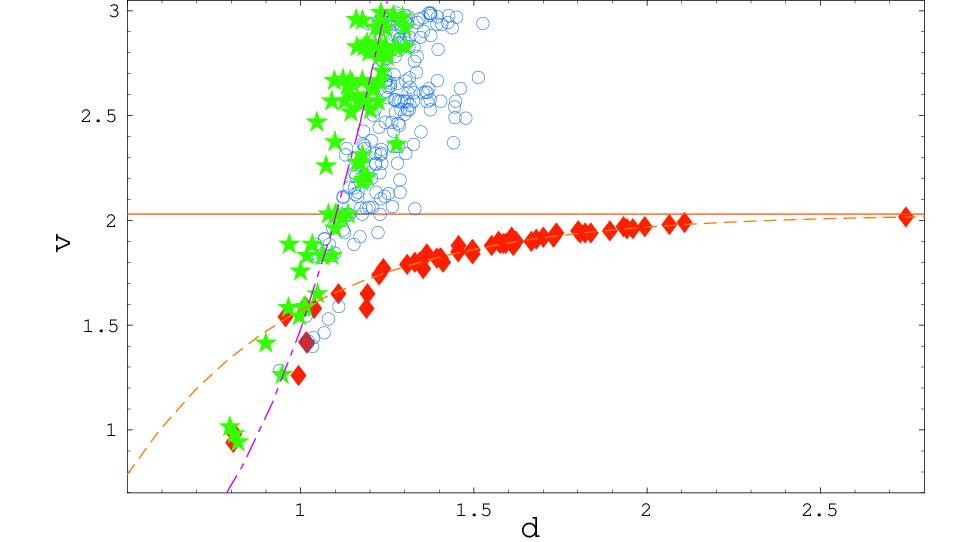,width=13cm}}
\caption{Diameter vs Volume}
\mycaption{Diameter $d$ versus volume $v$ for 79 manifolds 
($l_{min}\!>\!0.5$)(star), 184 manifolds 
($0.5\!>\!l_{min}\!>\!0.3$)(circle) and 53 manifolds 
obtained by performing Dehn surgeries on m003 (group B, diamond)
with the best-fit curves  
(\ref{eq:dvola}) (dashed-dotted curve) and (\ref{eq:volapp}) 
(dashed curve). 
The solid curve denotes the volume of m003. }

\label{fig:dvol}
\EF
\\
\indent
Finally, we look into the relation between the first eigenvalue and 
the volume. For CH manifolds with sufficiently large
$l_{min}$, (\ref{eq:fitB}) and (\ref{eq:dvola}) give 
\BE
v(k_1)=\pi(\sinh (g(k_1))-g(k_1)),~~~ g(k_1)=\f{4 \pi \alpha}{\beta \sqrt{
(k_1)^2-(k_1(\textrm{cusp}))^2}}.
\label{eq:volk1iso}
\EE
To be consistent with the Weyl's asymptotic formula which is valid for 
$k_1\!>\!>\!1$
\BE
k_1(v)
=\sqrt{\biggl ( \f{9 \pi^2}{v} \biggr )^{\f{2}{3}}+1},
\label{eq:k1Weyl}
\EE
the fitting parameters should satisfy  
$\alpha/\beta=3\cdot 4^{-5/6}\pi^{-2/3}\approx 0.44$ which well
agree with the numerically computed 
values $\alpha/\beta=0.69/(1.6-1.7)=0.41-0.43$ provided that 
$k_1(\textrm{cusp})=1$.  
One can see from figure \ref{fig:vk1} that 
both (\ref{eq:volk1iso}) and (\ref{eq:k1Weyl}) give a good estimate
of the first eigenvalue $k_1^2$
for globally ``slightly anisotropic'' 
manifolds with $v<3$ and $l_{min}>0.5$. 
\BF
\centerline{\psfig{figure=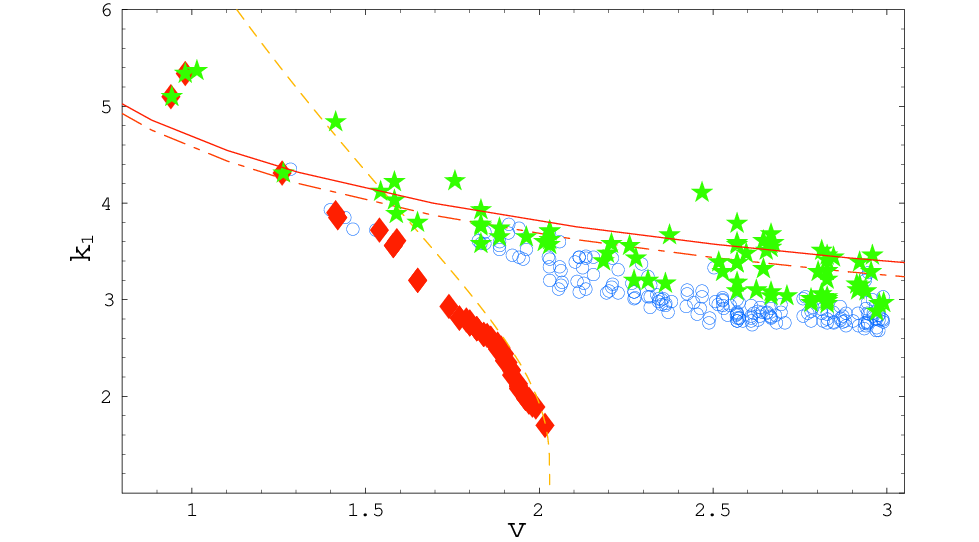,width=13cm}}
\caption{Volume vs First Eigenvalue}
\mycaption{Volume $v$ versus $k_1$ for 79 manifolds 
with $l_{min}\!>\!0.5$ (star), 184 manifolds 
with $0.5\!>\!l_{min}\!>\!0.3$ (circle) and 53 manifolds 
obtained by performing Dehn surgeries on m003 (group B, diamond)
with fitting curves 
(\ref{eq:volk1iso})(solid curve), (\ref{eq:k1Weyl})(dashed-dotted
curve), and (\ref{eq:k1vcusp})(dashed curve) 
where $k_1(\textrm{cusp})=1$ is assumed. 
}
\label{fig:vk1}
\EF
For manifolds with large $d$, $v$ and small $l_{min}$,
(\ref{eq:volk1iso}) and (\ref{eq:k1Weyl}) give
incorrect estimates for $k_1$. In that case, 
the effect of the curvature cannot be negligible. 
In contrast to compact flat spaces,
the volume of a sphere in the hyperbolic space
increases exponentially as the radius increases. Assuming 
that a relation (\ref{eq:volk1iso}) holds, 
for sufficiently globally ``isotropic'' CH manifolds (large 
$l_{min}$), \ti{$k_1$ is significantly larger than
that for compact flat spaces with the same volume even if one
assumes that $k_1(\textrm{cusp})\sim 0$}. 
However, for CH manifolds converging to the 
original cusped manifold $M_c$,
the formula (\ref{eq:volk1iso}) has to be modified.
If $l_{min}$ is sufficiently small and $d$ is large while
keeping the volume finite then $M$ has a ``thin'' part similar
to the neighborhood of a cusp. Then
one can use an asymptotic formula 
(\ref{eq:volapp}) instead of (\ref{eq:dvola}). 
\BE
k_1(v)=\sqrt{(k_1(\textrm{cusp}))^2+\f{4 \pi^2}{\beta^2
(d_0-\ln(1-v/v_c)/2)^2}},
\label{eq:k1vcusp}
\EE
where $v_c$ is the volume of $M_c$. 
As shown in figure \ref{fig:devk1minl}, 
$k_1$ shifts to a smaller value for
manifolds with smaller $l_{min}$ which 
have larger $d$. 
\BF
\centerline{\psfig{figure=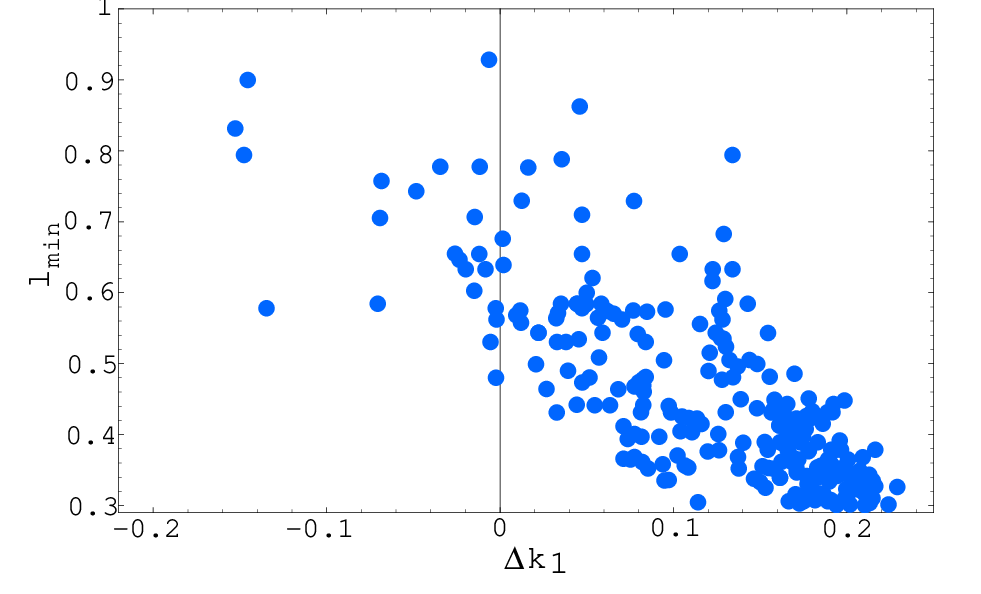,width=8.3cm}
\psfig{figure=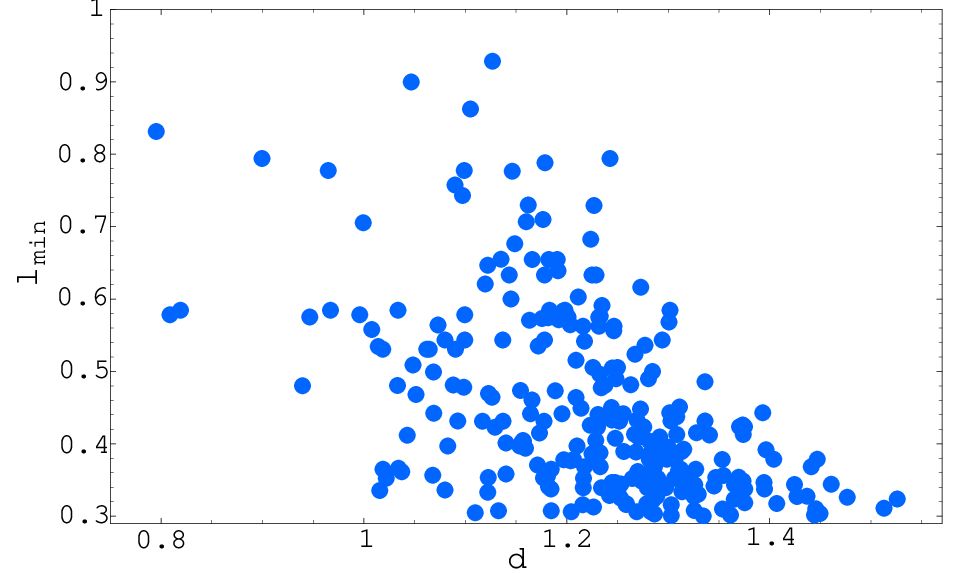,width=8.3cm}}
\caption{Length of Shortest Periodic Orbit vs $k_1$ or Diameter}
\mycaption{The length of the shortest periodic orbit 
$l_{min}$ versus  
the deviation $\Delta k_1$ from the fitting formula 
(\ref{eq:volk1iso}) where $k_1(\textrm{cusp})\!=\!1$ is assumed(left)
 and $l_{min}$ versus  
diameter $d$ (right) for 263 CH manifolds (group A).}
\label{fig:devk1minl}
\EF
In the limit
$M \rightarrow M_c$, the length of 
periodic orbits of $M$ also converges
to that of $M_c$ except for the shortest orbit whose length 
$l_{min}$ goes to zero. From the general Selberg 
trace formula (\ref{eq:GTF}), one can see that the 
wave which corresponds to the shortest orbit has a 
large amplitude $\sim 1/\cosh{l_{min}}$ with a long
wavelength $\sim 2 \pi/l_{min}$. Therefore, the presence
of a very short periodic orbit results in the deviation 
of the energy spectrum
at low-energy region (small $k$) from the asymptotic distribution. 
To confirm this,  the spectral staircase 
using the length spectra of m003(16,13) but the shortest periodic 
orbit is removed has been computed. Note that m003(16,13)
is very similar to the original cusped manifold m003
having $d=2.75$ and $l_{min}=0.0086$. As shown in 
figure \ref{fig:sdev}, the computed spectrum staircase agrees well with
Weyl's asymptotic formula. The computed spectrum may coincide
with the one for a manifold in which the ``thin'' part is cut off. 
\BF
\centerline{\psfig{figure=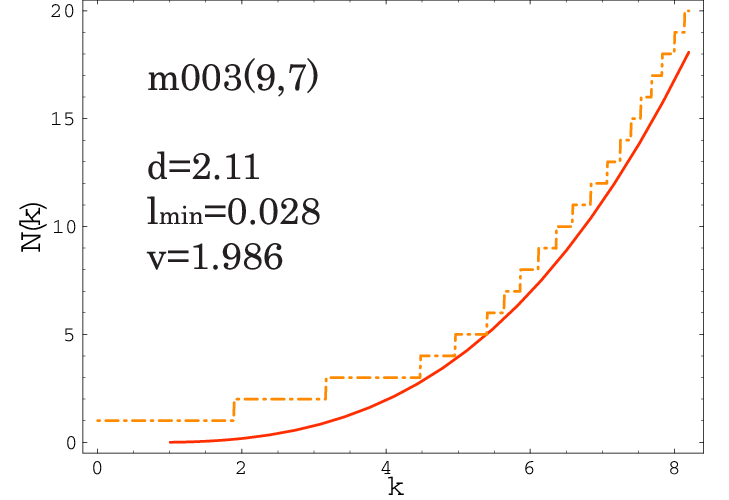,width=8.5cm}
\psfig{figure=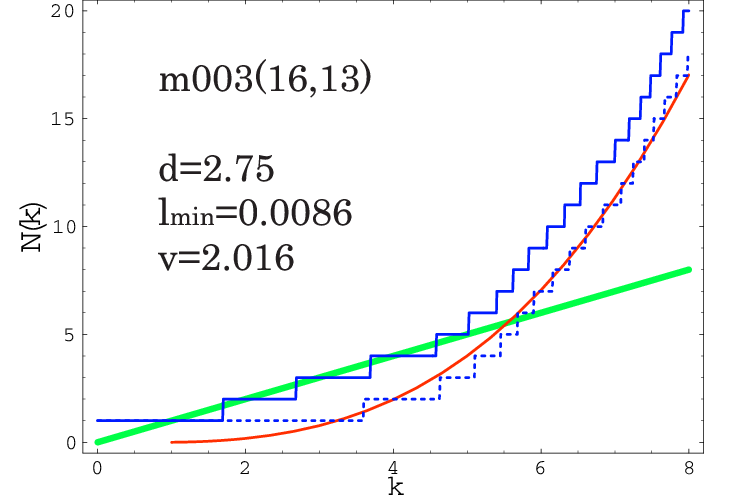,width=8.5cm}}
\caption{Deviation from Weyl's Formula} 
\mycaption{Spectral staircases for m003(9,7) (dashed-dotted staircase, left)
, m003(16,13) (solid staircase, right) and m003(16,13) without 
the shortest periodic orbit (dotted staircase, right) are shown 
in comparison with the corresponding Weyl's asymptotic
formula (solid curves) which have been numerically computed by
the POSM using all periodic orbits with length $l<7$.  
$N(k)=k$ perfectly fits the staircase of m003(16,13) 
for $k<5$ (thick line, right). }
\label{fig:sdev}
\EF
On the other hand, the spectral staircases for
m003(9,7) and m003(16,13) which have small $l_{min}$
deviate from the average part 
(=Weyl's asymptotic formula(\ref{eq:Weyl})) (figure \ref{fig:sdev}).  
These manifolds have a ``thin'' part which is 
virtually one-dimensional object.  Let us remind that
Weyl's asymptotic formula
for a n-dimensional compact manifold $M$ is given by
\BE
N(k)\sim \f{\omega_n v(M) k^n}{(2 \pi)^n},~~~~k>>1,
\label{eq:Weyln}
\EE
where $\omega_n=2 \pi^{n/2}/(n\Gamma(n/2))$ 
is the volume of the unit disk in the Euclidean 
n-space and $v(M)$ is the volume of $M$. If one assumes that 
the Weyl formula  still holds for small $k$, then 
the spectrum in the low-energy region
which corresponds to fluctuations on large scales for
these manifolds can be approximately 
described by the Weyl formula for $n=1$, $N(k)\propto k/\pi$.
For m003(16,13), it is numerically found that $N(k)=k$ 
provides a good fit for $k<5$.  
In the next section, we will measure the deviation from
the asymptotic distribution
using the low-lying eigenvalue spectra.
\section{Spectral measurements of global anisotropy}
Among many possibilities, we should choose
physically well-motivated quantities for measuring the global 
``anisotropy'' in geometry in terms of eigenvalue spectra.
First, we consider $\zeta$-function which is relevant to 
the CMB anisotropy. The 
angular power spectra for CH universes are 
approximately written as
\BE
C_l\sim \sum_{i=1}^{\infty} \f{F_l(k_i)}{k_i^3},
\EE
where $F_l(k)$ can be approximated as a polynomial function of $k$
for a given $l$. 
In order to measure the ``anisotropy'', we define the following 
parameter in terms of the $\zeta$-function,
\BE
\Delta(s)\equiv \zeta(s)/\zeta_w(s)= 
\sum_{i=1}^{\infty} k_i^{-2 s} \bigg / ~
\sum_{i=1}^{\infty} k_{wi}^{-2 s},~~k_i^{-2 s}\sim \f{F_l(k_i)}{k_i^3}
\EE
where $k_i^2$ are the eigenvalues of the 
Laplace-Beltrami operator on a CH 3-manifold $M$
with volume $v$
and  $k_{wi}^2$ are the eigenvalues obeying
Weyl's asymptotic formula 
with volume $v$ (\ref{eq:k1Weyl}). 
Note that the zero-mode $k_0=0$ is not
included in the summation. 
Here we only consider 
the case $s>1$ which ensures the convergence of the sum. 
The numerical result shows the clear difference
between the `slightly anisotropic'' manifold m003(-3,1) 
($d\!=\!0.82,l_{min}\!=\!0.58$)
, ``somewhat anisotropic'' manifold m003(9,7) 
($d\!=\!2.11,l_{min}\!=\!0.028$) and ``very anisotropic'' manifold
 m003(16,13) ($d\!=\!2.75,l_{min}\!=\!0.0086$)
(figure \ref{fig:dis}). The presence of the 
fluctuations on large scales in these ``anisotropic'' manifolds
shifts the corresponding $\zeta$-function to a larger value.
For the case in which the shortest periodic orbit is removed from the 
length spectra of m003(16,13), one can see that the spectrum
coincides with that obeying Weyl's asymptotic formula. 
\\
\indent 
Next,  we consider the spectral distance $d_s$ proposed by
Seriu which measures the degree of  semi-classical quantum 
decoherence between two universes having one massless scalar 
field\cite{Seriu},
\BE
d_s[M,\tilde{M}]\equiv\f{1}{2}\sum_{i=1}^\infty \ln \f{1}{2} \biggl (
\f{k_i}{\tilde{k}_i}+\f{\tilde{k}_i}{k_i} \biggr ),
\EE
where $k_i^2$ and $\tilde{k}_i^2$ are the eigenvalues of the
Laplace-Beltrami operator on a compact n-manifold 
$M$ and $\tilde{M}$, respectively. Here we choose eigenvalues
$k_{wi}^2$ as $\tilde{k}_i^2$. In practice we introduce
a cutoff in the summation. It is numerically found that 
the summation converges rapidly for the 3 examples, namely, 
m003(-3,1), m003(9,7) and m003(16,13). The contribution of
the first several terms dominates the summation (figure
\ref{fig:dis}). The result implies that a universe 
having a spatial geometry m003(9,7) or m003(16,13)
semiclassically decoheres with a universe having a spatial geometry 
m003(-3,1) (figure \ref{fig:Dirichlet}).  
\BF
\centerline{\psfig{figure=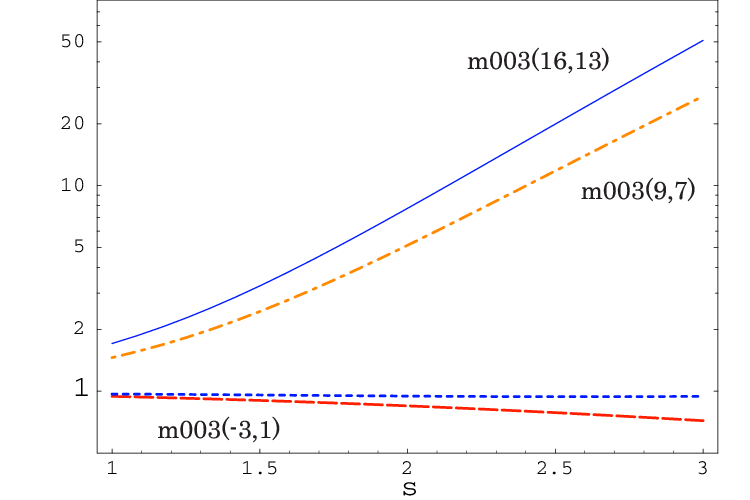,width=8.5cm}
\psfig{figure=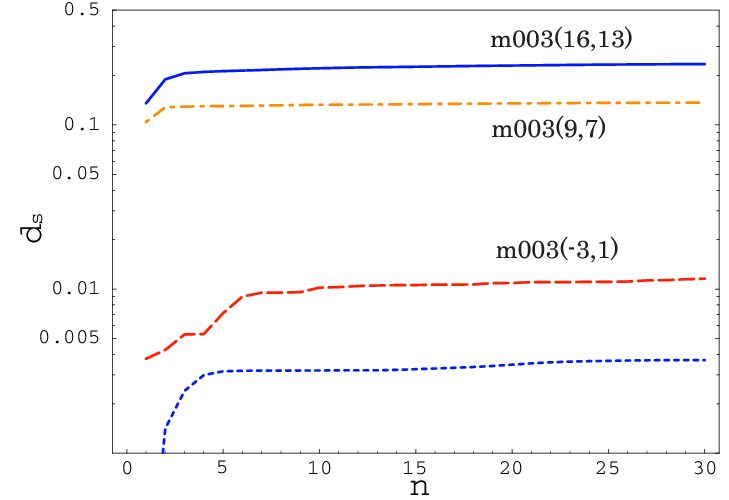,width=8.5cm}}
\caption{Spectrum Measurements} 
\mycaption{The figure
in the left shows  
$\Delta (s)$=$\zeta(s)/\zeta_w(s)$ for 3 examples of CH 3-manifolds
,m003(9,7) (dashed-dotted curve), m003(16,13) (solid curve),
m003(-3,1) (dashed curve) and m003(16,13) where the shortest 
periodic orbit is removed (dotted curve). The figure in the right shows
$d_s(n)=\f{1}{2}\sum_{i=1}^n \ln \f{1}{2} (
k_i/k_{wi}+k_{wi}/k_i )$ for the same examples 
where $n$ is the cutoff number in the summation.}     
\label{fig:dis}
\EF
\BF
\begin{center}
\centerline{\psfig{figure=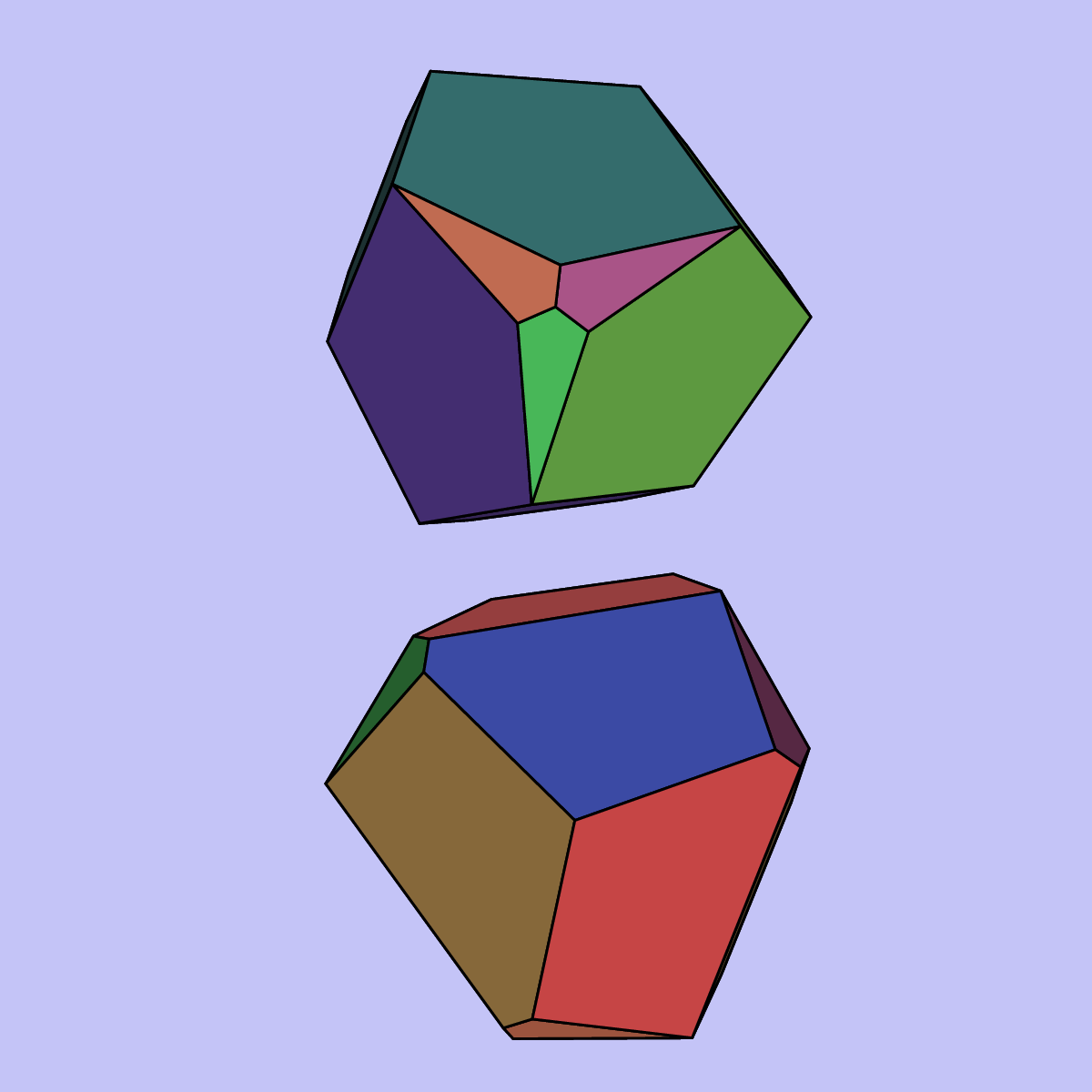,width=7.5cm}
\psfig{figure=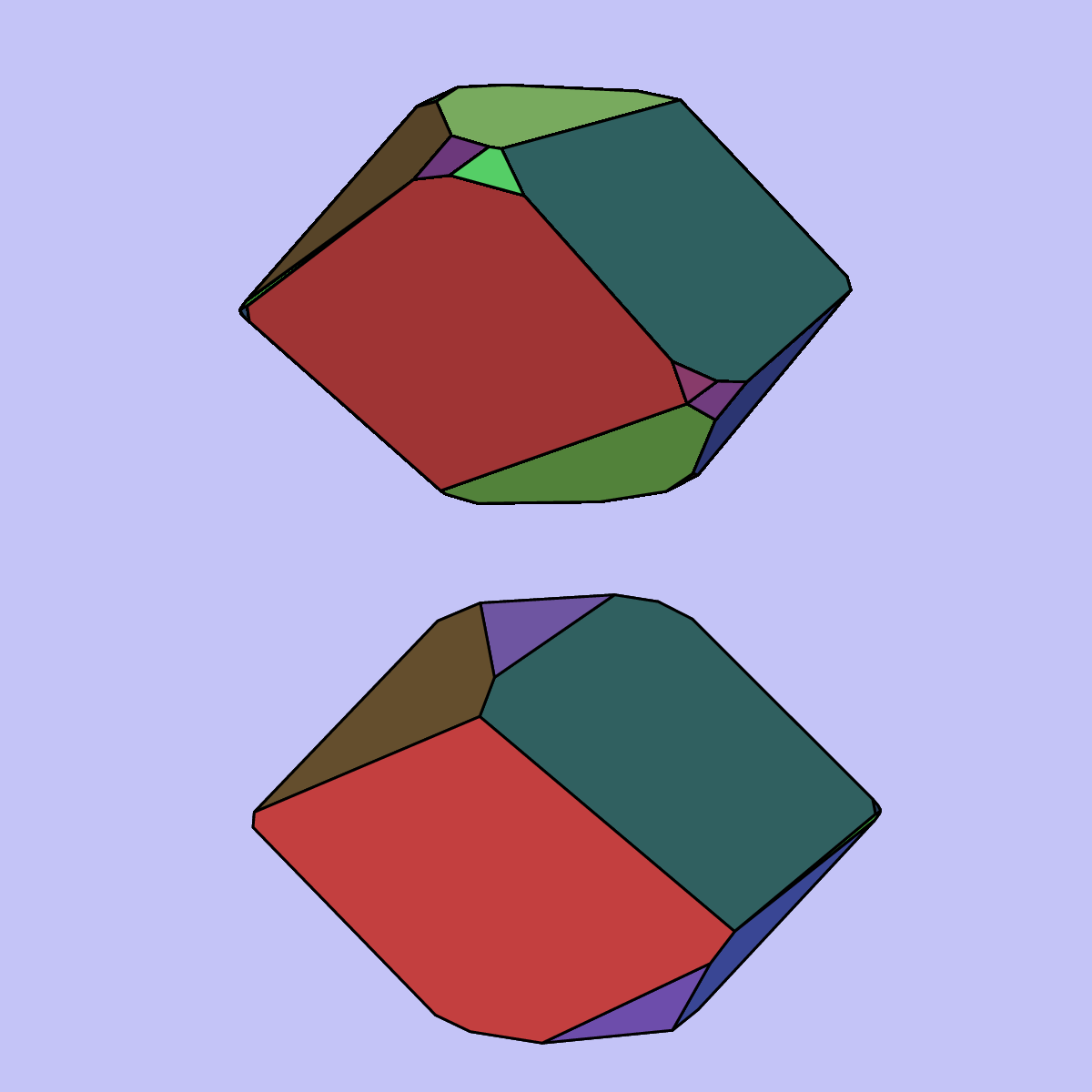,width=7.5cm}}
\caption{Anisotropy in Geometry}
\mycaption{Plots of a Dirichlet domain of the 
Weeks manifold m003(-3,1) (left) 
and that of m003(16,13) (right) viewed from two opposite directions 
in the Klein coordinates. The Dirichlet domain of m003(16,13)
is quite similar to that of the original cusped manifold m003
(see figure \ref{fig:fig8}). m003(16,13) has a ``thin'' part
which is similar to the neighborhood of a cusp. The colors on the faces
correspond to the
identification maps. }
\label{fig:Dirichlet}
\end{center}
\EF
\chapter{CMB Anisotropy}
\thispagestyle{headings}
\bq
\ti{The most incomprehensible thing about the universe is that it is
comprehensible.}
\vskip 0.1truecm
\rightline{(Albert Einstein, 1879-1955)}
\eq
\vspace{0.5cm}
The surprising discovery of the cosmic microwave background (CMB)
by Penzias and Wilson in 1964 \cite{Penzias65}
provided firm evidence that our universe started from
a hot big-bang.
The CMB is considered as the fossil of the 
photons as long ago as redshift $z\sim 10^3$ when the scale
factor was $10^{-3}$ of the present value and the temperature was hot as
$T=3\times10^3 K$. The almost perfect blackbody spectrum in the CMB 
implies that the radiation and matter were once in good thermal 
contact because of the interaction between photons and 
electrons through Compton scattering when the universe was very hot.
As the temperature decreased gradually owing to the cosmic expansion, 
the electrons have started to
combine with protons. Eventually, the 
density of free electrons became too low
to maintain the thermal contact and photons decoupled with matter.
During the subsequent cosmic expansion the wavelengths of the decoupled
photons are stretched which leads to the low temperature $2.7K$ in the
background today.
\\
\indent
The COBE satellite launched in 1989 discovered a
slight imperfection in the CMB:the temperature anisotropy
at the level of one part in $10^{-5}$ on angular scales $\sim 10^o$ and
roughly agree with the scale-invariant ($n=1$) initial spectrum
\cite{Smoot92}.
The temperature anisotropy provides not only the information
about the physical conditions at redshifts $z\sim 10^3$
but also provides an important 
probe of the spatial geometry on the present horizon scales.
On superhorizon scales, which are much larger than the Jeans scale, 
the interaction between
radiation and matter is negligible. Therefore the
photon-baryon fluid can be seen as a single fluid. 
The anisotropy is determined by only gravitational effects.
There are two kinds of effects which account for the 
generation of the CMB anisotropy\cite{Sachs,HSS}.
One is the so-called ordinary Sachs-Wolfe
(OSW) effect. 
When the photons climbs out the gravitational potential
at the last scattering, the temperature of the photons are redshifted.
Because the intrinsic temperature fluctuations
at the last scattering is specified by the  
gravitational potential if the initial condition is given, 
the temperature anisotropy is associated with
the spatial fluctuation of the 
gravitational potential at the last scattering.
Another one is the so-called integrated Sachs-Wolfe
(ISW) effect. If the gravitational potential decays
while the photons are climbing out the potential, 
the expected redshift would be reduced. The decay occurs
when the fluctuation scale enters the sound horizon while
the radiation is still dominant and when the curvature term or the
cosmological constant dominates the total energy.
\BF
\centerline{\psfig{figure=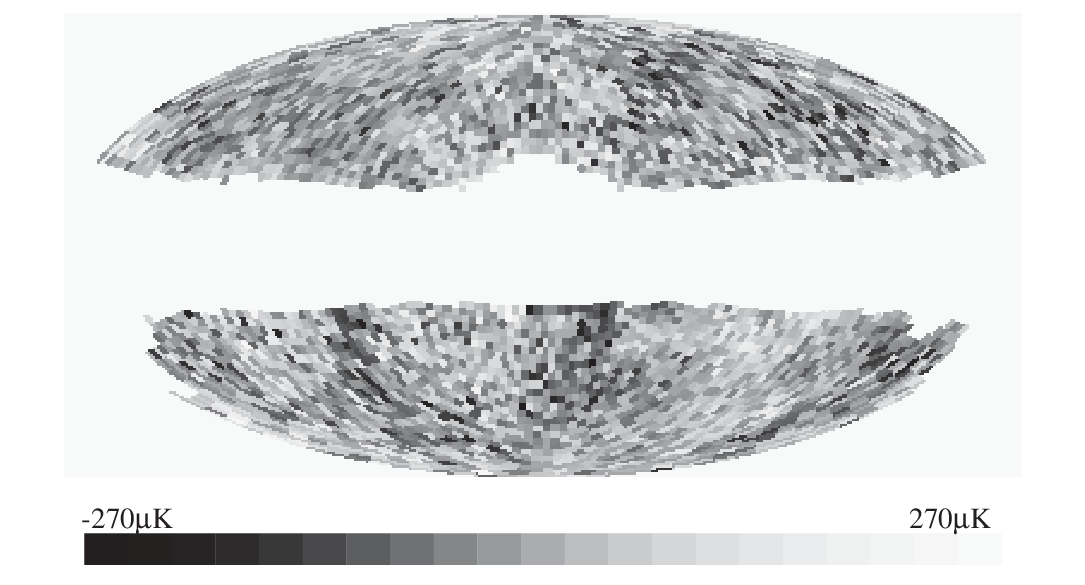,width=14cm}}
\caption{COBE-DMR Anisotropy Sky Map}
\mycaption{Temperature anisotropy detected by the COME-DMR experiment
is at the level of $\Delta T/T\sim 10^{-5}$. The plotted sky map 
of temperature fluctuations $\Delta T$ (in the Aitoff projection) has been
obtained by averaging data at frequencies 31GHz, 53GHz, and 90GHz
\cite{COBE}. 
The monopole and dipole components have been removed. }
\label{DMR}
\EF
\\
\indent
It is of crucial importance to test
whether the CMB anisotropy in spatially multiply connected models
is consistent with the COBE data. The observable effect of the 
non-trivial spatial topology 
becomes much prominent for models with smaller volume
since apparent fluctuations beyond the size of the 
fundamental domain are strongly suppressed.  
In other words, the periodical boundary conditions restrict 
the wave numbers $k$ to have discrete values
that are equal to or larger than
the that of the first non-zero mode $k_1$ (the first eigenvalue).
As we have seen,  $k_1$ is related with the diameter, the shortest
length of periodic geodesics and volume
of the space. 
\\
\indent
If the CMB anisotropy is completely determined by the OSW effect 
then the  ``mode-cutoff'' in the eigenmodes implies the 
suppression on the large-angle power spectrum that are statistically isotropic. 
Fluctuations on large angular scales which corresponds
to fluctuations on large physical scales beyond
the actual size of the space are strongly suppressed.
In fact, for closed flat toroidal
models without the cosmological constant,
the COBE data constrain the size of the fundamental domain 
$L>0.8 R_\ast$ where $R_\ast$ is the 
comoving radius of the last scattering surface
and $L$ is the side of the cube
\cite{Sokolov93,Staro93,Stevens93,Oliveira95a,SLS99}.
It should be emphasized that the constraint itself 
is not enough stringent to completely ``rule out'' 
closed flat toroidal models since it corresponds to $N<8$ where $N$ is 
the number of copies of the fundamental domain 
within the observable region at present.
\\
\indent
In contrast, for low matter density models, the constraint could be
considerably milder than the locally isotropic and homogeneous 
flat (Einstein-de-Sitter) models
since a bulk of large-angle CMB fluctuations 
can be produced by the ISW
effect which is the gravitational blueshift
effect of the free streaming photons by the decay of the gravitational 
potential\cite{CSS98b,Inoue00}. 
As the gravitational potential decays in either 
$\Lambda$-dominant epoch or curvature dominant epoch, 
the free streaming photons  
with large wavelength (the light travel time across the wavelength 
is greater than or comparable to
the decay time) that climbed a potential well at the last scattering 
experience blueshifts due to the contraction of the comoving space
along the trajectories of the photons.
Because the angular sizes of the fluctuations produced at late time 
are large, the suppression of the fluctuations on scale larger than the 
topological identification scale does not lead to a
significant suppression of the large-angle power if the 
ISW effect is dominant. In other words, the relevant fluctuations
which have been produced at late time $z=1\sim3$ are ``small''
enough compared with the size of the fundamental domain that 
they are not affected by the ``finiteness'' in the size of the 
spatial hypersurface.
In fact, recent works 
\cite{Aurich99,Aurich00,Inoue0,Inoue2,CS00} have shown that 
the large-angle power ($2\!\le\!l\!\le\!20$) is completely consistent with
the COBE DMR data for closed hyperbolic (CH) models whose spatial
geometry is described by 
a small CH orbifold, the Weeks and Thurston CH manifolds 
with volume $0.36, 0.94$ and $0.98$ in unit of the cube of the
curvature radius, respectively. Note that the Weeks manifold is the
smallest and the Thurston manifolds is the second smallest 
in the known CH manifolds.
\\
\indent
These results are clearly at odds with the previous constraints
\cite{Bond1,Bond2} on CH models
based on pixel-pixel correlation statistics.
It has been claimed that the statistical analysis using only 
the power spectrum is not sufficient since it
can describe only isotropic 
(statistically spherically symmetric) correlations.
This is true as long as one considers fluctuations observed at 
a particular point. Because any CH manifolds are globally
anisotropic, expected fluctuations would be also statistically 
globally anisotropic at a particular point. 
In order to constrain CH models, it is necessary to 
compare the expected fluctuation patterns at 
all the places with the data
since CH manifolds are globally 
inhomogeneous. 
However, it should be emphasized that the stringent 
constraints obtained in the previous analyses \cite{Bond1,Bond2}
are only for CH models
at a \ti{particular observation point} $Q$ where the injectivity 
radius is locally maximum for 24 \ti{particular orientations}. 
$Q$ is rather a special point at which 
some of the mode functions have a symmetric structure.  
It is often the case that the base point $Q$ becomes a fixed point of
symmetries of the Dirichlet fundamental domain or the manifold.
As we have seen, the mode functions for closed hyperbolic spaces
are well described by a set of Gaussian numbers which corresponds
to a choice of the position and orientation.  Therefore,
we expect that \ti{the likelihood should be highly dependent on the
choice of the place and orientation of the observer} which had been
ignored in the previous analysis. 
\\
\indent
In this chapter, we first derive the evolution of 
the linear perturbations expressed in the 
gauge-invariant formalism. Next, we solve the Boltzmann equation
on the perturbed background for computing
the CMB anisotropy. 
Then we investigate the angular power spectra for 
estimating the suppression on the large angular scales. Finally we 
carry out the Bayesian analysis using the COBE DMR 4-year 
data incorporating the effects of inhomogeneity and anisotropy
of the background geometry for testing closed flat and hyperbolic
models with low matter density.
\section{Linear perturbation theory}
The observable physical quantity should not be
dependent on choice of gauge. However, it is also true that 
a different time slice gives a different value of fluctuation.
In other words, observers with different dynamics would
observe different universes. If the physical state of the observer
is fixed, then no freedom is allowed. For instance, the synchronous  
gauge is a natural choice for the present free falling observer. 
If the observer were not free falling but 
at the rest frame of the total matter, then the total matter
gauge is a useful choice. Since the choice of the physical 
state of the imaginary observer in the past is arbitrary,
physically, none of gauge choices has priority 
over others. Mathematically speaking, the gauge freedom arises due to 
the freedom of correspondence between a perturbed manifold $M$ and 
a reference (locally homogeneous and isotropic) manifold $\tilde{M}$. 
\\
\indent
However, it is useful to choose a gauge in which we can ``easily''
understand the evolution of perturbation.  
Newtonian gauge is useful in analyzing propagation
of photons while the description of matter becomes simpler in the
total matter gauge. All we have to do is to choose a completely fixed gauge 
in which physical interpretation and evolution of the fluctuations 
are ``simplest''.    
\\
\indent
The line element of the perturbed FRWL metric takes the form
\BE
ds^2 = -a^2(\eta) \Bigl[(1+2 A) d\eta^2+B_{j} d\eta
dx^j-(\gamma_{ij}+2H_{L}\gamma_{ij}+2 H_{Tij})dx^idx^j\Bigr]~,
\EE
where 
\BE
\gamma_{ij}dx^i dx^j= \cases{-K^{-1}(d \chi^2+sinh^2\chi(d\theta^2+
sin^2\theta d\phi^2)& K$<$0\cr
(dx^1)^2+(dx^2)^2+(dx^3)^2 & K=0~.\cr}
\EE
$A$,$B_j$,$H_L$,and $H_{Tij}$ are functions of space and time that
describe a perturbation and $H_{Tij}$ denotes a
three-dimensional traceless tensor.  
It is convenient to decompose the metric perturbation variables in the 
following way. First, a 3-vector $B_j$ can be 
decomposed as the sum of the first 
derivative of a longitudinal 3-scalar $B_L$ and a transverse 3-vector
$B_{Tj}$.
\BE
vector~~~~~ B_{j}={\cal D}_jB_L+B_{Tj}~,~~~~~{\cal D}_{j}B_T{}^j=0,
\label{eq:vector}
\EE
where ${\cal D}_j$ is defined as the three dimensional covariant
derivative with metric $\gamma_{ij}$. Similarly, a 3-tensor $H_{ij}$ 
can be decomposed as the sum 
of a longitudinal 3-scalar $H_L$, the second derivative of a tracefree 
3-scalar $H_T$, the first derivative of
a transverse 3-vector $H_{Lj}$ and a transverse tracefree 3-tensor
 $H_{TTij}$ as
\BE
tensor~~~~~~~~H_{ij}=\gamma_{ij}H_L+\Bigl[{\cal D}_i{\cal
D}_j-\gamma_{ij}\frac{{\cal D}^2}{3}\Bigr]H_T+
{\cal D}_i H_{Tj}+{\cal D}_j H_{Ti}+H_{TTij}\label{eq:tensor}. 
\EE
For locally isotropic and homogeneous background geometry,
the Einstein equations are divided into a decoupled set of equations
each one of which consists of only one type of 
perturbation such as ``scalar'', 
``vector'' or ``tensor'' type perturbation\footnote{If the perturbed
spatial geometry  
is locally homogeneous but anisotropic, we have coupled equations
between different types of 
perturbations so that the decomposition is no longer 
convenient \cite{Kodama}.} Consequently, we can argue separately 
for each type of components. 
\\
\indent
The infinitesimal gauge transformation 
\BE
\eta \rightarrow \tilde{\eta}=\eta+T(x),~~~~x^j \rightarrow \tilde{x}^j=
x^j+L^j(x),
\EE    
transforms the metric as
\BEA
\tilde{g}^\mu{}_\nu(\tilde{\eta}, \tilde{x}^i)&=&\frac{\partial
\tilde{x}^\mu}{\partial x^\alpha}\frac{\partial
{x}^\beta}{\partial \tilde{x}^\nu}~ g^\alpha{}_\beta(\tilde{\eta}-T, \tilde{x}^i-L^i)
\nonumber
\\
\nonumber
\\
&=& g^\mu{}_\nu+ g^\alpha{}_\nu (\delta x^\mu)_{,\alpha}-
g^\mu{}_\beta (\delta x^\beta)_{,\nu}-g^\mu{}_{\nu,\lambda}\delta
x^\lambda,
\label{eq:gtransform}
\EEA
which gives the following transformation law, 
\BEA
scalar~ type~~~~~~~~~~~\tilde{A} &=& A-T'-\frac{a'}{a} T,
\nonumber
\\
 \tilde{B}_L&=&B_L+L_L'-T,
\nonumber
\\
  \tilde{H}_L&=&H_L-\frac{{\cal D}^2 L_L}{3}-\frac{a'}{a} T,
\nonumber
\\
  \tilde{H}_T&=&H_T-L,
\\
\nonumber
\\
vector~ type~~~~~~~~\tilde{B}_{Tj} &=& B_{Tj}+L'_{Tj},
\nonumber
\\
\tilde{H}_{Tj}&=&H_{Tj}-L_{Tj},
\\
\nonumber
\\ 
tensor~ type~~~~~\tilde{H}_{TTjk} &=& H_{TTjk},
\EEA
where $'$ denotes $\partial/\partial \eta$.
\\
\\
\indent
The perturbed energy momentum tensor is written as
\BEA
T^o{}_\nu &=&(\tilde{P}+\tilde{\rho})\tilde{u}^o \tilde{u}_\nu
+\T{P}\delta^o_\nu
\nonumber
\\
T^i{}_j &=&(\tilde{P}+\tilde{\rho})\tilde{u}^i \tilde{u}_j+\tilde{P} \delta^i{}_j+\delta
 P^{~i}_{T~j}{},
\EEA
where 
\BEA
\tilde{P}=P+\delta P_L,~~~\tilde{\rho}=\rho+\delta\rho,~~~\tilde{u}^\mu=
u^\mu+\delta u^\mu,
\nonumber
\\
u^\mu=(a^{-1},0,0,0),~~~~u_\mu=(-a,0,0,0),~~~\sum_{i=1}^3 
\delta P_T^{~i}{}_{~i}=0.
\EEA
$P$ is the averaged pressure, $\rho$ is the averaged energy density, 
$\delta P_L$ is the longitudinal pressure perturbation and 
$\delta P^{~i}_{T~j}$ is the anisotropic pressure of the fluid. Note that 
we have chosen coordinates for which the averaged velocity of the fluid 
vanishes. 
\\
\indent
Let us introduce perturbation variables for the density and stress,
\BEA
\pi_L\equiv\frac{\delta P_L}{P},~~~\pi^{~i}_{T~j}\equiv\frac{\delta 
P^{~i}_{T~j}}{P},~~~~\delta\equiv\frac{\delta \rho}{\rho}, 
\EEA
and for the three-dimensional velocity,
\BE
V^j\equiv a\frac{dx^j}{dt}=\frac{dx^j}{d\eta}=\frac{\delta u^j}{u^o}.
\EE
Since the three dimensional velocity is defined on the space with
metric $\gamma_{ij}$, the index of $V^j$ is raised or lowered
with $\gamma^{ij}$, $\gamma_{ij}$, respectively.
\\
\\
\indent
In terms of these variables, the energy momentum tensor is written as
\BEA
T^o{}_o&=&-\rho(1+\delta),~~~~~~T^o{}_j=(\rho+P)(V_{j}-B_{j}),
\nonumber
\\
T^j{}_o&=&-(\rho+P)V^j,~~~T^j{}_k=P(\delta^j{}_k+\pi_L 
\delta^j{}_k+\pi^{~j}_{T~k}).
\EEA
Like the metric perturbation, the perturbation variables for matter
are decomposed into three different types of components. The transformation 
law is described as 
\BEA
scalar~type~~~~~~~~~~~\tilde{\delta}&=& \delta+3(1+w)\frac{a'}{a} T,
\nonumber
\\
\tilde{V}_L&=&V_L+L_L',
\nonumber
\\
  \tilde{\pi}_L&=&\pi_L+3c_s{}^2\frac{(1+w)}{w}\frac{a'}{a} T,
\nonumber
\\
  \tilde{\pi}_T&=&\pi_T,
\\
\nonumber
\\
vector~type~~~~~~~~\tilde{V}_{Tj} &=& V_{Tj}+L'_{Tj},
\nonumber
\\
\tilde{\pi}_{tj}&=&\pi_{Tj},
\\
\nonumber
\\ 
tensor~type~~~~~\tilde{\pi}_{TTjk} &=& \pi_{TTjk},
\EEA
where $w\equiv P/\rho$, and $c_s^2\equiv dP/d\rho$.
\\
\\
\indent
For simplicity,  we will consider only the scalar type perturbations.
It should be noted that the degree of apparent 
freedom of the metric perturbation is four and the degree of gauge 
freedom is two, so that
the degree of actual freedom of the metric perturbation is two 
for the scalar type perturbations.
\\
\\
\indent
In linear theory, each normal mode evolves independently. Therefore, 
the mode expansion is a very useful method in analyzing the evolution
of perturbed quantities. From now on, we follow the notations by Kodama 
and Sasaki \cite{Kodama}. A harmonic (=a mode function of the
Laplacian) is defined as a function 
which satisfies the Helmholtz equation
\BE
({\cal D}^2+k^2)u_{\mathbf{k}}=0,
\EE
\\
The line element for the scalar type 
perturbation can be written in terms of harmonics as
\BEA
ds^2 = -a^2 \Bigl[(1+2 A u) d\eta^2+
B_L u_j~ d\eta~dx^j-
\nonumber
\\
~~~~~~~~~~~~~~~~~~~~
((1+2H_{L}u)\gamma_{ij}+2 H_T u_{ij})dx^idx^j\Bigr], 
\label{eq:scalar}
\EEA
where $u_i$ and $u_{ij}$ are defined as
\\
\BE
u_j\equiv -k^{-1}{\cal D}_j u,~~~~~u_{ij} \equiv k^{-2}({\cal D}_i 
{\cal D}_j-\frac{{\cal D}^2}{3}\gamma_{ij})u.
\EE
Note that we have omitted the labels for the wave number
$\mathbf k$ in Eq.(\ref{eq:scalar}) since 
there is no mode coupling. 
The energy momentum tensor $T^\mu{}_\nu$ is given by
\BEA
T^o{}_o&=&- \rho(1+\delta u),~~~~~~~T^o{}_j=(\rho+P)(V_L-B_L)u_j,
\nonumber
\\
T^j{}_o&=&-(\rho+P)V_L u^j,~~~T^j{}_k=P(\delta^j{}_k+\pi_L u 
\delta^j{}_k+\pi_T u^j{}_k).
\EEA
\indent
We have to make a choice of gauge in which the physical interpretation and 
evolution are simplest. The Newtonian gauge, defined as $B_L=H_T=0$
giving a time slicing in which the expansion is isotropic, is
convenient gauge in analyzing the propagation of photons. On the other
hand, the evolution of matter and metric becomes simplest in the
total matter gauge, defined as $V_T=B_L$, $H_T=0$, giving a time slicing 
in which the total matter is at rest. We choose a hybrid choice of
representation for fluctuations. For the metric fluctuation, we
choose perturbations defined in the Newtonian gauge, namely 
the Newtonian curvature $\Phi$ and potential $\Psi$, which are written
in terms of quantities in an arbitrary gauge as
\BEA
\Phi &\equiv& {\cal R}-k^{-1}\Bigl(\frac{a'}{a}\Bigr) \sigma_g,
\\
\Psi &\equiv& A-k^{-1}\Bigl(\frac{a'}{a}\Bigr) \sigma_g-k^{-1} \sigma_g', 
\EEA 
where 
\BE
{\cal R} \equiv H_L+\frac{1}{3}H_T,~~~~\sigma_g \equiv k^{-1} H_T'-B_L.
\EE
Note that in the Newtonian gauge, $\Phi=H_L$, and $\Psi=A$.
\\
\indent
For matter perturbations, we choose the four variables as 
\BEA
V &\equiv& V_L-\frac{H_T'}{k},
\\
\Delta &\equiv& \delta+ 3(1+w) \frac{a'}{a}(V_L-B_L)k^{-1},
\label{eq:delta}
\\
\Gamma &\equiv& \pi_L-\frac{c_s^2}{w} \delta,
\\
\Pi &\equiv& \pi_T,
\EEA   
where $V$ is the velocity of the fluid in the gauge with $H_T=0$, 
$\Delta$ is the density perturbation in the gauge with
$B_L=V_L$ where the total matter is at rest, and $\Pi$ is the
anisotropic pressure of the fluid. $\Gamma$ can be considered as 
the entropy perturbation because it can be written as 
\BEA
\Gamma &=&\frac{\delta P-c_s^2 \delta \rho}{P}= \frac{4}{3}
\frac{1-3w}{1+w}~ s,
\nonumber
\\
s &=& \frac{\delta (s_{\gamma}/n)}{s_{\gamma}/n},
\EEA
where $s_\gamma$ is the entropy density of photons, and
$n$ is the number density of  non-relativistic particles. 
In other words, $\Gamma$ is proportional to the fluctuation 
in the ratio of photon number density to the matter density since the 
entropy density of the photon is proportional to the photon 
number density. 
\\
\indent
Now, we are ready to write the perturbed Einstein equations.
First, the evolution of background is described as
\BEA
G^o{}_o ~~~\Rightarrow~~~~~~~~~~~~~~~\Bigl (\f{a'}{a} \Bigr)^2 + K 
&=& \f{8 \pi G}{3} \rho a^2, \label{eq:00}
\\
G^i{}_i ~~~\Rightarrow ~~ - \f{2a''}{a}+\Bigl(\f{a'}{a}
\Bigr)^2 - K 
&=& 8 \pi G P a^2, \label{eq:ii}
\EEA
where $G_{\mu \nu}$ denotes the Einstein tensor.
From the perturbed Einstein tensor $\delta G_{\mu \nu}$,
we obtain the generalized Poisson equations,
\BEA
\delta G^o{}_o, ~~\delta G^o{}_i~~~~ \Rightarrow ~~~~~~
(k^2-3K)\Phi &=& 4 \pi G a^2 \rho \Delta, \label{eq:Poisson1}
\\
\delta G^i{}_i ~~~ \Rightarrow ~~~~~~~~
k^2(\Psi + \Phi) &=& -8 \pi G a^2 P \Pi. \label{eq:Poisson2}
\EEA
From the energy conservation $T^{o \mu}{}_{: \mu}=0$, we 
obtain the continuity equation 
\BE 
\Delta '-3w \f{a'}{a} \Delta = -(1- \f{3 K}{k^2})(1+w)kV-2(1-
\f{3K}{k^2}) \f{a'}{a}w \Pi, \label{eq:continuity}
\EE
and from the momentum conservation $T^{i \mu}{}_{: \mu}=0$,
we have the Euler equation, 
\BE
V'+\f{a'}{a}V = \f{c_s^2}{1+w}k \Delta+k \Psi+\f{w}{1+w} k
\Gamma - \f{2}{3}(1-\f{3K}{k^2})\f{w}{1+w} k \Pi.\label{eq:Euler}
\EE
\\
\indent
We derive the evolution equation for the Newtonian 
curvature $\Phi$ assuming that the anisotropic
pressure is negligible $\Pi=0$. In this case, the Newtonian 
potential coincides with the Newtonian curvature except
for the sign $\Phi=-\Psi$. Let us consider 
the perturbed Einstein equation for $\del G^{i}{}_0$,
\BE
-\Bigl( \f{a'}{a} \Bigr )\Phi-\Phi'=4 \pi
G a^2(P+\rho)k^{-1}V,
\label{eq:i0}
\EE
Note that this can be also derived from the background equations and 
the continuity and the generalized Poisson equations.
From the derivative of (\ref{eq:i0}) we have
\BEA
4 \pi Ga^2(P+\rho)k^{-1}(V'+{\cal H}V)
=
-({\cal H}\Phi)'-\Phi''+{\cal H}^2\Phi+\Phi'{\cal H}
\nonumber
\\
+({\cal H}\Phi+\Phi')(P'+\rho')(P+\rho)^{-1},
\label{eq:10}
\EEA
where ${\cal H}=a'/a$. On the other hand, the Euler and the Poisson
equations yield
\BE
V'+{\cal H}'V=\f{kw\Gamma}{1+w}+\f{k c_s^2}{1+w}\f{(k^2-3K)\Phi}
{4 \pi Ga^2 \rho}-k\Phi,
\label{eq:7}
\EE
and the background equations give
\BE
4\pi Ga^2(P+\rho)=-{\cal H}'+{\cal H}^2+K
\label{eq:12}
\EE
and
\BE
\rho '=-3 {\cal H}(P+\rho).
\label{eq:13}
\EE
From (\ref{eq:10}),(\ref{eq:7}),(\ref{eq:12}) and 
(\ref{eq:13},) we finally obtain the second order 
ordinary differential equation for the Newtonian curvature
\BE
\Phi''+3 {\cal{H}}(1+c_s^2)\Phi'+c_s^2k^2\Phi+
(2 {\cal{H}}'+(1+3 c_s^2)({\cal{H}}^2-K))\Phi=-4 \pi G a^2 P \Gamma.
\label{eq:phi-evolution}
\EE
Let us consider the adiabatic case ($\Gamma=0$) and assume
that the energy consists of two components: radiation and 
non-relativistic matter for which 
the sound speed is written in terms of energy density 
as $c_s^2=4 \rho_r/(12 \rho_r+9\rho_m)$. First we consider
the early universe when the radiation is dominant $(P=\rho/3)$
For $\eta<<1$, the effect of curvature is negligible.
The background equations yield the evolution
of the scale factor as $a=a_r \eta$ where $a_r$ is a constant.
Then (\ref{eq:phi-evolution}) is written as
\BE
\Phi''+\f{4}{\eta}\Phi'+\f{k^2}{3}\Phi=0,
\EE 
which has an analytical solution
\BE
\Phi(\eta)=\eta^{-3}\{(\omega \eta \cos(\omega \eta)
-\sin(\omega \eta))C_1+(\omega \eta \sin(\omega \eta)
-\cos(\omega \eta))C_2\},
\label{eq:analytical}
\EE
where $\omega=k/\sqrt{3}$ and $C_1$ and $C_2$ are constants which
depend on $k$. Around $\eta=0$, 
(\ref{eq:analytical}) can be expanded as
\BE
\Phi(\eta)=\eta^{-3}\{C_2+\f{C_2}{2}\omega^2\eta^2-\f{C_1}{3}\omega^3\eta^3
-\f{C_2}{8}\omega^4 \eta^4+{\cal O}((\omega \eta)^5) \}.
\label{eq:expansion}
\EE
Thus, for the long-wavelength modes ($\omega \eta <<1$),
the non-decaying mode ($C_2=0$) has a constant amplitude 
and the time derivative at the initial time vanishes,
\BE
\Phi\sim \textrm{const},~~~~~\Phi'(0)=0.
\EE
During the radiation-matter equality era, the 
equation of state changes from $w=1/3$ to 
$w=0$. Then the amplitude of the long-wavelength mode 
changes by a factor of $9/10$. 
Using an explicit form for 
the scale factor for hyperbolic models in the matter
dominant epoch,
\BE
a=\f {\Omega_o}{2(1- \Omega_o)}~\bigl(\cosh(\sqrt{-K} \eta)-1\bigr),
\label{eq:scalefactor}
\EE
the non-decaying mode of $\Phi$
in the matter dominant epoch for hyperbolic models is given by
\BE
\Phi(\T{\eta})={\Phi}(\T{\eta}_{ini})
\f{5(\sinh^2 \T{\eta}-3 \T{\eta}\sinh\,\T{\eta}+4 \cosh\,\T{\eta}-4)}
{(\cosh\,\T{\eta}-1)^3},
\label{eq:Ncurvature}
\EE
where $\T{\eta} \equiv \sqrt{-K} \eta$ and $\T{\eta}_{ini}$
denotes the initial time normalized by the present curvature.
In the curvature or $\Lambda$ dominant
epoch, the amplitude decays as $1/a$. 
\\
\indent
Using the Newtonian curvature $\Phi$, the total density fluctuation
can be readily obtained from the Poisson equation (\ref{eq:Poisson1}).
In the radiation and matter dominant 
epochs,  $\Delta$ evolves as $a^2$ but the amplitude freezes
in the curvature or $\Lambda$ dominant epoch. 
As for the velocity of the total matter, the continuity equation
(\ref{eq:continuity}) gives a solution
\BE
V=a^{-1}(\eta)C+ a^{-1}(\eta)\int d \eta~ a(\eta) \Bigl( \f{c_s^2}{1+w} 
k \Delta-k\Phi \Bigr),
\EE
where $C$ is a constant. The decaying mode corresponds
to $a^{-1}(\eta)C$ which can be omitted for the non-decaying mode. 
In the radiation and matter dominant epochs, 
the amplitude of $V$ increases as $a$ and as $a^{1/2}$ respectively, 
but it decays in the curvature or $\Lambda$ dominant epoch. 
\section{Sachs~-Wolfe effect}
\indent
The fundamental equation of radiative transport is derived from 
the Boltzmann equation, which is written as
\BE
\f{df}{dt}={\cal C}~[f],
\EE  
where $f=f(\x,p,t)$ is the distribution function and ${\cal C}~[f]$ denotes
the collision terms.
We assume that the photons propagate freely after the decoupling
($i.e.$~no reionization is considered), and hence we can neglect
the collision terms (this is a good approximation for
superhorizon perturbations even before the decoupling time), 
\BE
\f{df}{dt}=\f{\del{f}}{\del x^{\mu}} \f{dx^{\mu}}{dt}+
\f{\del{f}}{\del p^{\mu}} \f{dp^{\mu}}{dt}=0,
\EE 
which is also known as the Liouville equation.
\\
\indent
Although the phase space density of photons is conserved along
their trajectories, each of the configuration space density
and the momentum space density changes with time and space because the
gravitational force affects the photon trajectories. 
As we shall see later, the redshift of photons can be interpreted as 
the change in the momentum space density which is affected by the
background metric, gradients in the perturbative potential, and 
dilation effects caused by the curvature perturbation. 
\\
\indent
First of all, we need to fix the gauge in order to 
solve the geodesic equation coupled with the Liouville equation
on the perturbed FRWL space-time.
We choose the Newtonian gauge in which the physical interpretation of
the dynamics of photons is simplest. The line element takes the form, 
\BE
ds^2 = -(1+2 \Psi({\mathbf{x}},t)) dt^2+
a^2(1+2\Phi({\mathbf{x}},t))\gamma_{ij}dx^idx^j.
\EE
The geodesic equation for the photon is
\BE
\f{d^2 x^{\mu}}{d \lambda^2}+\Gamma^\mu_{\alpha \beta}
\f{dx^{\alpha}}{d \lambda} \f{dx^{\beta}}{d \lambda}=0, 
~~~~~~~p^\mu=\f{d x^\mu}{d \lambda},
\EE
where $\lambda$ is an affine parameter. Since the photon 
momentum satisfies the relation,
\BE
\f{p^i}{p^o}=\f{d x^i}{d t},
\EE
the geodesic equation can also be written as
\BE
\f{d p^\mu}{d t}=g^{\mu \nu}\Bigl(\f{1}{2} 
\f{\del g_{\alpha \beta}}{\del x^\nu}
-\f{\del g_{\nu \alpha}}{\del x^\beta}\Bigr)\f{p^\alpha p^\beta}{p^o}.
\label{eq:geodesic}
\EE
Now, the Liouville equation is expressed as 
\BEA
\f{df}{dt}=\f{\del f}{\del x^i}\f{d x^i}{d t}+
\f{\del f}{\del p}\f{d p}{d t}+\f{\del f}{\del\gamma^i}
\f{d \gamma^i}{d t}=0,\label{eq:Liouville}
\EEA
where 
\BE
p^2 \equiv (p^o)^2,~~~ \gamma^i \equiv a (1+\Phi)\f{p^i}{p}.
\EE
Note that $\gamma^i$ is a three-dimensional vector
defined on the space with metric $\gamma_{ij}$.
Since static curvature effects are unimportant
in determining the redshift, it will be assumed
that the background metric is flat without loss of generality
(\textit{i.e.}\,$\gamma_{ij}=\delta_{ij}$).
The geodesic equation (\ref{eq:geodesic}) 
with scalar type perturbations gives
\BE
\f{1}{p} \f{d p^o}{dt}=- \f{\del \Psi}{\del t}+\f{da}{dt}\f{1}{a}
(\Psi-1)-\f{\del\Phi}{\del t}-2~\f{\del \Psi}{\del x^i}~
\f{\gamma^i} {a}.
\label{eq:geo}
\EE 
With the definition of $p$, we obtain the photon energy equation
\BE
\f{1}{p} \f{d p}{dt}=-\Biggl(\f{da}{dt}\f{1}{a}+\f{\del\Phi}{\del t}
+\f{\del\Psi}{\del x^i}\f{\gamma^i}{a}\Biggr).\label{eq:photonenergy}
\EE
The first term in the right-hand side of Eq.(\ref{eq:photonenergy}) 
is interpreted as a redshift by the
isotropic expansion of the universe. The second term represents 
the dilation effect caused by the fluctuation of the spatial
curvature. The third term can be considered as a gravitational
redshift caused by the gradient in the potential. 
It is known that the CMB spectrum is well described
by a blackbody spectrum, which obeys the distribution
\BE
f(p)=(2 \pi)^{-3} (e^{p/T}-1)^{-1}.
\EE  
Then the fractional shift in frequency from the gravitational effect
does not depend on frequency.  In other words, the 
blackbody distribution will not
change its form,
\BEA
f'(p')&=&\f{1}{(2 \pi)^3} \f{1}{e^{(p'-\delta p)/T)}-1},
\nonumber
\\
&=&\f{1}{(2\pi)^3} \f{1}{e^{p'/T'}-1},
\EEA
where
\BE
T'=T+\delta T=T(1+\delta p/ p).
\EE
Because the energy density of photons satisfies $\rho_\gamma \propto
T^4$ in the blackbody radiation, the temperature fluctuation
$\Theta$ is defined as
\BE
\Theta \equiv \f{\delta T}{T}= \f{1}{4}\f{\delta \rho_\gamma}{\rho_\gamma}.
\label{eq:Theta}
\EE
Note that $\rho_\gamma$ is a global averaged quantity (background
value), which is averaged over the ensemble and $\boldsymbol{\gamma}$ while
$\T{\rho}_\gamma \equiv \rho_\gamma+\delta \rho_\gamma$ is a 
quantity which is locally averaged over the ensemble, 
\BEA
\T{\rho}_\gamma&=&2\int f(p)~4 \pi p^2~dp,
\nonumber
\\ 
&=&\f{1}{\pi^2} \int \f {p^3}{e^{p/T}-1} ~dp,
~~~~~T=T(\eta,\x,\boldsymbol{\gamma}).
\EEA
Substituting the total time derivative of p in the geodesic equation
(\ref{eq:geodesic}) into Eq.(\ref{eq:photonenergy}),
and integrating over $p$, we obtain in the zero-th order,
\BE
\f{\del \rho_\gamma}{\del t}= -4 H \rho_\gamma,~~~~~ 
(H\equiv \dot a/a\label{eq:free})
\EE
which represents the free-streaming of photons 
$\rho_\gamma \propto a^{-4}$, and in the first order, 
\BE
4 \pi^2 \rho_\gamma \f{d \Theta}{dt}
=4\pi^2 \rho_\gamma \Biggl[
\f {\del \Phi}{\del t}+\f{\del \Psi}
{\del x^i} a \gamma^i\Biggr].\label{eq:fluc}
\EE  
The latter equation can be written as
\BE
\f{d}{d \eta} \bigl[\Theta+\Psi \bigr] \bigl[
\eta, \mathbf{x},\boldsymbol{\gamma}\bigr]
=\f{\del \Psi}{\del \eta}-\f{\del \Phi}{\del \eta},
\EE 
where $dt=ad\eta$. Integrating over the time, the 
temperature fluctuation is finally written as 
\BE
\Theta(\eta_o,{\mathbf{x}}_o, \boldsymbol{\gamma}_o)+ \Psi(\eta_o, {\mathbf{x}}_o)
=\Theta(\eta_\ast,{\mathbf{x}}_\ast, \boldsymbol{\gamma}_\ast)+ 
\Psi(\eta_\ast, {\mathbf{x}}_\ast)
+\int _{\eta_\ast}^{\eta_o} \Bigl(\f {\del \Psi}{\del \eta}
- \f {\del \Phi}{\del \eta}\Bigr) d \eta,\label{eq:TA}
\EE
where $\eta_\ast$ is the decoupling time, and $\eta_o$ is 
the present time. The contribution of intrinsic temperature
fluctuations and potential redshifts $[\Theta+\Psi][\eta_\ast,
\mathbf{x}_\ast,\gamma_\ast]$ is called the ordinary Sachs-Wolfe
(OSW) effect. The contribution of the remaining integrated term is
called the integrated Sachs-Wolfe (ISW) effect \cite{Sachs}. 
For adiabatic perturbations, when the curvature perturbation
begins to decay before the present time, the ISW effect dominates
over the OSW effect. This is the case if the density 
parameter $\Omega_o$ is much smaller than the unity .
\\
\indent
Now, we estimate the temperature fluctuations on superhorizon scales
at the decoupling time. 
Using Eq.(\ref{eq:Theta}), the temperature fluctuation can be written
in terms of $k$-component of the 
Newtonian density fluctuation of radiation(gauge invariant
form)\footnote{More rigorously, $\Theta_\K$ must be understood as the 
monopole ($l=0$) component $(\Theta_\K)_0$ of the density fluctuations. 
On superhorizon scales, the contribution of higher multipoles of
density fluctuations at the decoupling time is 
negligible in estimating the present
anisotropies \cite{HS95}. }
\BE
\Theta_\K=\f{\delta^N_\gamma}{4}.
\EE
This can also be written in terms of quantities in total matter gauge as
\BE
\Theta_\K=\f{\Delta_\gamma}{4}-\f{a'}{a}\f{V_T}{k},
\EE
where $\Delta_\gamma$ is the photon density fluctuation 
in the total matter gauge. Let us 
assume $\Pi=0$ and the effect of the radiation at the last
scattering is negligible.  Substituting 
the explicit form of the adiabatic perturbation 
(\ref{eq:Ncurvature}) 
for hyperbolic models into the Euler equation (\ref{eq:Euler}), 
the non-decaying mode of the velocity of the total matter is 
given by  
\begin{equation}
V(\T{\eta})=-\f{5 k \Phi(\T{\eta}_{\ast})}
{\sqrt{-K}} 
\f{2 \T{\eta}+\T{\eta}\,\cosh~\T{\eta}-3 \sinh\,
\T{\eta}}{(\cosh\,\T{\eta}-1)^2},
\label{eq:VV}
\end{equation}
where $\T{\eta}_{\ast}$ is the (normalized) last scattering time.
On the other hand, from the continuity equation (\ref{eq:continuity}) 
the energy density $\Delta$ is explicitly written as 
\BEA
\Delta(\T{\eta})
&=&\f{1}{K}\bigl(1-\f {3K}{k^2} \bigr) k^2 \Phi(\T{\eta}_{\ast})
f(\T{\eta}),
\nonumber
\\
f(\T{\eta})&=&-\f{5}{2}\biggl(\rm{cosech}\,\f{\T{\eta}}{2}\biggr)^3 
\Bigl(- \T{\eta}\,\cosh\,\f{\T{\eta}}{2}+2
\sinh\,\f{\T{\T{\eta}}}{2}\Bigr)-\f{5}{3} 
\label{eq:Delta2}
\EEA
From (\ref{eq:scalefactor}), (\ref{eq:VV}) and (\ref{eq:Delta2}), 
we have 
\BE
\Theta_\K(\T{\eta})
= \Bigl(\f{2}{3}-{\cal O}(\T{\eta}^2)\Bigr )\Phi(\T{\eta}_\ast),
~~~~~\T{\eta}\ll 1.
\EE 
Up to the first order in $\T{\eta}$, we obtain
\BE
\Theta_\K(\T{\eta}_\ast) \approx \f{2}{3} \Phi(\T{\eta}_\ast). 
\label{eq:relation2}
\EE
Because $\eta_{\ast}\sim 10^{-2}$ for the typical range of cosmological 
parameters, (\ref{eq:relation2}) gives a good approximation.  
Substituting Eq.(\ref{eq:relation2}) into Eq.(\ref{eq:TA}), we
obtain the explicit formula for the temperature fluctuation, 
\BE
\Theta_\K(\eta_o,{\mathbf{x}}_o, \gamma_o)+ \Psi(\eta_o, {\mathbf{x}}_o)
=-\f{1}{3} \Phi(\eta_\ast {\mathbf{x}}_\ast )
-2\int _{\eta_\ast}^{\eta_o} \f {\del \Phi}{\del \eta} d \eta.
\label{eq:ApproxTA}
\EE
It should be noted that one should employ the full form of equation 
(\ref{eq:TA}) instead of (\ref{eq:ApproxTA})
for much rigorous calculation including the effect of the 
radiation contribution at the last scattering and the anisotropic 
pressure. However, for superhorizon scale temperature perturbation,
(\ref{eq:ApproxTA}) gives a fairly good approximation provided that
the matter dominates the energy at the last scattering time. 
\pagebreak

\section{Angular power spectrum}
In order to analyze the angular dependence of the temperature
anisotropy in the sky it is 
convenient to expand the anisotropy in terms of spherical harmonics
$Y_{lm}$ as
\BE
\f{\Delta T}{T}(\n)=\sum_{lm} a_{lm} Y_{lm}(\n),
\EE
where $\n$ is an unit vector along the line of sight. Then the
temperature 2-point correlation is given by
\begin{equation}
\Biggl\langle \f{\Delta T}{T}(\n) \f{\Delta T}{T}(\n')\Biggr\rangle
= \sum_{l,m,l',m'} \bigl\langle a_{lm} a^*_{l'm'} \bigr\rangle
Y_{lm}(\n) Y^*_{l'm'}(\n'),
\end{equation}
where $\langle \rangle$ denotes the ensemble average taken over the initial
conditions and the observation positions and orientations.
If the temperature fluctuations obeys the Gaussian statistic,
it can be fully characterized by the angular power spectrum $C_{lm}$,
which is defined as
\BE
\bigl\langle a_{lm} a^*_{l'm'}
\bigr\rangle=C_{lm}\delta_{ll'}\delta_{mm'},~~~~~ C_{lm} \equiv 
\bigl\langle |a_{lm}|^2 \bigr\rangle.\label{eq:cl}
\EE
If the Gaussian fluctuation is isotropic then
the angular power spectrum has no 
dependency on $m$ and can be written as $C_l$, which is 
best estimated by simply averaging over m,
\begin{equation}
\T{C_l}=\f{1}{2l+1} \sum_{m=-l}^l |a_{lm}|^2.
\end{equation}
The fact that we have only a finite number of $2l+1$ samples
for estimating $C_l$ is called $cosmic~ variance$ 
\cite{Cosmic.Variance1,Cosmic.Variance2} which sets a limit
on how precisely we can measure the angular power spectrum.
$\T{C}_l$ obeys a $\chi^2$ distribution with $2l+1$ degree of freedom.
The fractional uncertainty in the estimate of $C_l$ is given by
\begin{equation}
\f {\sqrt{Var(\T{C_l})}}{\T{C_l}}=\sqrt{\f{2}{2l+1}}.
\end{equation} 
For example, there is a 63\% uncertainty in the 
quadrapole $C_2$, 31\% in $C_{10}$,
and 7.1\% in $C_{200}$. 
\subsection{Closed flat models}
Let us now consider cosmological models where the spatial geometry is
represented as a flat 3-torus obtained by gluing 
the opposite faces of a cube by three translations. Then the wave numbers
of the square-integrable eigenmodes of the Laplacian are restricted to
discrete values $k_i=2 \pi n_i/L$, ($i=1,2,3$) where $n_i$'s run
over all integers. Because the background geometry is
globally anisotropic, the temperature anisotropy is no
longer statistically isotropic. Therefore, the effect of 
the non-diagonal elements $l\ne l'$ or $m\ne m'$ in 
the correlation is non-negligible in general.
The $SO(3)$ invariant quantity in the 2-point correlation
consists of only diagonal elements as 
\BE
\hat{C}_l=\f{1}{2l+1} \sum_{m=-l}^l <|a_{lm}|^2>
\EE
where $\langle \rangle$ denotes an ensemble average taken over the initial
conditions. For brevity, we also call $\hat{C}_l$ as 
the angular power spectrum. $\hat{C}_l$'s
describe the 2-point correlations for an isotropic 
ensemble since the non-diagonal elements in the 2-point correlation 
vanish if they are averaged over $SO(3)$ transformations.
Note that $\hat{C}_l$'s are not independent each other since the 
fluctuations form a non-Gaussian random field for an isotropic 
ensemble. Thus $\hat{C}_l$'s still give us information
about the 2-point correlations although they are non-Gaussian
and have a larger cosmic variance.     
\\
\indent
Let us choose the Euclidean 
coordinates such that the z-axis is perpendicular to
one of the face and the origin of the coordinates is located at 
the center of cube. Then the angular power spectrum 
is written as
\BEA
\hat{C}_l&=&\sum_{\K\ne 0} \f{8 \pi^3 {\cal{P}}_\Phi(k)F_{kl}^2}{k^3 L^3},
\nonumber
\\
F_{kl}&=&\f{1}{3}
\Phi_t(\eta_\ast) j_l(k(\eta_o\!-\!\eta_\ast))
\!+2 \int_{\eta_\ast}^
{\eta_o}\!\!\!\!\!\!d \eta\, 
\f{d\Phi_t}{d \eta}j_l(k(\eta_o\!-\!\eta)),
\label{eq:psT3}
\EEA
where ${\cal{P}}_\Phi(k)$ is the initial power spectrum for $\Phi$
, $k\equiv\sqrt{k_1^2+k_2^2+k_3^2}$ and $j_l$ denotes the spherical
Bessel function and $\Phi_t=\Phi/\Phi(0)$. From now on, we assume
the scale-invariant 
Harrison-Zel'dovich spectrum (${\cal{P}}_\Phi(k)
\!=\!k^{n-1}\!=\!const.$) as the initial power spectrum as predicted
by the standard inflationary scenarios.
\BF
\centerline{\psfig{figure=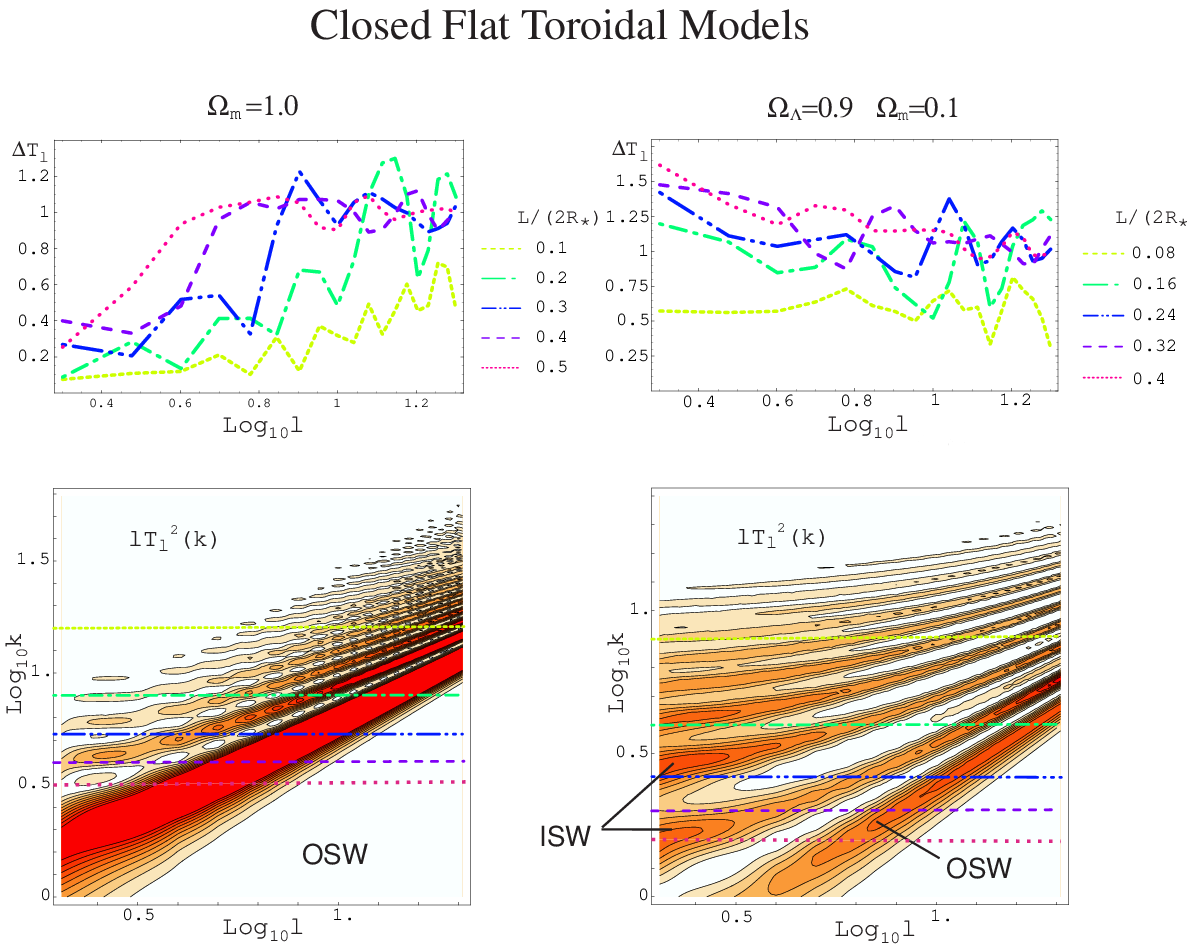,width=16cm}}
\caption{Suppression in Large-angle Power for Toroidal Models}
\mycaption{A large-angle suppression in 
$\Delta T_l\equiv \sqrt{l(l+1)\hat{C}_l/(2
\pi)}$ (all the plotted values are 
normalized by $\Delta T_{20}$ with infinite volume) 
occurs for the ``standard'' toroidal model 
$(\Omega_m,\Omega_\Lambda)\!=\!(1.0,0)$
at $l\!<\!l_{cut}\!\sim\! 2\pi R_\ast/L-1$ 
while such a prominent suppression is
\ti{not} observed for the toroidal model with 
($\Omega_m, \Omega_\Lambda$)=(0.1,0.9). 
The transfer function $T_l(k)$ describes how each $k$-mode
contributes to the power for a particular angular scale $l$ 
(lower figures). The corresponding first eigenvalues $k_1$ are plotted
as horizontal lines.
The unit of $k$ is $H_0$. }
\label{fig:TL}
\EF
Let us estimate the angular scale below which the power spectrum
is suppressed owing to the mode-cutoff $k_{cut}=2 \pi/L$. 
First of all, we consider the transfer function $T_l(k)$
\BE
\f{2l+1}{4 \pi}\hat{C}_l= \int T_l^2(k){\cal P}(k) \f{dk}{k},
\EE
which describs how the spatial information is contained in the angular 
power. For a given $k$, the contribution to the angular power
on smallest angular scales comes from the OSW effect
where the transfer function can be written in terms of the 
spherical Bessel function as $T^{OSW}_l(k)=j_l(k\eta_0)$. 
Because $j_l(x)$'s have the first peaks at $x\sim1+l$, the 
angular cutoff $l_{cut}$ is determined by $l_{cut}=2 \pi \eta_0/L-1$.
For instance, $l_{cut}\sim 5$ for $L=\eta_0=2 H_0^{-1}$. 
Thus the large-angle suppression scale is determined by the
largest fluctuation scale at the last scattering. On smaller angular scales 
$l>l_{cut}$, the second peak in the power corresponds to the
fluctuation scale of the second eigenmode at the last scattering. 
This behavior is analogous to the acoustic oscillation where the
oscillation scale is determined by the sound horizon at the last scattering.
On angular scales larger than $l_{cut}$ 
one observes another oscillation feature
in the power for models with a smaller cell 
where $k_{cut} \eta_0$ is sufficiently
large, which is apparently determined by the behavior of the 
first eigenmode. From the asymptotic form 
\BE
j_l(x)\sim \f{1}{x} \sin(x-l \pi/2),~~~x>>\f{l(l+1)}{2},
\EE
one notices that the angular scale of the oscillation 
for asymptotic values \\
$k_{cut} \eta_0\!>\!>\!l(l+1)/2$ is $\Delta l=2$. 
For intermediate values of $k_{cut}\eta_0$, $\Delta l$ take much 
larger values (see figure \ref{fig:TL}).
\\
\indent
So far, we have studied the effect of the non-trivial topology 
on the OSW contribution only, which is sufficient for constraining 
the topology of the ``standard'' toroidal model with $\Omega_{tot}=1$.
However, for low matter density models, 
one cannot ignore the (late) ISW contribution which is
generated by the decay of the gravitational potential at the 
$\Lambda$-dominant epoch. The crucial point is that 
they are produced at \ti{later} time $1+z\sim
(\Omega_\Lambda/\Omega_0)^{1/3}$ well after the last scattering.
Although it is impossible to generate 
fluctuations beyond the size of the cell (in 3-dimensional sense),
that does not necessarily mean that any   
fluctuations on large angular scales (in 2-dimensional sense)
cannot be produced.
Suppose that a fluctuation is produced at a point nearer to us,
then the corresponding angular scale 
becomes larger if the background geometry 
is flat or hyperbolic. 
Therefore, one can expect that the
suppression on large angular scales owing to the mode-cutoff
is not stringent for low matter density models. 
As shown in figure \ref{fig:TL}, the angular powers for a model with
$(\Omega_\Lambda, \Omega_m)\!=\!(0.9,0.1)$ are almost flat.
In contrast to the ``standard'' model, the transfer function
for low matter density models 
distributes in a broad range of $k$, which implies the additional late production of the fluctuations
which contribute to the angular power. 
Surprisingly, in low matter density models with small volume,  
a slight excess power due to the ISW effect is \ti{cancelled out}
by a moderate suppression owing to the mode-cutoff which leads to
a flat spectrum. However, as observed in the ``standard'' toroidal 
model, the power spectra have prominent oscillating features. The 
oscillation scale for $l<l_{cut}$ is 
determined by the first eigenmode. The peaks in the angular power  
correspond to the first SW ridge and the first and the second 
and other ISW ridges. Is such an oscillating feature 
already ruled out by the current observation?
We will see the results of our Bayesian analyses for testing 
the goodness-of-fit to the COBE data in the next section.
\subsection{Closed hyperbolic models}

Assuming that the initial fluctuations obey the Gaussian statistic, 
and neglecting the tensor-type perturbations, 
the angular power spectrum $\hat{C}_l$ for closed hyperbolic models 
can be written in terms of the expansion coefficients $\xi_{\nu l m}$
as
\BE
(2\,l+1)\,\hat{C}_l
=\sum_{\nu,m}\f{4 \pi^4~{\cal P}_\Phi(\nu) }
{\nu(\nu^2+1)\textrm{Vol}(M)}~|\xi_{\nu l m}|^2 |F_{\nu l}|^2 ,
\EE
where
\BE
F_{\nu l}(\eta_o)
\!\!\equiv\!\!\f{1}{3}
\Phi_t(\eta_\ast) X_{\nu l}(\eta_o\!-\!\eta_\ast)
\!+\!\! 2 \!\!\int_{\eta_\ast}^
{\eta_o}\!\!\!\!\!\!d \eta\, 
\f{d\Phi_t}{d \eta}X_{\nu l}(\eta_o\!-\!\eta).\label{eq:CMBcorCH}
\EE
Here, $\nu\!=\!\sqrt{k^2-1}$, ${\cal P}_\Phi(\nu) $ 
is the initial power spectrum, and
$\eta_\ast$ and $\eta_o$ are the
last scattering and the present conformal time, respectively. 
$X_{\nu l}$ denotes the radial eigenfunctions of the universal
covering space and
$\xi_{\nu l m}$ denotes the expansion coefficients,
$\Phi_t=\Phi/\Phi(0)$ and $\textrm{Vol}(M)$ is the volume of the 
space.
The (extended) Harrison-Zel'dovich spectrum corresponds to 
${\cal P}_\Phi(\nu)\!=\!\nu^{n-1}\!=\!Const.$ which we shall use as the initial 
condition. Using Weyl's asymptotic form 
one can easily show that in the limit $\nu \rightarrow \infty$, 
(\ref{eq:CMBcorCH}) coincides with that for the 
infinite counterpart provided that $<|\xi_{\nu l m}|^2>\propto \nu^{-2}$.
Because we assume that the 
normalization factor in $<|\xi_{\nu l m}|^2>$
does not depend on the volume of the space, the volume factor
is necessary for the manifold-invariant form which cancels out the 
volume factor term in Weyl's formula. However, it might
not be necessary if $<|\xi_{\nu l m}|^2>$
is proportional to $1/\textrm{Vol}(M)$. 
As we have seen in chapter 3, the ratio 
of the normalization factor in $<|\xi_{\nu l
m}|^2>$ is 1.03 for the Weeks and the Thurston manifolds
which is close to the inverse ratio in the volume 1.04.
\\
\indent
In contrast to the previously studied 
closed flat toroidal models, closed hyperbolic spaces are 
globally inhomogeneous. Therefore,
the angular power spectrum depends on the place in the space. 
It seems that evaluation of the full spatial dependence is intractable. 
Fortunately, we can understand the spatial dependence  
in terms of the pseudo-Gaussian random property of the eigenmodes.
As we have seen in chapter II, each set of expansion coefficients
$\xi_{\nu lm}$ at a particular place with particular orientation of
the coordinates can be well approximated as a ``realization'' of 
random Gaussian numbers. Hence, uncertainty in the power
is well described by the statistical property of the random Gaussian
numbers. The uncertainty in the power owing to the globally
inhomogeneous geometry can be called the \ti{geometric variance}.
Because we do not have any information about the position in the 
spatial hypersurface, any particular places do not have 
priority over others. 
However, there are some 
places at which one observes symmetries of the background
geometry.  
For instance, in the Thurston manifold having a $D2$ symmetry group
there are three axes that are fixed by a $\pi$ rotation.
If we were in a position near these axes by chance then 
we would observe $Z2$ symmetry in the sky where the 
pseudo-Gaussian behavior breaks down since there are non-trivial
correlations in the expansion coefficients. 
In general, the chance that we actually observe any
symmetries (elements of the isometry group of the manifold or orbifold
or of their finite sheeted covers) is expected to be 
very low since a fixed point by an element of the isometric group
is either a part of a line (for instance, an axis of a
rotation) or an isolated point (for instance, a center of an antipodal map).
On the other hand, the symmetric structures can
be a crucial sign which implies the non-trivial topology.
\BF
\centerline{\psfig{figure=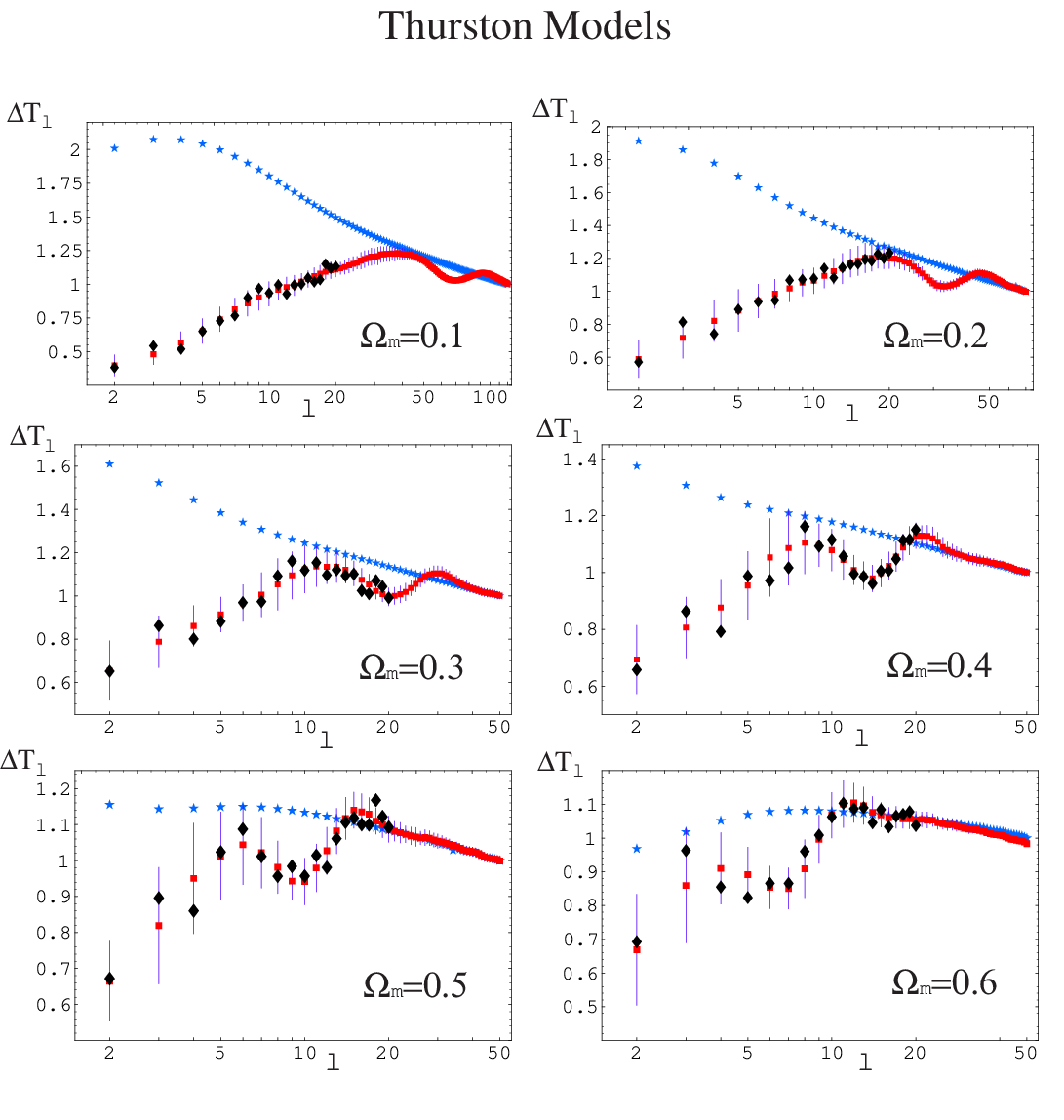,width=14cm}}
\caption{Geometric Variance}
\mycaption{The geometric variance in the angular power spectra 
$\Delta T_{l}\!=\!\sqrt{l(l+1)\hat{C}_{l}/(2 \pi)}$ (normalized 
by $\Delta T_{l(max)}$, $l(max)=100,70,50$ for 
$\Omega_m=0.1,0.2,0.3-0.6$, respectively)
for the Thurston models (volume=0.98).
The ensemble averaged values (boxes) 
are compared with those at a place 
where the injectivity radius is locally maximal (diamonds)
and those for the infinite hyperbolic models (stars)
for different matter density at present.
The cosmological constant is not included. 
Two-sigma geometric variance is
shown in vertical lines which has been 
obtained by $500$ random realizations for the expansion coefficients.
The initial fluctuations are assumed to be adiabatic and
obey Gaussian distribution specified by the extended 
Harisson-Zel'dovich power spectrum. Since  
we neglected the 'early' ISW effect, 
on intermediate scales ($l>20$) the plotted powers 
are slightly suppressed than the expected values. }
\label{fig:GeomVar}
\EF
On large angular scales, the contribution to the power only 
comes from the eigenmodes with long wavelength. Therefore, the degree
of freedom which is equivalent to the number of terms in the
summation that gives the power is small. In figure (\ref{fig:GeomVar}),
one can notice a large uncertainty in the power 
on large angular scales. However, the geometric variance is 
always smaller than the ``initial'' variance owing to the 
uncertainty in the initial conditions since the degree of freedom
($k,m$) is larger than $2l+1$. Therefore, the net cosmic variance 
(initial variance $+$ geometric variance) cannot be significantly
greater than the values for the infinite counterparts.
On small angular scales, the contribution to the power 
comes from a number of eigenmodes which causes an increase in the 
degree of freedom and the geometric variance 
becomes small. In that case,
one cannot distinguish the closed model with the infinite counterpart.
When the power coincides with that for the infinite
counterpart, the geometric variance becomes negligible.
Thus we have two kinds of uncertainty:one is the initial
variance arising from the uncertainty in the initial conditions;
the other one is the geometric variance arising from the
uncertainty in the position of the observer owing to the 
global inhomogeneity in the geometry. The latter effect is negligible on 
small scales compared with the size of the fundamental domain.
\\
\indent
Compared to flat $\Lambda$ models, the ISW contribution 
is much significant for hyperbolic (open) models since the
curvature dominant epoch comes $1+z\sim (1-\Omega_0)/\Omega_0$
earlier than the $\Lambda$ dominant epoch $1+z\sim 
(\Omega_\Lambda/\Omega_0)^{1/3}$. Therefore, the suppression
in the large-angle power owing to the mode-cutoff is 
less stringent compared with closed flat $\Lambda$ models.  
\BF
\centerline{\psfig{figure=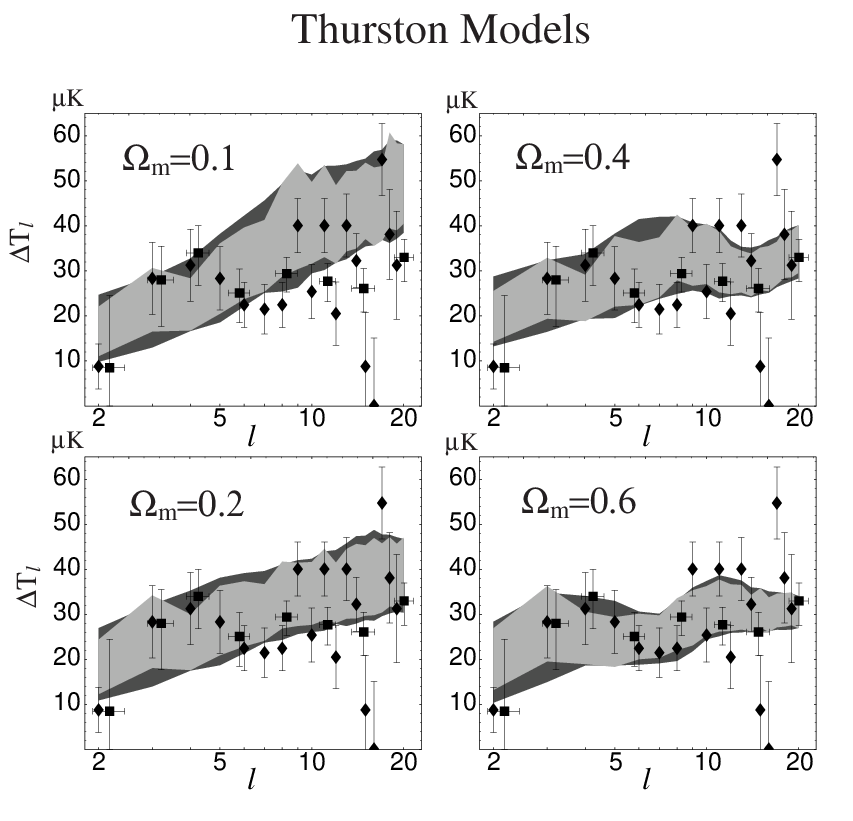,width=14cm}}
\caption{Comparison with COBE data}
\mycaption{Plots of the angular power 
$\Delta T_{l}\!=\!\sqrt{l(l+1)\hat{C}_l/(2 \pi)}$ 
for the Thurston models with different 
matter density parameters in comparison with the 
COBE-DMR data analyzed 
by Gorski (diamonds) and Tegmark (boxes).
The light-gray band denotes
one-sigma ``initial'' variance around the power averaged over 
initial conditions for a particular choice of the observation point
(where the injectivity radius is locally maximal) and 
the darkgray band represents the net variance of 
one-sigma ``initial'' variance and 
one-sigma ``geometric'' variance around the power averaged over 
both initial conditions and positions.  }
\label{fig:COSMIC1426}
\EF
\BF
\centerline{\psfig{figure=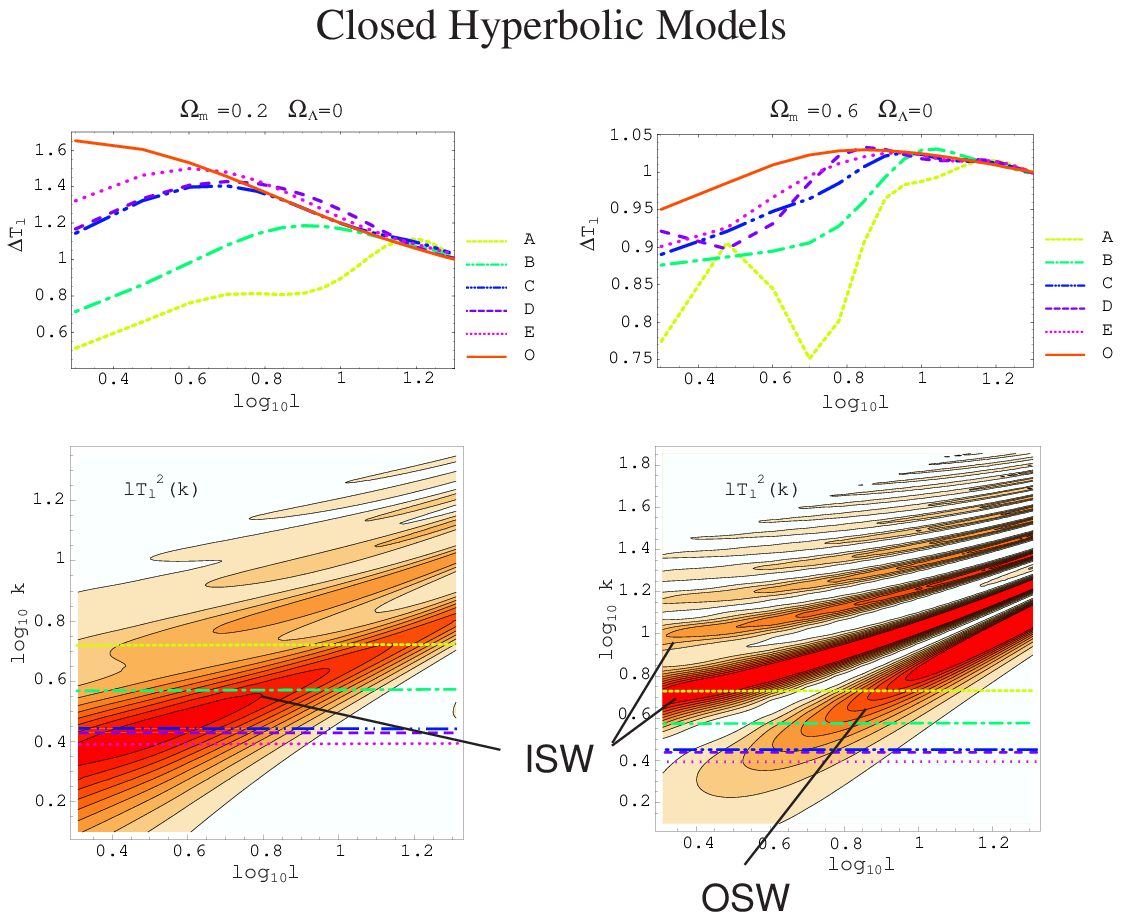,width=17cm}}
\caption{Suppression in Large-angle Power for CH models} 
\mycaption{Suppression in large-angle power 
 $\Delta T_l\equiv \sqrt{l(l+1)\hat{C}_l/(2
\pi)}$ for 5 closed hyperbolic
models with different volume 
(name,volume)=A:(m003(3,-1)(Weeks),0.94), B:(m010(-1,3),1.9),
C:(m082(-2,3),2.9), D:(m288(-5,1),3.9) E:(s873(-4,1), 4.9)
in comparison with the infinite counterpart (denoted as O). 
All the plotted values are normalized by 
$\Delta T_{20}$ for the infinite hyperbolic model.
First eigenvalues $k_1$ for the five models are
plotted as horizontal lines (lower figures). 
The unit of $k$ is equal to the inverse of the present
curvature radius.}
\label{fig:TL5CHM}
\EF
Similar to $\Lambda$ models, for a given angular scale 
the transfer function for hyperbolic
low matter density models distributes in a broad range of $k$
owing to the additional production of the fluctuations in the
late epoch.
As shown in (\ref{fig:TL5CHM}) for $\Omega_0=0.2$ the contribution
from the ISW effect dominates the large-angle power on large angular
scales $l<20$. Therefore, the mode-cutoff owing to the finite size of the 
fundamental domain does not lead to a significant suppression
even for small CH models like the Weeks or the Thurston models.
The angular cutoff scale $l_{cut}$ is determined by the angular scale 
of the fluctuations of the first eigenmode at the last scattering.
The cutoff scale is approximately given by the volume of the 
space $v$ and the smallest non-zero wavenumber $k_1$ 
\BE
l_{cut}=\f{k_1}{4} \sqrt{v^{-1}((\sinh(2(R_\ast+r_{ave}))
-\sinh(2(R_\ast-r_{ave}))-4 r_{ave}))}
\EE
where $v=\pi(\sinh(2r_{ave})-2r_{ave})$ and $R_\ast$ is the comoving
radius of the last scattering surface. For the Weeks manifold
($v=0.94$ and $k_1=5.27$), $l_{cut}=26$ for $\Omega_0=0.2$ and
$l_{cut}=7$ for $\Omega_0=0.6$. On angular scales $l> 2 l_{cut}$,
the power for a closed model coincides with that for 
the infinite counterpart. Compared with closed flat 
models the ``topological oscillation'' on scales $l_{cut}<l<2
l_{lcut}$ is not prominent. The oscillation feature in the power is 
purely determined by the first eigenmode. The peaks in the power
corresponds to the OSW ridge and the ISW ridges.  
\\
\indent
As shown in figure (\ref{fig:COSMIC1426}), the quadrapole
in the COBE data is very low and there is a peak at $l \sim 4$.
A Thurston model with $\Omega_m=0.6$ has the first peak in this 
scale. Therefore, the fit to the observed power is much better than do 
any FRWL models. For the infinite hyperbolic models, the ISW 
contribution leads to the excess power on large angular scales
which is not compatible with the behavior of the observed
power on very large angular scales. Interestingly, the angular
suppression owing to the mode-cutoff effectively reduces the 
excess power as observed in closed flat $\Lambda$ models. 
Thus we expect that the COBE constraint for CH models
are less stringent than the ``standard'' $\Omega_m=1$ 
closed flat toroidal models which are claimed to have been
``ruled out''.  In the next section, we will statistically 
test the likelihoods of closed models with non-trivial topology
using the COBE-DMR data.  
\pagebreak
\section{Bayesian analysis}
In general, the covariance in the temperature  
at pixel $i$ and pixel $j$ in the sky map can be written as
\BE
M_{ij}=\langle T_i T_j\rangle =\sum_{l} \langle a_{lm}a_{l'm'} \rangle W_l W_{l'}Y_{l m}(\hat{n_i})
Y_{l' m'}(\hat{n_j})+ \langle N_i N_j\rangle 
\label{eq:M}
\EE
where  $a_{lm}$ is an expansion coefficient with respect to a
spherical harmonic $Y_{lm}$, $\langle \rangle$ denotes an ensemble
average taken over all initial conditions, positions and
orientations of the observer, 
$T_i$ represents the temperature at pixel $i$, $W_l^2$  is the
experimental window function that includes the effect of 
beam-smoothing and finite pixel size,
$\hat{n_i}$ denotes an unit vector towards the center of pixel $i$
and $ \langle N_i N_j \rangle $ represents a noise covariance between 
pixel $i$ and pixel $j$. 
If the temperature fluctuations form a 
homogeneous and isotropic random Gaussian
field then the covariance matrix can be written 
in terms of the power spectrum $C_l$ as
\BE
M_{ij}=\f{1}{4 \pi}\sum_{l}(2l +1) W_l^2 C_l P_l
(\hat{n_i}\cdot \hat{n_j})+ \langle N_i N_j \rangle
\label{eq:MI}
\EE
where $P_l$ is the Legendre function. 
Then the probability distribution function of 
the pixel temperature $\vec{T}$ for the Gaussian field is 
\BE
f(\vec{T}|C_l)=\f{1}{(2 \pi)^{N/2} \det^{1/2}M(C_l) } 
\exp \Biggl (\f{1}{2}\vec{T}^T\cdot M^{-1}(C_l) \cdot \vec{T}\Biggr),
\EE
where $N$ is the number of pixels. Bayes's theorem states that
the probability distribution function of a set of parameters
$\vec{A}$ given the data $\vec{T}$ is
\BE
f(\vec{A}|\vec{T})\propto f(\vec{T}|\vec{A}) f(\vec{A}).
\EE
If we assume a uniform prior distribution, \ti{i.e.} 
taking $f(\vec{A})$ to be constant,
the probability distribution function of a power spectrum $C_l$
is then 
\BE
\Lambda(C_l|\vec{T})\propto\f{1}{\det^{1/2}M(C_l) } 
\exp \Biggl (\f{1}{2}\vec{T}^T\cdot M^{-1}(C_l) \cdot \vec{T}\Biggr). 
\label{eq:LL}
\EE
In the following analysis, we use the inverse-noise-variance-weighted
average map of the 53A,53B,90A and 90B  
COBE-DMR channels. To remove the emission from the galactic
plane, we use the extended
galactic cut (in galactic coordinates) \cite{Banday}.
After the galactic cut, best-fit monopole and dipole are removed  
using the least-square method.
To achieve efficient analysis in computation,  we  
compress the data at ``resolution 6'' $(2.6^o)^2$  
pixels  into one at ``resolution 5'' $(5.2^o)^2$ 
pixels for which there are 1536 pixels in the
celestial sphere and 924 pixels surviving the extended galactic cut.
The window function is given by $W_l=G_l F_l$ where $F_l$ are the Legendre
coefficients for the DMR beam pattern \cite{Lineweaver} and $G_l$ 
are the Legendre
coefficients for a circular top-hat function with area equal to the
pixel area which account for the pixel smoothing effect 
(which is necessary for ``resolution 5'' pixels since the 
COBE-DMR beam FWHM is comparable to the pixel size) \cite{Hinshaw}.
To account for the fact that we do not have useful information
about monopole and dipole anisotropy, the likelihood must be
integrated over $C_0$ and $C_1$ in principle. However, in practice  
we set $C_0=C_1=100 mK^2$ which renders the likelihood insensitive to
monopole and dipole moments of several $mK$.  We also assume
that the noise in the pixels is uncorrelated from pixel to pixel 
which is found to be a good approximation\cite{Tegmark-Bunn}.
\subsection{Closed flat models}
Before applying the method to the flat toroidal models 
one must be aware that a flat 3-torus we are considering is globally
anisotropic although it is globally homogeneous.
In contrast to the standard infinite models, the 
fluctuations form an anisotropic Gaussian field for a fixed orientation
if the initial fluctuation is Gaussian. 
In other words, for a given $l$, a set of 
$a_{lm}$'s ($-l \! \le \!m\!\le\! l$) are not $2l+1$ independent 
random numbers. In order to see this, we write a plane wave 
in terms of eigenmodes in the spherical coordinates,
\BE
e^{i \K\cdot\x}=\sum_{l m}  b_{k l m}~ j_l(k x) Y_{lm}(\n), ~~~
b_{k l m}=4 \pi (i)^l~ Y^\ast_{l m} (\hat \K),
\EE
where $\hat \K$ denotes the unit vector in the direction of $\K$.
Then we have
\BE
a_{lm}=\sum_\K \Phi(k)~~b_{k l m} F_{k l}
\EE
where $ F_{k l}$ is given by (\ref{eq:psT3}). Because
$a_{lm}$ is linear in $\Phi(k)$, it is Gaussian. However,
they are not independent since 
$b_{k l m}$'s are proportional to 
spherical harmonics. The statistical 
isotropy is recovered iff
the size $L$ of the cube becomes infinite where $\hat \K$ takes
the whole values for a given $k$. If one marginalizes the
likelihood with respect to the $SO(3)$ transformation, or
equivalently, the orientation of the observer,
the distribution function of $a_{lm}$ becomes non-Gaussian
since $a_{lm}$ is written in terms of a sum of a product of 
a Gaussian variable times a variable that is determined by 
a spherical harmonic $Y_{lm}(\hat \K)$ where $\hat \K$ 
is a random variable which obeys a uniform distribution.
Thus in principle, in order to compare a whole set of 
fluctuation patterns over the isotropic ensemble one should
use the likelihood function which is different
from the Gaussian one. 
\\
\indent
Nevertheless, 
we first carry out a Bayesian analysis using the Gaussian likelihood
function (\ref{eq:LL}) which depends only on $C_l$'s (which is 
$SO(3)$ invariant) 
in order to estimate the effect of the mode-cutoff
imposed by the periodic boundary conditions. 
\\
\indent
From figure \ref{fig:LLres5F} one can see that the constraint
on the size of the cell is less stringent for the low matter density
model as expected from the shape of the power spectrum. 
The effect of the suppression of the large-angle power is
not prominent unless the cell size is sufficiently smaller
than the observable region ($L<0.8H_0^{-1}=0.08\times 2R_\ast$ 
for $\Omega_m=0.1$). However, the likelihood function varies rapidly 
as the size increases since the power is jagged in $l$.
Unfortunately, the peaks in the power at $l\sim 6$ and $l\sim 11$ 
which correspond to the first and the second ISW ridge give
a bad fit to the COBE data for $L\sim 1.7H_0=0.17\times 2R_\ast$.
However, the parameter range which gives a good fit to the data 
is wider for the low matter density model. 
It should be emphasized that for both flat models
there is a parameter region in 
which the fit is much better than the infinite counterpart.
For example, the peaks at $l\sim 4$ and $l\sim 9$ in the angular 
power of the low matter density model with 
$L\sim 2.8H_0^{-1}$ give a much better fit to the COBE data
than the infinite counterpart. 
In fact, the quadrapole component in the COBE data is very low 
and the angular power is peaked at $l\sim 4$. 
For infinite flat $\Lambda$-models with
a scale-invariant initial spectrum such features
can be a problem since the ISW contribution gives an excess power
on large angular scales.
\BF
\centerline{\psfig{figure=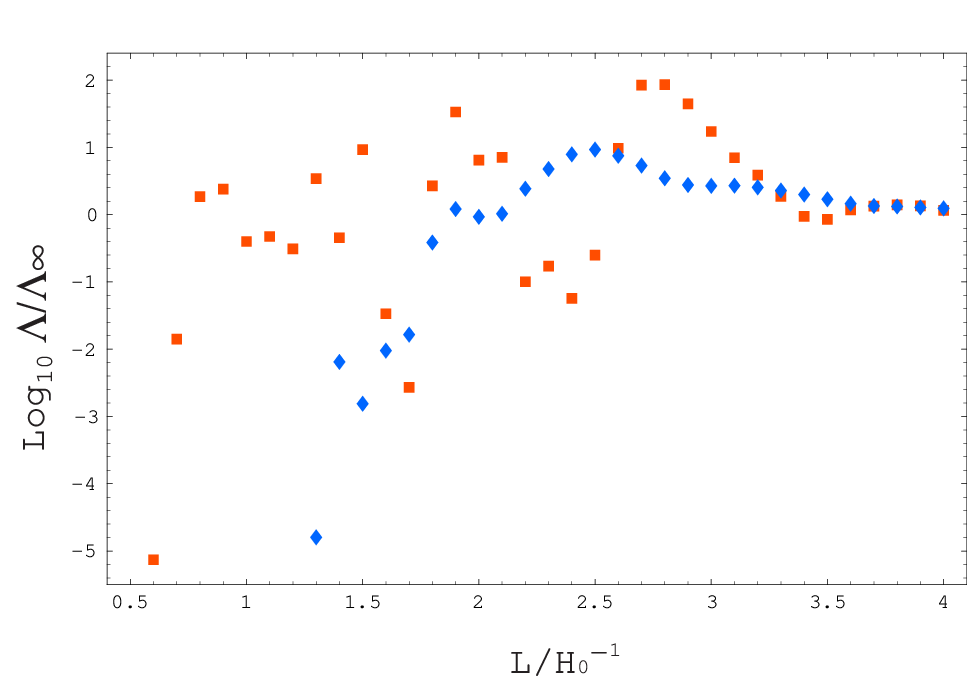,width=12cm}}
\caption{Relative Likelihoods of Toroidal Models I}
\mycaption{The likelihoods of the toroidal models 
with $(\Omega_m,\Omega_\Lambda)=(1.0,0)$ (diamond) 
and (0.1,0.9) (box)
relative to the infinite models with the same density parameters
are plotted. 
The likelihoods are marginalized over the quadrapole normalization
$Q\equiv(5C_2/(4 \pi))^{1/2}$. An isotropic Gaussian approximation has 
been used for computing the likelihoods. The COBE-DMR data is
compressed to 924 pixels at ``resolution 5''.  
}
\label{fig:LLres5F}
\EF
\\
\indent
Next, we carry out a full Bayesian analysis 
in which all the elements of $ \langle a_{lm}a_{l'm'} \rangle $ are included.
Because of the limit in the CPU power we further compress
the data at ``resolution 5'' pixels to ``resolution 3''$(20.4^o)^2$
pixels in galactic coordinates for which there are 60 pixels surviving the 
extended galactic cut. Although the information on smaller angular scales 
$l>10$ is lost, they still provide us a sufficient 
information for discriminating the effect of the non-trivial topology
since it is manifest only on large angular scales.
The computation has been done in a similar manner as previous 
analysis except for the covariance matrix for which we use (\ref{eq:M})
instead of (\ref{eq:MI}). 
The likelihoods $\Lambda$ are computed for a 
total of 2000 random orientations 
for each model. 
The approximated likelihoods which depend on only 
the power spectra are also computed for comparison. 
\BF
\centerline{\psfig{figure=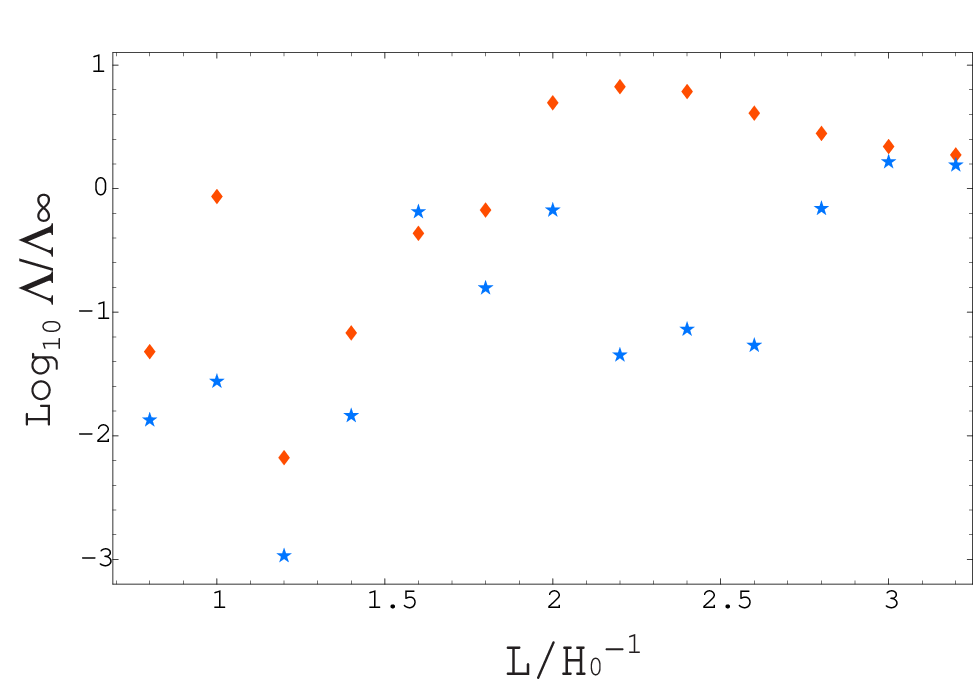,width=7.5cm}
\psfig{figure=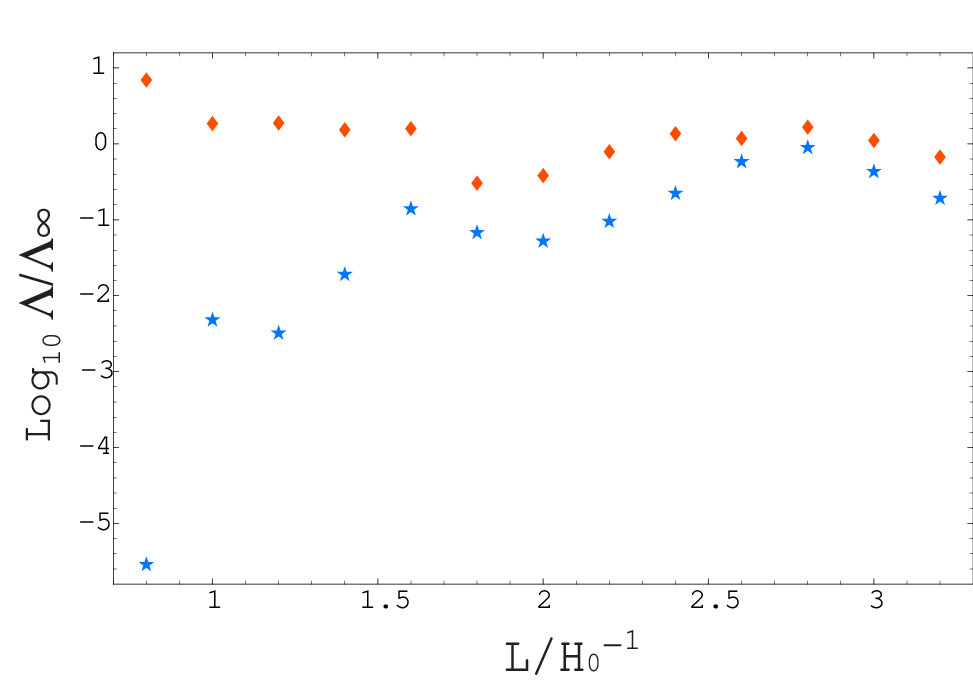,width=7.5cm}}
\caption{Relative Likelihoods of Toroidal Models II}
\mycaption{The likelihoods of the toroidal models 
relative to the infinite models with the same density parameters
marginalized over 2000 orientations (star) and that using only
the power spectrum (diamond). 
The inverse-noise-variance-weighted
average map of the COBE-DMR data is compressed to 60 pixels
(resolution 3).    
All the likelihoods are also marginalized over the quadrapole normalization
$Q\equiv(5C_2/(4 \pi))^{1/2}$. 
}
\label{fig:LLres3F}
\EF
As shown in figure \ref{fig:LLres3F}, the discrepancy between the 
likelihood marginalized over the orientation of the observer 
and the approximated likelihood is prominent for models with small volume
in which the effect of the non-trivial topology is significant. 
Let us estimate the size of the cell 
for which the effect becomes insignificant.
For the ``standard'' model with $\Omega_m\!=\!1.0$, the volume of
the observable region is $4 \pi R_\ast^3/3\!\sim\! 34$ which gives 
the critical scale $L_c\!\sim\!3.2 H_0^{-1}$ for which the volume of the
cell is comparable to the volume of the observable region at present.  
For the low matter density model 
a significant amount of large-angle fluctuations are produced 
at the $\Lambda$-dominant epoch $z\!\sim\!1$. 
Therefore, one should compare the volume of the 
sphere with radius $\eta(z\!=\! 0)-\eta(z\!=\!1)\!\sim\! 0.9 H_0^{-1}$
to the volume of the cell instead of the observable region
which gives $L_c\!\sim\!1.5 H_0^{-1}$.   
We can see from figure \ref{fig:LLres3F} that these 
estimates well agree with the numerical result. 
\\
\indent
We can give two explanations for the discrepancy 
although they are related each other.
One is the non-Gaussianity of fluctuations 
and another one is the correlations between $a_{l m}$'s 
owing to the global anisotropy in the background geometry.
\\
\indent
Suppose a random variable $Z=XY$ in 
which $X$ obeys a distribution function $E(X)$ which is even and
$Y$ obeys a distribution function $F(Y)$. Then the  
distribution function G of $Z$ is given by 
\BE
G(Z)=\int \f{E(Z/Y)}{|Y|}F(Y) dY,
\EE
which is apparently even. 
Because the fluctuations are written in terms of a sum of 
products of Gaussian variable  $\Phi(k)$ (with zero average) 
times a non-Gaussian variable $Y_{lm}(\hat \K)$, the skewness
in the distribution function of $a_{lm}$ marginalized over the
orientation is zero although the kurtosis is non-zero. 
\\
\indent
The correlations in $a_{lm}$'s are the consequence 
of a gap between the degree of freedom of $(\hat \K)$
and $(l,m)$. The degeneracy number in a $k$-mode 
(=the number of the direction $\hat \K$) is 
much less than the number of relevant ``quantum numbers''
$(l,m)$\footnote{Here we assumed that for each degenerated 
mode specified by $\hat \K$, $\Phi(k)$  
are independent Gaussian variables.} if the scale
(=$2\pi/k$) is comparable 
to the length of the side $L$.
Taking an ensemble average over the initial condition, we have
\BE
\langle a_{lm}a_{l'm'} \rangle =\sum_{\K \ne 0} \f{8 \pi^3}{k^3L^3} {\cal{P}}_\Phi(k)
Y_{lm}(\hat \K)Y_{l'm'}(\hat \K)
F_{k l}F_{k l'},
\label{eq:aa}
\EE
where $F_{kl}$ is given by (\ref{eq:psT3}). The sum does not
vanish in general when $(l,m)\ne(l',m')$ which is the consequence
of the global anisotropy in the background geometry. 
However, if one takes an average of
(\ref{eq:aa}) over $\hat \K$, one finds that  
all off-diagonal elements vanish because of the orthogonality 
of the spherical harmonics.
Similarly, one can consider 4-point correlations
$\langle a_{l_1 m_1}a_{l_2 m_2}a_{l_3 m_3}a_{l_4 m_4} \rangle $. In this case,
all the off-diagonal elements $(l_i,m_i)\ne(l_j,m_j)$ do not
necessarily vanish even if one takes an average over $\hat\K$.  
Thus the non-Gaussianity for the flat toroidal models 
contrasts sharply with the one for the compact hyperbolic models
in which the pseudo-random Gaussian property of the 
expansion coefficients $b_{klm}$ (which are obtained by
expanding the eigenmode in terms of eigenmodes in the universal covering 
space)\cite{Inoue1,Inoue3} renders off-diagonal elements always
vanish if an average is taken over the position of the
observer (although the kurtosis is non-zero). 
The difference can be attributed to the property of eigenmodes.
In a less rigorous manner the property of the eigenmodes 
which are projected onto a sphere with large radius 
can be stated as follows:
\ti{Projected eigenmodes of compact hyperbolic spaces 
are ``chaotic'' whereas those of compact flat spaces  
are ``regular''}. 
\\
\indent
As for the constraint on the size of the cell, 
we set a slightly severe condition for the 
low matter density model since the likelihood of the infinite counterpart
is $\sim 10^{-1}$ of that of the infinite ``standard'' model due
to a slight boost on the large angular scales caused by the 
ISW effect.  Together with the previous analysis using the data on 
the ``resolution 5'' pixels, 
the conditions on the relative likelihood 
$\log_{10}(\Lambda/\Lambda_\infty)\!>\!-2$
and $\log_{10}(\Lambda/\Lambda_\infty)\!>\!-1$ yield
$L\ge 0.40R_\ast$ and $L\ge 0.22 R_\ast$ 
for the ``standard'' toroidal model $(\Omega_m,\Omega_\Lambda)=(1.0,0)$ 
and the low matter density toroidal model 
$(\Omega_m,\Omega_\Lambda)=(0.1,0.9)$,
respectively. Here $\Lambda_\infty$ denotes
the likelihood of the infinite counterpart with the same 
density parameters. The maximum number $N$ of images of the cell  
within the observable region at present is 8 and 49 for the
former and the latter, respectively. 
Note that the constraint 
on the ``standard'' toroidal model is consistent with
the previous result\cite{Sokolov93,Staro93,Stevens93,Oliveira95a}.
\subsection{Closed hyperbolic models}
For CH models, the temperature fluctuations for a homogeneous and
isotropic ensemble can be written
as a sum of modes which consist of 
a product of :a Gaussian random
number $\Phi_\nu(0)$ originated from quantum fluctuations; a Gaussian
pseudo-random number $ \xi_{\nu l m}$ 
originated from geometry. Thus the amplitude of 
temperature fluctuations in $(l,m)$ space is written as
\BE
a_{l m}=\sum_{\nu} \Phi_\nu(0) \xi_{\nu l m}F_{\nu l}
\label{eq:almCH}
\EE
which are non-Gaussian. 
In contrast to flat models, correlations between
$2l+1$ variables for a given $l$ are very small
since the case for which one observes any symmetric structure 
in the temperature fluctuations is rare. 
\\
\indent
First of all, we consider the effect of suppression on large
angular power. To do so, we fix the initial conditions as 
$\Phi^2_\nu(0)=Const./(\nu (\nu^2+1))$ so that they are not random
number but satisfying the standard assumption that the power spectrum
is described by the extended Harrison-Zel'dovich spectrum. 
Then the fluctuations approximately form a Gaussian random field 
owing to the pseudo-Gaussianity of eigenmodes and the 
property of fluctuations is fully specified by the corresponding 
angular power $C_l$. As in the previous sections, we use
the inverse-noise-variance-weighted
average map of the 53A,53B,90A and 90B  
COBE-DMR channels at 924 pixels ($(5.2^o)^2$)
surviving the extended galactic cut. 
\BF[t]
\centerline{\psfig{figure=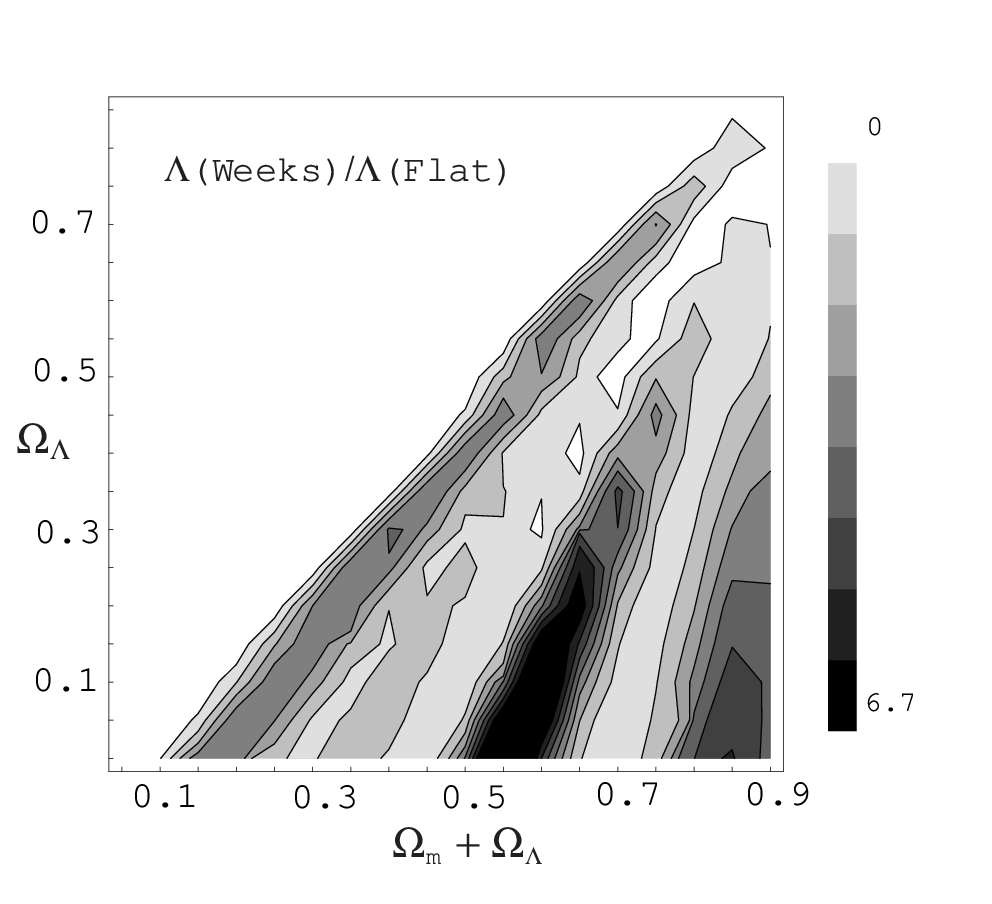,width=8cm}
\psfig{figure=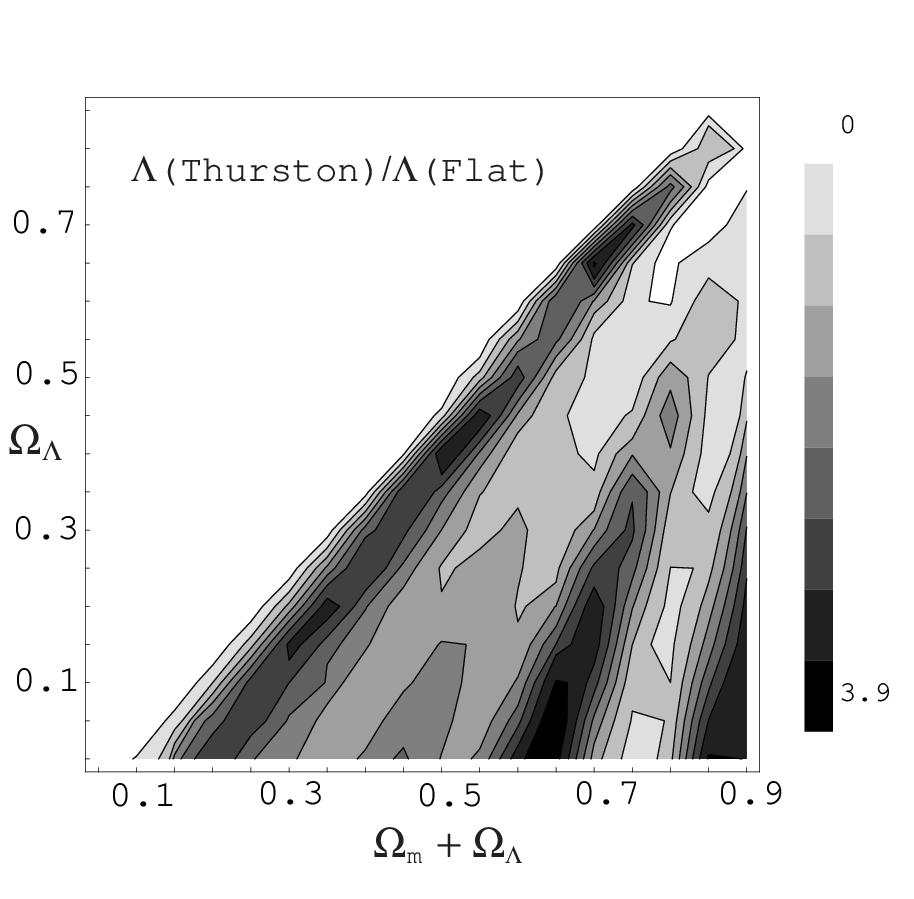,width=7.4cm}}
\caption{Relative Likelihoods of Smallest CH Models}
\mycaption{Plots of the ratio of approximated likelihoods for two smallest 
CH models (Weeks and Thurston) to a likelihood for the Einstein-de Sitter model
($\Omega_m=1.0$) with scale invariant initial spectrum ($n\!=\!1$).
All the likelihoods are marginalized over the normalization of the 
power. Here we have assumed ``fixed initial fluctuations''
$\Phi^2_\nu(0)=Const./(\nu (\nu^2+1))$ for CH models.} 
\label{fig:LLWT}
\EF
\BF
\centerline{\psfig{figure=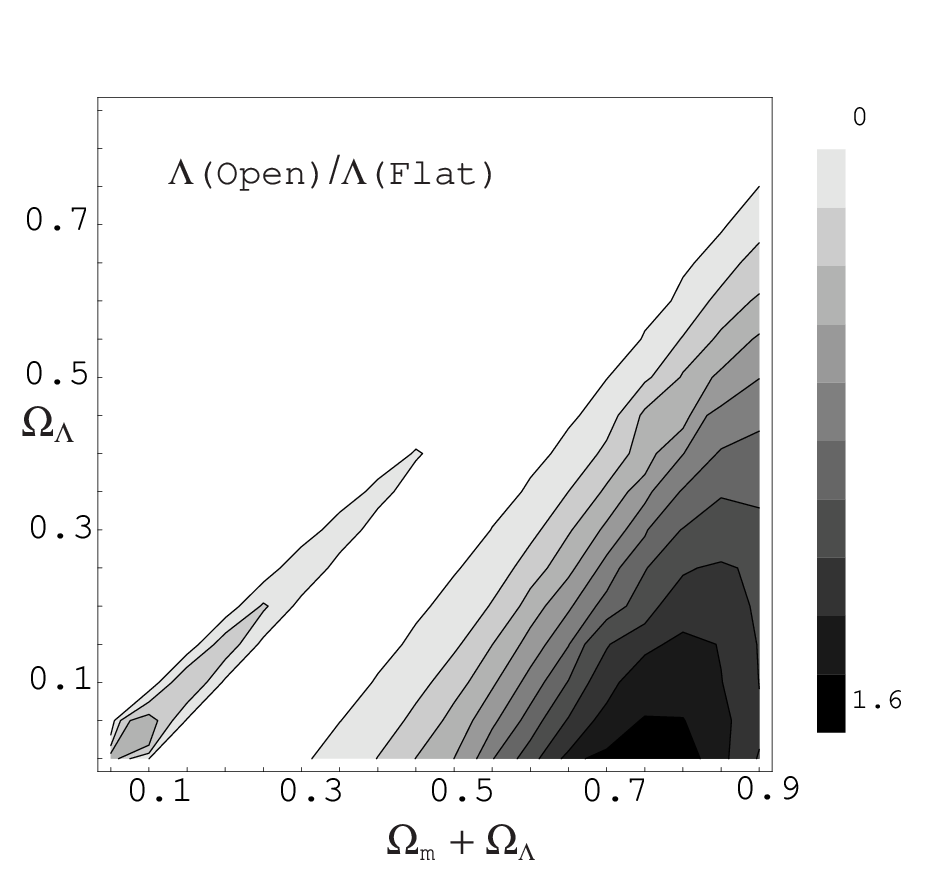,width=7.5cm}
\psfig{figure=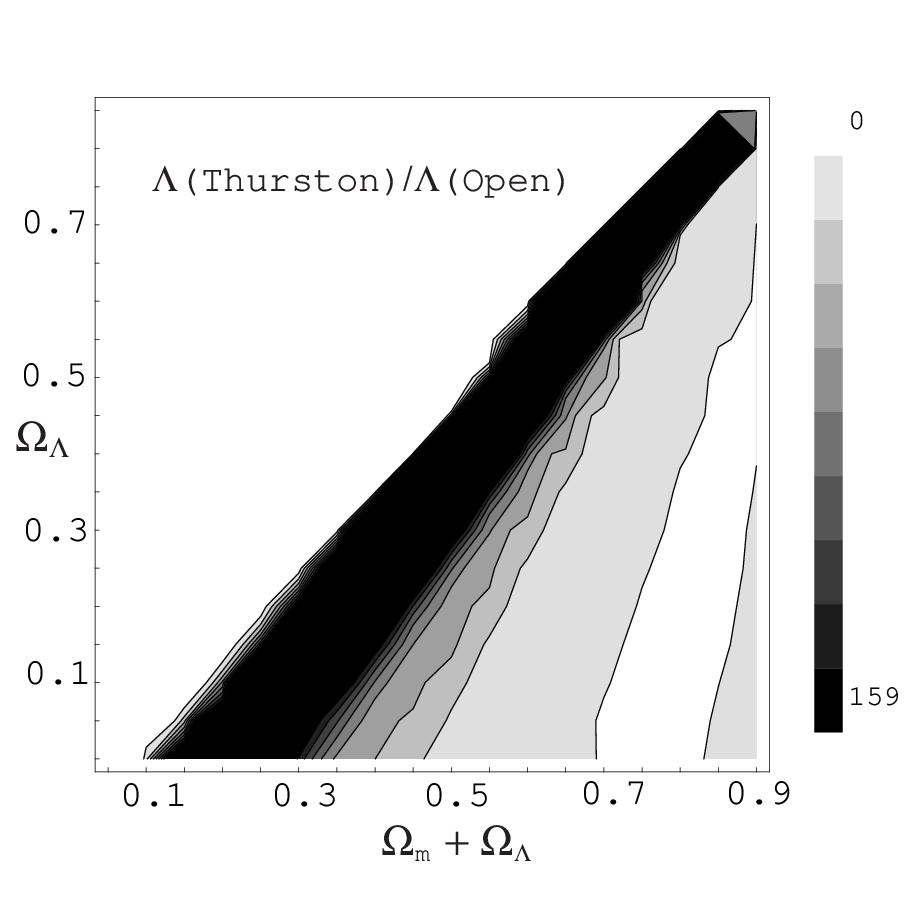,width=7.5cm}}
\caption{Relative Likelihoods of Open and Thurston Model}
\mycaption{Plots of the ratio of approximated likelihoods 
for the infinite hyperbolic ``open'' models 
($n\!=\!1$) to one for the Einstein-de Sitter model ($n\!=\!1$) (left)
and likelihoods for the Thurston models relative to one
for infinite hyperbolic models (right).
All the likelihoods are marginalized over the normalization of the 
power. We have assumed ``fixed initial fluctuations''
$\Phi^2_\nu(0)=Const./(\nu (\nu^2+1))$ for the Thurston models.
The slight improvement in the likelihood of infinite hyperbolic models
with $\Omega_m<0.1$ is caused by the absence of the supercurvature
modes. For the Thurston models with $\Omega_m=0.1 \sim 0.3$ the 
likelihoods are significantly improved.
} 
\label{fig:LLTO}
\EF
As is expected from the shape of the angular powers,
for a wide range of parameters ($\Omega_m+\Omega_\Lambda>0.1$), 
the likelihoods for the smallest (for manifolds) CH models  
are better than one for the Einstein-de Sitter models
with the scale-invariant spectrum ($n=1$) where $\Delta T_l=
(l(l+1)C_l/2 \pi)^{1/2}$ is almost constant in $l$.
One can see that the fit to the COBE data is much better than
the Einstein-de Sitter models for three parameter regions:1.
$\Omega_m= 0.5 \sim 0.7$ with small $\Omega_\Lambda$ for which
the angular power is peaked at $l\sim 4$   
which corresponds
to the first ISW ridge of the transfer function
;2. $\Omega_m=0.85\sim0.9$ where the angular scale $l\sim 4$
corresponds to the OSW ridge;3. $\Omega_m \sim0.2$
where the slope of the power on large angular scales $l<10$
fits well with the data. 
\\
\indent
For low matter density models, the ISW effect leads to an excess
large-angle power. 
Therefore the likelihoods of low matter density models 
$\Omega_m=0.1 \sim 0.3$
with infinite volume are relatively small because of the 
low quadrapole moment in the COBE data.
However, for small CH models,
the excess power owing to the ISW effect is mitigated by 
suppression owing to the mode-cutoff. 
Therefore, likelihoods for small CH models with low matter density 
are significantly improved compared with the infinite counterparts
(figure \ref{fig:LLTO}). 
\\
\indent
As we have seen in chapter 3, the property of eigenmodes 
is associated with the volume and the ``anisotropy'' which 
can be characterized by the length of the shortest periodic geodesic 
$l_{min}$ or diameter $d$. 
For instance, as can be seen in figure (\ref{fig:LLWT}) the 
likelihoods of the Weeks models 
(volume$\!=\!0.94$, $l_{min}\!=\!0.585$ and $d\!=\!0.82$) are similar 
to those of the Thurston models with similar geometrical property
(volume$\!=\!0.98$, $l_{min}\!=\!0.578$ and $d\!=\!0.81$). 
If $l_{min}$ is comparable to the ``average'' radius $d_{ave}$ defined
as the radius of a sphere with the same volume of the CH space then 
the statistical property of CMB anisotropy is described by Weyl's 
formula. 
\\
\indent
For CH models with large volume, the likelihoods will converge
to those of the infinite counterparts although the convergence rate 
depends on the cosmological parameters.
One can see in figure (\ref{fig:LL27CHM}) that 
the conspicuous difference for $\Omega_m\!=\!0.1-0.3$ still persists
for volume$\sim 6$ whereas such difference is not observed for
nearly flat cases ($\Omega_t\!=\!\Omega_m+\Omega_\Lambda\!=\!0.9$).
The difference depends on the number $N$ of the copies of the
fundamental domain in the observable region at present.
Suppose that $\Omega_\Lambda\!=\!0$ then the comoving radius of the
last scattering surface in unit of the present curvature radius 
$R_{curv}$ is approximately given by
\BE
R_{LSS}=R_{curv} \cosh^{-1}(2/\Omega_m-1).
\EE
The comoving volume of the ball inside the last scattering surface 
in hyperbolic space is
\BE
v=\pi R^3_{curv}(\sinh (2 R_{LSS}/R_{curv})-2 R_{LSS}/R_{curv})
\EE
For example, $v\!\sim\!490 R^3_{curv}$ for 
$(\Omega_m,\Omega_\Lambda)\!=\!(0.2,0)$ 
whereas $v\!\sim\! R^3_{curv}$ for 
$(\Omega_m,\Omega_\Lambda)\!=\!(0.9,0)$. Apparently, in nearly flat
cases, the volume of the spatial hypersurface 
should be smaller than $R^3_{curv}$ for  
detecting the effect of the non-trivial topology.
However, inclusion of the cosmological constant leads to an increase  
in $R_{LSS}$ because of a slow increase in 
the cosmic expansion rate in the past. Thus it allows
a large $N$, for instance,
$N$ is 8.7 for a Weeks model with 
$\Omega_\Lambda\!=\!0.7$ and $\Omega_m\!=0.2\!$ 
whereas $N\!=\!1.2$ for one with 
$\Omega_\Lambda\!=\!0$ and $\Omega_m\!=0.9\!$. 
\BF
\centerline{\psfig{figure=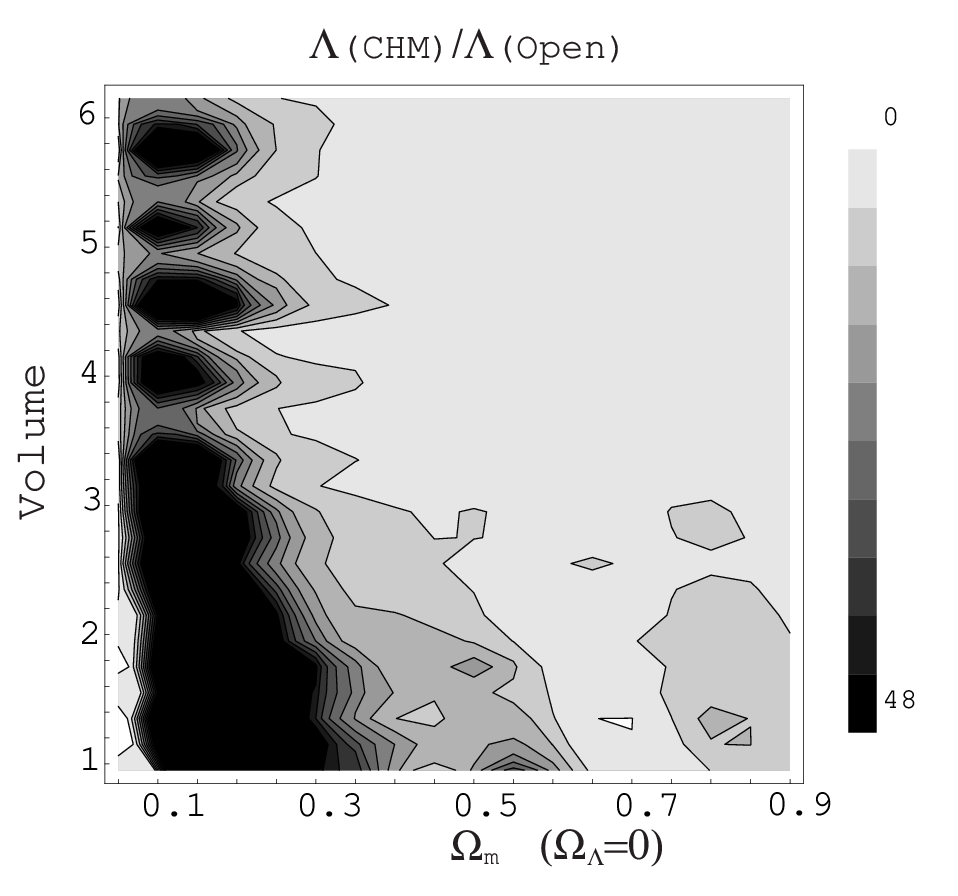,width=7.7cm}
\psfig{figure=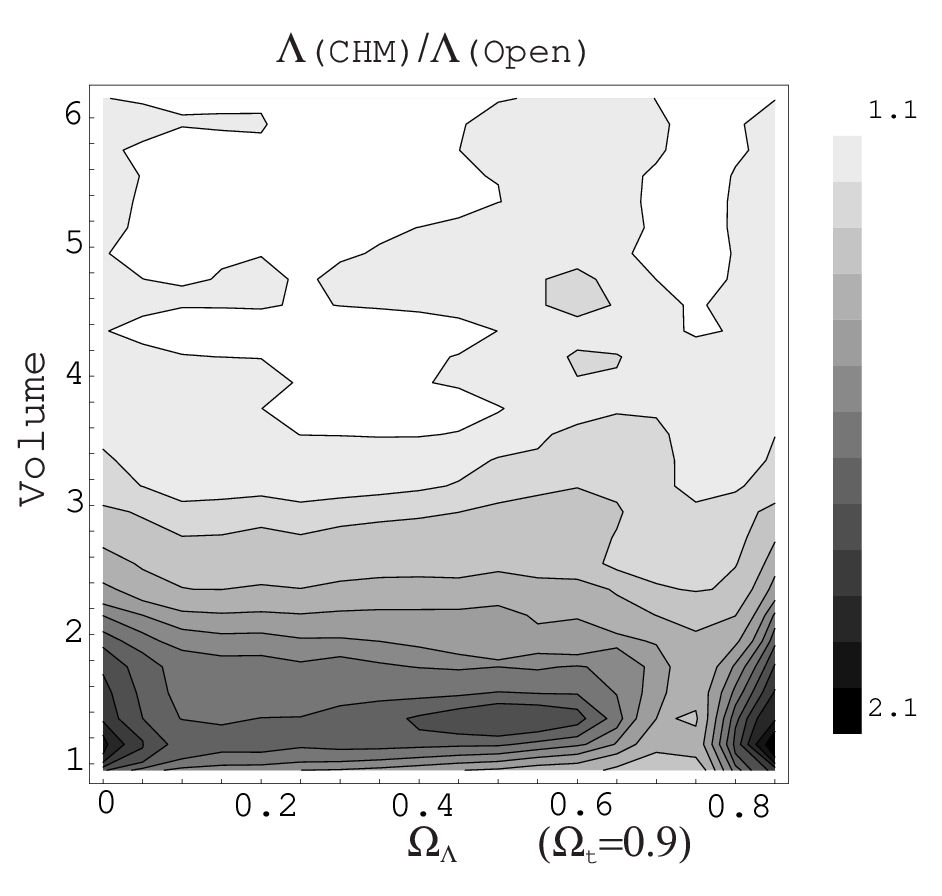,width=7.5cm}}
\caption{Relative Likelihoods of 27 CH Models}
\mycaption{Plots of the ratio of approximated likelihoods for 27  
CH models with volume ($0.94-6.15$) 
to a likelihood for the Einstein-de Sitter model
($\Omega_m=1.0$) with scale invariant initial spectrum ($n\!=\!1$).
All the likelihoods are marginalized over the normalization of the 
power. Here we have assumed ``fixed initial fluctuations''
$\Phi^2_\nu(0)=Const./(\nu (\nu^2+1))$ for CH models.} 
\label{fig:LL27CHM}
\EF
\BF
\centerline{\psfig{figure=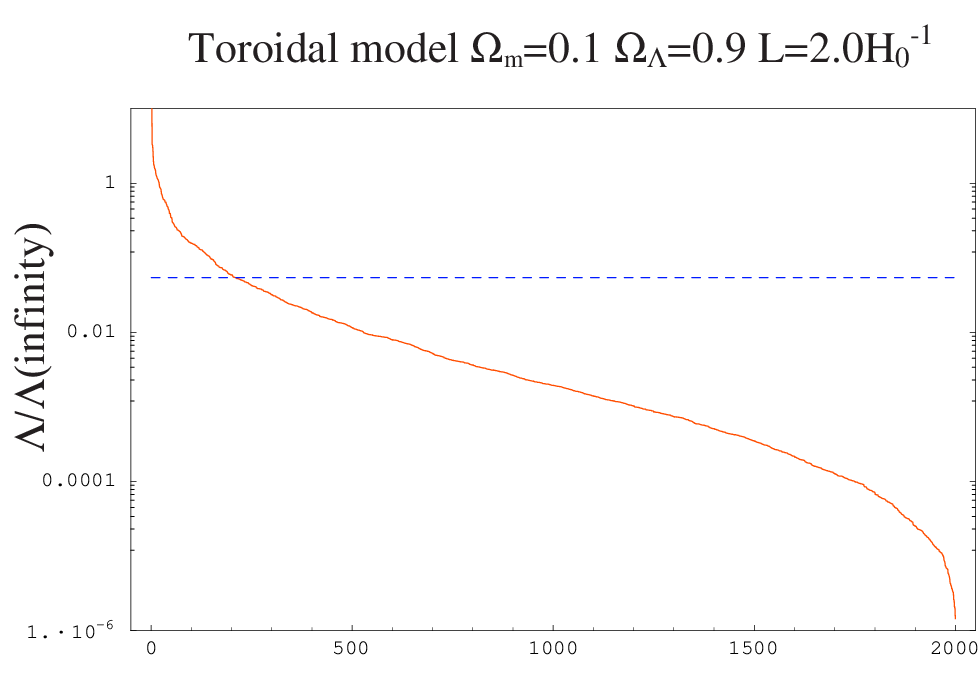,width=11cm}}
\EF

\BF
\centerline{\psfig{figure=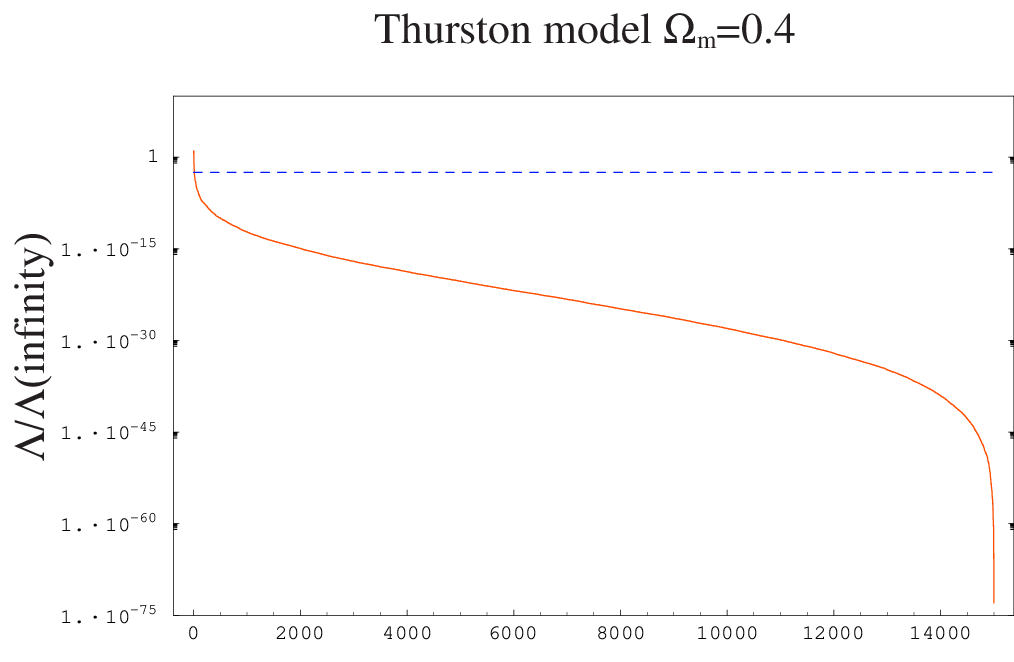,width=11.5cm}}
\caption{Relative Likelihoods of Toroidal and Thurston Models} 
\mycaption{Plots of the ratio of ``rigorous'' likelihoods 
incorporating the effect of inhomogeneity and anisotropy of the
background geometry for two closed multiply
connected models relative to one for the 
infinite counterparts in ascending 
order(full curve) for 2000 random realizations of orientation:
closed flat toroidal model with $\Omega_m=0.1,\Omega_\Lambda=0.9,
L=2.0 H_0^{-1}$ (left) and 
1500 random realizations of position and 10 realizations of orientation
;a Thurston model
with $\Omega_m=0.4$ (right).
The dashed lines denote the ensemble averaged values.
The number of the copies of fundamental domain inside
the last scattering surface is 64.7 for the closed
flat toroidal model and 72.3 for the Thurston model.}
\label{fig:LLND}
\EF
Next, we consider the effect of the non-diagonal elements which we
have neglected so far. 
The likelihood for homogeneous and isotropic ensemble is obtained by
marginalizing the likelihoods all over the positions $\x_{obs}$ 
and the orientations $\alpha$ of the observer,
\BE
\Lambda=\int\!\!\sqrt{g}~ d\x_{obs} d\alpha \Lambda(\x_{obs},\alpha).
\EE
We assume a constant distribution for the volume element 
$\sqrt{g} d\x_{obs}$ and $d\alpha$ which is the volume element 
of a Lie group $SO(3)$ with Haar measure.   
Assuming that the initial fluctuations are Gaussian, the 
likelihood $\Lambda(\x_{obs},\alpha))$ is given by 
(\ref{eq:M}) and (\ref{eq:LL}) where 
$\langle a_{lm}a_{l'm'} \rangle $ is written in terms of the expansion
coefficients $\xi_{\nu l m}$ and the initial fluctuation $\Phi_\nu$ as
\BE
\langle a_{lm}a_{l'm'} \rangle \propto \f{{\cal{P}}_\Phi(\nu)}{\nu(\nu^2+1)}
\xi_{\nu l m} \xi_{\nu l' m'}F_{\nu l} F_{\nu l'},
\EE
where $\langle \rangle$ denotes an ensemble average
taken over the initial conditions, ${\cal{P}}_\Phi(\nu)$ is the initial
power spectrum and $F_{\nu l}$ describes contribution from 
the OSW effect and the ISW effect, respectively.
$\xi_{\nu l' m'}$'s are functions of $\x_{obs}$ and $\alpha$.
As a typical example we choose the Thurston manifold $m004(-5,1)$ 
which is the second smallest known CH manifold.  
In order to compute likelihoods, we use a compressed 
data at ``resolution 3''$(20.4^o)^2$
pixels in galactic coordinates for which there are 60 pixels surviving the 
extended galactic cut as in closed flat cases. 
\\
\indent
In comparison with the flat toroidal model with low matter density
($\Omega_\Lambda=0.9,\Omega_m=0.1, N=65$) with similar value in N 
the likelihoods of
a Thurston model ($\Omega_\Lambda=0,\Omega_m=0.4, N=72$)
distribute in a wide range of values(figure \ref{fig:LLND}).
For the toroidal model, the likelihoods are larger than the average
for 11 percent out of the total of 2000 orientations which are
randomly chosen. On the other hand, for the Thurston model, only 0.09
percent of the total of 1500 positions and 10 orientations are larger
than the average. We have also computed likelihoods of the 
Thurston model for 5100 positions and 40 orientations. Then the 
percentage has reduced to 0.02 from 0.09. The log of ratio
of the likelihood marginalized over 5100 positions and 40 orientations
to the one of the infinite hyperbolic model ($\Omega_m\!=\!0.4$)
is $\textrm{Log}_{10}(\Lambda/\Lambda(\infty))=-1.4)$. The maximum value for
the likelihood is $\textrm{Log}_{10}(\Lambda(\max)/\Lambda(\infty))=3.6)$.
Thus the fit to the COBE data at a particular
place with a particular orientation is usually not good
for the Thurston model but it does not constrain the model.
For a particular set of position and orientation likelihoods are much
\textit{better} than the standard infinite counterpart. 
\\
\indent
The likelihood analyses in \cite{Bond1,Bond2} are based on correlations 
for a particular choice of position $Q$ where the 
injectivity radius is locally maximal with 24 orientations.  
It should be emphasized that the 
choice of $Q$ as an observing point is very special one. For instance 
it is the center of tetrahedral and Z2 symmetry of the Dirichlet domain 
of the Thurston manifold 
although they are not belonging to 
the symmetry of the manifold \cite{Inoue3}. 
In mathematical literature it is standard to choose $Q$ 
as the base point which belongs to a ``thick'' part of the 
manifold since one can expect many symmetries. 
However, considering such a 
special point as the place of the observer cannot 
be verified since it is inconsistent with the 
Copernican principle. Because any CH models are globally inhomogeneous,
one should compare fluctuation patterns expected 
at \ti{every place} on the space. On the other hand, 
the previously studied closed flat models in which the spatial geometry is 
obtained by gluing the opposite faces by three translations are
globally homogeneous\footnote{As described in chapter 1, other closed
flat spaces are globally inhomogeneous.} for which one does not need to
compute fluctuations at different positions. 
Any CH models cannot be constrained until all the 
possible fluctuations are compared to the data.
\\
\indent
The result is not surprising if one knows the pseudo-random
behavior of eigenmodes on CH spaces. 
For each choice of the position and orientation of the
observer, a set of expansion coefficients $\xi_{\nu l m}$ of eigenmodes 
is uniquely determined (except for the phase factor), 
which corresponds to a ``realization'' of independent random 
Gaussian numbers. By taking an average over the position and the
orientation, the non-diagonal terms proportional to 
$\langle \xi_{lm}\xi_{l'm'} \rangle, l \ne l', m \ne m'$ vanish. In other
words, a set of anisotropic patterns all over the place in the CH
space comprises an almost isotropic random field. Consider two realizations 
$A$ and $B$ of such an isotropic random field. The chance you would
get an almost similar fluctuation pattern for $A$ and $B$ would be 
very low but we do have such an occasion. Similarly, the likelihood at a
particular position with a certain orientation is usually very low
but there \ti{are} cases where the likelihoods are considerably high.
Thus we conclude that the COBE constraints on small CH models are
less stringent compared with the ``standard'' flat toroidal model with 
($\Omega_m\!=\!1.0$) as long as the Gaussian pseudo-randomness of the 
eigenmodes holds. 
\chapter{Search for Topology}
\thispagestyle{headings}
\bq
\ti{Clever persons cannot fall in love. 
Love is blind. To become a scientist, one must love
Mother Nature. She would open her heart to only those who
truly love her.}     
\vskip 0.1truecm
\rightline{(Torahiko Terada, 1878-1935)}
\eq
Suppose that the spatial hypersurface of the universe is multiply connected
on scales smaller than the comoving distance to the last scattering
surface, then, in principle we would be able to 
detect the signature of the non-trivial
topology of the spatial geometry which would appear as 
the periodicity or non-Gaussianity in the CMB sky map.
\\
\indent
The periodic structure in the CMB sky map would certainly be the 
direct evidence of the non-trivial topology. Assuming that 
the temperature fluctuations are produced entirely at the last scattering
surface(determined by the OSW effect only), 
the periodical structure would appear as a pair of circles
in the sky on which the fluctuation patterns are identified\cite{CSS98a}.
In practice the effect of the periodical structure in the CMB
would appear as an anomalously large 
correlations on a particular set of directions in the sky 
\cite{Levin98}.
Suppose a sphere with thickness comparable or less than 
the typical correlation length of the 
relevant fluctuations in the universal covering space.
Then one can expect a number of copied ``thick'' spheres that intersect
the original one at a particular place. In other words, one would
expect a number of copies of a particular realization of fluctuations
in different directions in the sky. Note that the set of the 
points with large correlation does not necessarily belong
to the ``matched circles'' because of the thickness of the spheres.
\\
\indent
Another feature is the Non-Gaussianity in the CMB 
temperature fluctuations which is the consequence of 
the break of the global homogeneity or isotropy.
For a particular choice of the position and the orientation of
the observer, the temperature fluctuations form an anisotropic 
Gaussian random field. If one marginalizes the likelihood 
over the positions and the orientations, 
then the fluctuations are regarded as an isotropic non-Gaussian 
random field\cite{FM97}. Because we do not have any information of the 
position and the orientation of the observer we should
give the same chance for each set of position and orientation.
In contrast to cosmic string models, the non-Gaussianity
only appears on large-angular scales assuming that the spatial 
hypersurface is multiply connected on scale not significantly smaller
than the present horizon scale. Furthermore it does not depend on the 
type of the promordial perturbation(adiabatic or isocurvature).
In some non-standard (with multiple scalar fields) 
inflationary models, non-Gaussianity
in the initial fluctuations are also expected but with isocurvature
modes in most cases\cite{Bucher97}. It seems that 
one needs unnatural conditions
for the potential(not smooth) and the inflaton for generating 
adiabatic non-Gaussian fluctuations\cite{Salopek99}. The detection
of adiabatic non-Gaussianity will certainly be the positive sign of the 
non-trivial topology of the universe.  
\\
\indent
As we shall see, in order to search for the periodic patterns, one 
needs a sky map with high signal-to-noise ratio.  
Therefore, absence of such periodical structure in the COBE-DMR data 
in which the noise is comparable to the signal 
(on angular scales $\sim 10^o$) does not imply the 
trivial topology (=simply connectivity) of the spatial geometry.
In near future, satellite missions MAP and \ti{Planck}
will provide us CMB sky maps with high resolution and
high signal-to-noise ratio compared with COBE. 
Using such maps we may detect the imprint of the 
non-trivial topology if we live in a ``small universe''. 
\section{Circles in the sky}
Suppose a ``bubble'' $B$ in the form of a 2-sphere with small radius 
in a multiply connected constantly curved space $M$.
As one inflates $B$ in $M$,
$B$ will eventually meet itself 
when the radius is equal to a half
length of a closed piece-wise geodesic curve 
(which cannot be contracted to a point by
continuous deformation). Let us further assume that $B$ is able to 
cross itself without any deformation. Then 
the self-intersection becomes a circle $C$ when inflated further.
For an observer sitting at the center of $B$, the shortest geodesic 
distance from the center $O$ to a point on $C$ is equivalent to 
the radius. Therefore, the length of a geodesic segment $s1$
that connects $O$ and a point $p$ on $C$ is equal to that of another 
geodesic segment $s2$ that also connects $O$ and a point $p$ on $C$.
In other words, we can see a pair of circles in two different 
directions which are actually two images of a 
circle $C$ at the \ti{same time}.  
Now let us consider the last scattering surface as $B$.
If the injectivity radius is less than the 
radius of the last scattering surface, then one can find
a pair (or pairs) of circles around which the temperature 
fluctuation patterns are identical.
In the universal covering space, one can consider
the copied last scattering surface $g(B)$ by the element $g$ of the
discrete isometry group. Then the matched circles correspond to the 
intersection of $B$ and $g(B)$ and that of 
$B$ and $g^{-1}(B)$(see figure \ref{fig:circle}). 
\BF[tpb]
\begin{flushleft}
\centerline{\psfig{figure=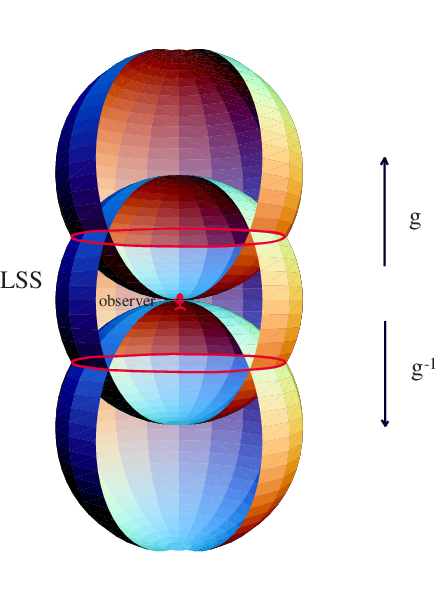,width=11cm}}
\caption{Matched Circles}
\end{flushleft}
\mycaption{For multiply connected models identical fluctuation 
patterns around a pair of circles might be observed. 
The circles correspond to the intersections of the 
last scattering surface and its clones. }
\label{fig:circle}
\EF 
It should be noted that the distance 
between the observer and the copied observer should
be less than the diameter of the last scattering surface
for detecting the matched circles.
\\
\indent
In order to statistically test the significance of the
periodical structure, we need to develop a statistical tool.
Let us choose the directions $n_1$ and $n_2$ which correspond to 
the centers of circles $C_1$ and $C_2$
with angular radius $\alpha$ in the sky. 
We now introduce a statistic for the correlation
of the temperature fluctuations $\Delta T$ along the circles 
which is known as a Pearson's (product-moment) correlation constant $r$    
\BE
r^2(j)=\f{(\sum_{i} \Delta T(n_{1i}) \Delta T(n_{2 i+j}))^2}
{\sum_i \Delta T^2(n_{1 i})\sum_k \Delta T^2(n_{2 k+j})}  
\EE 
where $n_{j 1},n_{j 2},\dots,n_{j N},n_{j N+1}\!=\!n_{j 1},j\!=\!1,2$
denotes the directions pointing to pixels around $C_1$ and $C_2$
(numbering the pixels clockwise for $C_1$ but counterclockwise
for $C_2$). From the Cauchy-Schwartz inequality, it is apparent that 
$r$ ranges in $[-1,1]$. $r=1$ corresponds to the perfect 
correlation while $r=-1$ corresponds to the perfect 
anti-correlation. In spatially orientable models, for a pair of 
matched circles we expect a nearly perfect correlation $r\approx 1$
if the noise is negligible. If we include
spatially unorientable spaces, we should also test for 
correlation in which pixels are numbered in the same orientation
(clockwise-clockwise) for both circles. 
\\
\indent
In actual experiments, we should treat the noise that 
might smear the clear detection of the periodicity. Let us assume
that the temperature fluctuations at each pixel $i$ for a circle is
written as a sum of 
independent Gaussian random numbers for signal 
$s$ and noise $\eta$ as
\BE
\Delta T(n_{i})=s_{i}+\eta_{i}
\EE
which obey the distribution 
\BE
p(s_,\eta_1,\eta_2)=\f{1}{(2 \pi)^{3/2}\sigma^2_\eta \sigma_s}
\exp \Biggl [-\f{1}{2}\Biggl( \f{s_1^2+s_2^2}{\sigma^2_\eta}
+\f{\eta^2}{\sigma^2_s} \Biggr ) \Biggr ].
\EE
where $\sigma^2_s$ and $\sigma^2_\eta$ denote the variance of 
the signal and the noise, respectively.  
Marginalizing over the signal, the distribution of 
the temperature fluctuations for a pair of matched circles 
$C_1$ and $C_2$ is given by the Gaussian distribution 
\BE
p(\Delta T_1,\Delta T_2)=
\f{1}{2 \pi \sigma^2 \sqrt{1-\rho^2}}
\exp \Biggl [
-\f{1}{2(1-\rho^2)}
\Biggr (
\f{\Delta T^2_1+\Delta T^2_2-2 \rho \Delta T_1 \Delta T_2}
{\sigma^2}
\Biggr )
\Biggr ],
\EE
where 
\BE 
\rho \equiv \f{\textrm {cov}(\Delta T_1,\Delta T_2)}
{\sigma _{\Delta T_1}\sigma_{\Delta T_2}}=
\f{\sigma^2_s}{\sigma^2_s+\sigma^2_\eta}=\f{\xi^2}{1+\xi^2}
\EE
and $\sigma^2\equiv \sigma^2_s+\sigma^2_\eta$, 
$\xi=\sigma_s/\sigma_\eta$.
Let $N_c$ be the total number of pixels in each circle. 
Then the distribution function of the correlation coefficient $r$
for a pair of matched circle is written as \cite{Maruyama}
\BE
p(r)=\pi^{-1} (N_c-1)(1-r^2)^{(N_c-3)/2}
(1-\rho^2)^{N_c/2}
\int_0^\infty
\f{d\beta}{(\cosh \beta- \rho r)^{N_c-1}}.
\label{eq:r} 
\EE
For sufficiently large $N_c$, 
\BE
z=\f{1}{2}\ln \f{1+r}{1-r} 
\EE
is known to obey the Gaussian distribution 
with average 
\BE
\langle z \rangle =\f{1}{2}\ln \f{1+\rho}{1-\rho}
\EE
and variance $\sigma^2_z=1/(N_c-2)$.
Therefore, the distribution (\ref{eq:r})
is peaked at
\BE
r_{\max}=\rho=\f{\xi^2}{1+\xi^2},
\EE
where $\xi$ denotes the signal-to-noise ratio. 
Thus the large signal-to-noise ratio implies that $p(r)$ is peaked 
at near $r=1$ which makes it easy to detect the
signature of the non-trivial topology.
As for the distribution of $r$ for unmatched circles
we have to consider the spatial correlations of the
temperature fluctuations. 
However, for pairs of circles which are sufficiently 
separated each other, the spatial correlations can be negligible. 
Approximating the temperature fluctuation at each pixel as a 
random Gaussian number ($\rho=0$), 
the distribution for unmatched circles 
can be obtained from (\ref{eq:r})  
\BE
p_0(r)=\f{(1-r^2)^{(N_c-3)/2} \Gamma (N_c/2)}
{\sqrt{\pi}\Gamma ((N_c-1)/2)}.
\EE 
In this case, the variance of $r$ is approximately given by
\BE
\textrm{var}(r)=\f{\sqrt{\pi} \Gamma((N_c-1)/2)}{N_c \Gamma(N_c/2)}
\approx \f{\sqrt{\pi e}}{N_c}.
\EE
The distribution of $r$ for unmatched circles is peaked at zero
and the variance is inversely proportional to the number of pixels
$N_c$ on the circle. Thus an increase in $N_c$ will also make it 
easier to detect the imprint of the non-trivial topology.
\\
\indent
Let us now estimate the effect of the noise 
in actual experiments. First we consider the detectability by using
the COBE-DMR data for which the signal-to-ratio
is approximately $\xi\sim 2$ on angular scales $\Delta \theta\sim 10^o$. 
The number of pixels on a circle with angular radius $\alpha$
is given by $N_c(\alpha)=2 \pi \sin \alpha/\Delta \theta$.
Ignoring the galactic cut, for the COBE-DMR data with resolution 
$\Delta \theta=10^o$, 
$N_c(\alpha)\sim 36 \sin \alpha$ and the total number of pixels in the 
sky is $N_t\sim 4.1\times 10^2$. If one considers the galactic cut,
it reduces to $N_t\sim 3.0\times 10^2$. 
The total number of the circle with angular radius $\alpha$
is $N_t$ and the total number of pairs of
circles with angular radius $\alpha$ is 
$N_{pt}(\alpha)\sim N_c(\alpha)N_{t}(N_{t}-1)/2$. 
\BF[t]
\begin{flushleft}
\centerline{\psfig{figure=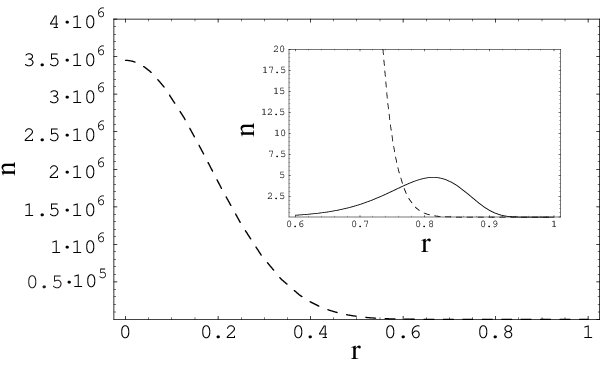,width=12.5cm}}
\caption{Detectability in COBE Data} 
\mycaption{Distribution $n$ for unmatched circle pairs (dashed curve)
and a matched circle pair (full curve) with angular radius 
$\alpha=70^o$  
as measured by the COBE-DMR (using pixels $(10^o)^2$). 
$n(r_0) dr$ gives the number of pairs of circles with
$r=r_0$   }
\end{flushleft}
\label{fig:CircleCOBE1}
\EF 
As shown in figure \ref{fig:CircleCOBE1}, one can see that 
the detection at most would be marginal for the COBE-DMR data. 
It should be noted that 
the above estimate is done under idealistic circumstances.
In reality, we should also consider physical effects
such as the ISW effect and the Doppler effects which might smear
the signal of the periodicity. For low matter density models, the   
ISW effect significantly reduces the signal-to-noise ratio 
(if we regard the ISW contribution as the ``noise'') on 
large angular scale $\sim 10^o$. 
\\
\indent
Next, we estimate the detectability by the future satellite
missions MAP(launched in 2001) and \ti{Planck}(launched in 2007). 
According to the 
technical information at \cite{MAP}, the MAP satellite will
measure the temperature fluctuations covering the full sky 
at five frequencies (22GHz, 30GHz, 40GHz, 60GHz and 90GHz)
with angular resolution (in FWHM of the Gaussian central beam) 
($0.93^o,0.68^o,0.53^o,0.35^o,<0.23^o$),respectively. 
The sensitivity at a pixel with angular scale $(0.3^o)^2$ 
is $\sim 35 \mu K$ for all the frequencies. If the Galactic
emission is negligible at high Galactic latitudes and frequency above 
40GHz then the three highest channels will give the sensitivity 
$\sim 20\mu K$ since the noise of the combined maps can be 
approximately given by $\eta=(\sum \eta^{-2}_i)^{-1/2}$ where
$ \eta_i$ denotes the noise for map $i$. On the angular scale 
$1^o$, corresponding to the acoustic peak, the expected sensitivity
will be $\sim 6\mu K$. Because the observed amplitude of the acoustic
peak at $l\sim 200$ is $60\mu K \sim 80
\mu K$ \cite{Boomerang,MAXIMA}, we have $\xi=10\sim 13$. 
\ti{Planck} will provide us sky maps with 
angular resolution $5$ to $10$ arcminutes\cite{Planck}. Certainly the 
signal-to-noise ratio will be much improved compared with the data 
supplied by MAP. Thus the prospect for the 
detectability is far brighter for the future satellite missions.
\\
\indent
However, one should be cautious about the physical effects
that will smear the clear periodical structure. If one uses
the filtered sky map on $1^o$ scale, the 'early' ISW effect owing to 
the decay of the gravitational potential at the radiation dominant epoch 
equality epoch and the Doppler effect owing to the 
dipole component of the temperature
fluctuations at the last scattering cannot be completely negligible
although they are subdominant.
Below the Jeans scale the photon pressure resists the
gravitational compression of the photon-baryon fluid leading to
driven acoustic oscillations. When the scale of the
fluctuations enters the sound horizon $r_s$ the gravitational potential 
starts to decay because of the resistance of the photon pressure to
the gravitational force leading to an enhancement in the amplitude
for the adiabatic modes. At the start of the oscillation, the
amplitude of the monopole and dipole increase with $R=3\rho_b/4\rho_\gamma$
(where $\rho_b$ and $\rho_\gamma$ denote the baryon density and the
photon density, respectively) owing to the reduction 
in pressure\cite{HS95}. 
As the gravitational potential starts to decay, the amplitude
of the monopole or the dipole decreases. However, the amplitude of
the dipole decreases rapidly compared to the monopole 
by a factor of $\sqrt{3}dr_s/d\eta=
(1+R)^{-1/2}$ where $\eta$ denotes the conformal time.
Let us estimate the difference in the amplitude. 
The radiation density is given by the present temperature $T$
normalized by $2.7K$ and the Hubble parameter $h$  
as $2.38\times10^{-5} h^{-2} T^4_{2.7}$\cite{Early}. Hence for 
$\Omega_b=0.06$, $h=0.75$ and $z(\textrm{recombination})=1300$,
the factor $(1+R)^{-1/2}$ is $\sim 0.8$. However, one should also consider 
the effect of the zero point shift of the oscillation for the
monopole owing to the gravity since the zero point shift 
increases the amplitude of the observed temperature
fluctuations. Note that the shift is \ti{absent} for the dipole
since it is approximately given by the time derivative of the 
monopole. 
Assuming that $R$ is constant in time, 
the zero point shift for the
temperature fluctuation $\Theta_0+\Psi$ is $-R\Psi$\cite{HueD}.
In reality, one should also consider the effect 
of the evolution of $R$. If one assumes that the zero point shift
is equal to the amplitude of the temperature fluctuation plus the 
gravitational potential at the last
scattering$(\Theta_0+\Psi)(\eta_*)\approx \Psi(\eta_*)/3$, then the 
relative amplitude of the monopole plus
the gravitational potential to the dipole (for idiabatic case)
is approximately equal to $\sim 2$. Thus the Doppler effect is not
completely negligible.
\BF[tpb]
\begin{flushleft}
\centerline{\psfig{figure=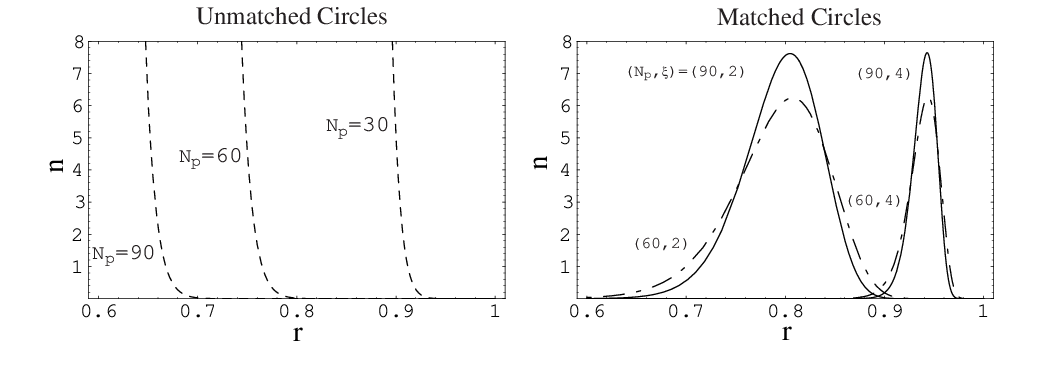,width=17cm}}
\caption{Detectability in MAP Data}
\mycaption{Distribution $n$ as measured by the MAP (using pixels
$(1^o)^2$) for unmatched circle pairs with 
the total number of pixels on the circle $N_p=30,60,90$ (dashed curve)
and a matched circle pair with $N_p=90$(full curve) and 
$N_p=60$(dashed-dotted curve). $\xi$ denotes the signal-to-noise ratio.
The pair of numbers in the figure in the right represent $N_p$ and 
$\xi$. $n(r_0) dr$ gives the number of pairs of circles with
$r=r_0$}
\label{fig:MAP}
\end{flushleft}
\EF 
\\
\indent
As shown in figure \ref{fig:MAP}, we need approximately 
60 pixels on the circle for $\xi=4$ whereas 90 pixels for
$\xi=2$ in order to detect one pair of matched circles 
by using the expected sky map with resolution $(1^o)^2$ supplied 
by MAP. The decrease in the signal-to-noise ratio
implies that it is necessary to have a pair of matched circles 
with large angular radius. Thus if the observed peak 
in the angular power is actually relevant  
to the acoustic oscillations, then it is possible to  
detect the non-trivial topology provided that the injectivity
radius at the observation point is smaller than the 
radius of the last scattering and the angular radius of the 
circles are sufficiently large. 
\section{Non-Gaussian signature}
Let us write the temperature fluctuations
in the sky in terms of real spherical harmonics $Q_{lm}$ and real
expansion coefficients $b_{lm}$ as
\BE
\f{\Delta T}{T}=\sum_{lm} b_{lm} Q_{lm}.
\EE
If one assumes the spatial homogeneity and
isotropy on the background geometry then the temperature fluctuations
in the sky form an isotropic Gaussian random field, in other words,
the distribution of the expansion coefficients $b_{lm}$ is given by 
\BE
f(b_{lm})=\f{1}{\sqrt{2 \pi C_l}}\exp(-b^2_{lm}/2 C_l). 
\EE
\\
\indent
However, in locally isotropic and homogeneous models which are
spatially multiply connected, Gaussian temperature fluctuations 
for a particular choice of position and orientation of teh observer 
are no longer isotropic since the spatial hypersurface 
is not globally isotropic. 
In other words, the variance of $b_{lm}$s may depend on $m$ or
$b_{lm}$s may not be independent each other for a given $l$.  
Marginalizing the distribution over the position and the
orientation the temperature fluctuations 
becomes non-Gaussian. Assuming that the initial fluctuations are
Gaussian, then the distribution has vanishing skewness
but non-vanishing kurtosis since $a_{lm}$ is 
written as a sum of a product of 
a Gaussian variable times an expansion coefficient of an eigenmode
which can be considered as a ``random'' variable. 
Non-Gaussianity owing to the break of the global isotropy and homogeneity
in the background geometry can be a prominent signature of the spatial 
non-trivial topology.
\\
\indent
In order to test for Gaussianity, it is necessary to study the
ensemble average of higher-order correlations 
($N$-point correlations $N\!>\!2$). Assuming vanishing
average, a Gaussian distribution can be defined as one 
for which the odd correlations all vanish but the even correlations
all factorize into products of 2-point correlations.
\\
\indent
As we have seen, a temperature fluctuation
in locally homogeneous and isotropic background geometry
is written as a sum of a product of two factors each one of which 
corresponds to the initial condition and the geometric property, 
respectively. For CH models, the both factors can be described 
by random Gaussian numbers.   
Let us calculate the distribution function $F(Z,\sigma_Z)$ 
of a product of two independent random numbers $X$ and $Y$ 
that obey the Gaussian (normal) distributions 
$N(X;0,\sigma_X)$ and $N(Y;0,\sigma_Y)$, respectively,
\BE
N(X;\mu,\sigma)\equiv \f{1}{\sqrt{2 \pi} \sigma}
\textrm{e}^{-(X-\mu)^2/2\sigma^2}.
\EE
Then $F(Z\!=\!XY,\sigma_Z)$ is readily given by
\BEA
F(Z,\sigma_Z)&=&2\int_0^\infty N(Z/Y,0,\sigma_X)N(Y,0,\sigma_Y)
\f{dY}{Y}
\nonumber
\\
&=& \f{1}{\pi \sigma_X \sigma_Y}K_{0}\Bigl ( \f{|Z|}{\sigma_X \sigma_Y}
\Bigr ),
\EEA
where $K_{0}(z)$ is the modified Bessel function. The average of $Z$
is zero and the standard deviation satisfies 
$\sigma_Z=\sigma_X \sigma_Y$. As is well known,
$K_{0}(z)$ is the Green function of the diffusion equation with 
sources distributed along an infinite line.  Although $K_{0}(z)$ is 
diverged at $z\!=\!0$ its integration over 
$(-\infty,\infty)$ is convergent. 
\begin{figure}[tpb]
\centerline{\psfig{figure=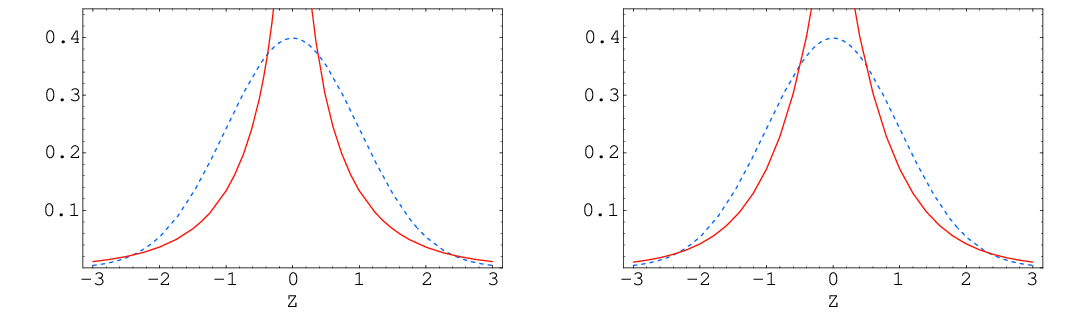,width=15.5cm}}
\caption{Non-Gaussian Distribution}
\mycaption{On the left, the distribution function 
$F(Z,1)$ 
for a product of two random Gaussian numbers is plotted in solid
curves. On the right, the distribution function 
$G(Z,1)$ (1$\sigma\!=\!1$) 
of a sum of two random variables that obey  
$F(Z,1/\sqrt{2})$. The dashed curves represent the
Gaussian distribution $N(Z;0,1)$.
}
\label{fig:GtimesG}
\end{figure}
From the asymptotic expansion of the modified
Bessel function
\BE
K_0(z)\sim \sqrt{\f{\pi}{2 z}}\textrm{e}^{-z}\Biggl[1-\f{1^2}{1!8z}
+\f{1^2\cdot 3^2}{2!(8z)^2}-\f{1^2\cdot 3^2\cdot 5^2}{3!(8z)^3}+\ldots
\Biggr],~~~ z>>1,
\EE
one obtains in the lowest order approximation, 
\BE
F(Z,\sigma)\sim \f{1}{\sqrt{2 \pi \sigma
|Z|}}\textrm{e}^{-|Z|/\sigma},~~~Z>>1.
\EE 
Thus $F(Z,\sigma)$ is slowly decreased in $Z$ 
than the Gaussian distribution 
function with the same variance in the large limit
while it has a prominent peak around zero.
The skewness is zero but the kurtosis $\sigma_4$ 
is positive
\BE
\sigma_4=\int_{-\infty}^\infty Z^4 F(Z,1) dZ -3=6.
\EE
For fluctuations on large angular scales 
only the eigenmodes with large wavelength 
($\!\equiv\!2 \pi/k$) can contribute to the sum. Owing to the finiteness of 
the space, the number of eigenmodes which dominantly contribute to 
the sum is finite. Therefore, the fluctuations become
distinctively non-Gaussian. On small angular scales 
the number of eigenmodes that contribute to the
sum becomes large and the distribution function converges to
the Gaussian one as the central limit theorem implies.
One can see from figure \ref{fig:GtimesG} (right) that the distribution 
function $G(W,1)$ of $W\!=\!Z1+Z2$ where both $Z1$ and $Z2$ 
obey $F(Z,\sqrt{2})$ is much similar to the Gaussian distribution 
$N(Z,0,1)$ than $F(W,1)$ although divergence at $Z\!=\!0$
still persists.    
\subsection{Cosmic variance}
In order to test for Gaussianity a simple approach 
is to measure the variance of the 
power spectrum ${\hat C}_l$ which contains the information of
4-point correlations. 
\begin{figure}[tpb]
\centerline{\psfig{figure=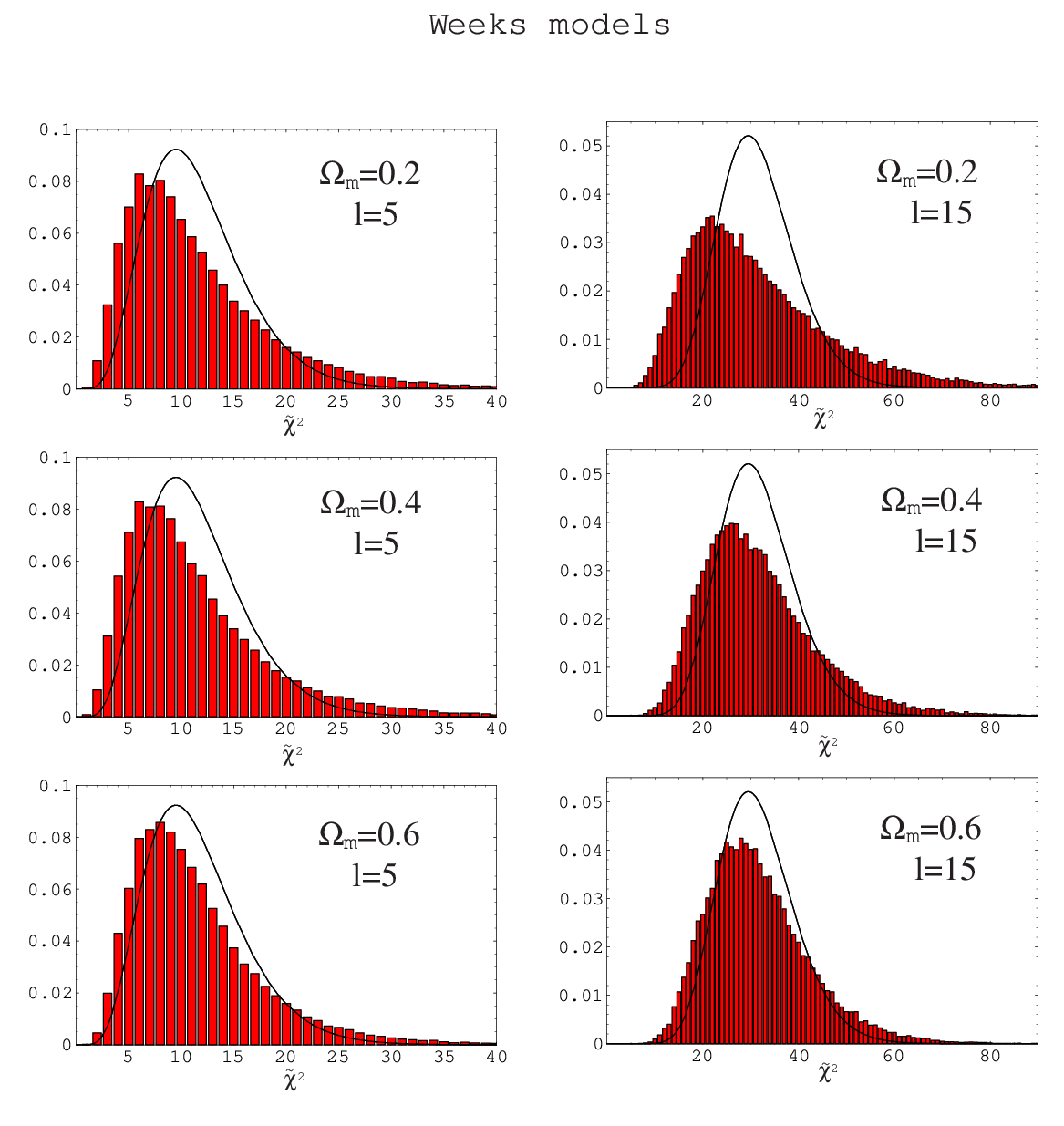,width=16cm}}
\caption{Deviation from $\chi^2$ Distribution}
\mycaption{Plots of the distributions of 
$\tilde{\chi}^2\! \equiv\!
(2l+1)\hat{C_l}/C_l$ for the Weeks models in comparison
with the $\chi^2$ distributions with $2l\!+\!1$ degree of 
freedom (solid curves).
The distributions are calculated using the eigenvalues
obtained by the DBEM and the Gaussian approximation 
for the expansion coefficients. The Monte-Carlo
simulations are carried out 
based on 200 realizations of the 
initial Gaussian fluctuations $\Phi_{\nu}(0)$,
and 200 realizations of the base points.}
\label{fig:CHISQW}
\end{figure}
If the expansion coefficients $b_{lm}$ of the 
temperature fluctuation in the sky are Gaussian, 
$\tilde{\chi}^2\! \equiv\! (2l+1)\hat {C_l}/C_l$
must obey the $\chi^2$ distribution with $2l\!+\!1$ degrees of freedom. 
In the case of CH models with small volume,
the distribution of $b_{lm}$ for a homogeneous and isotropic
ensemble has vanishing skewness but positive kurtosis. The 
non-Gaussian features in the distribution of 
$\tilde{\chi}^2$ appears as the slight shift of the 
peak of the distribution to the center(zero) and
the slow convergence to zero for large amplitudes 
(figure \ref{fig:CHISQW}).
\begin{figure}[tpb]
\centerline{\psfig{figure=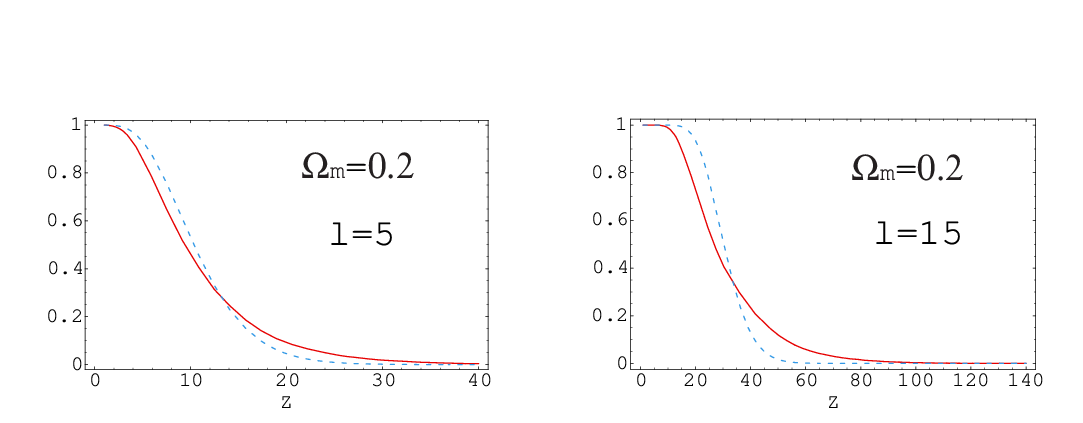,width=15.5cm}}
\caption{Cumulative Distribution}
\mycaption{Plots of $1-P(Z)$ ($P(Z)$ is the cumulative 
distribution function) which gives the probability of observing
$X\ge Z$.  The solid curves correspond to  
$1-P(\tilde{\chi}^2)$ for the
Weeks model $\Omega_m\!=\!0.2$, $l\!=\!5$ (left) and $l\!=\!15$ (right). 
The dashed curves correspond to $1-P(\chi^2)$ of the Gaussian model.} 
\label{fig:Prob0.2W}
\end{figure}
\begin{figure}[tpb]
\centerline{\psfig{figure=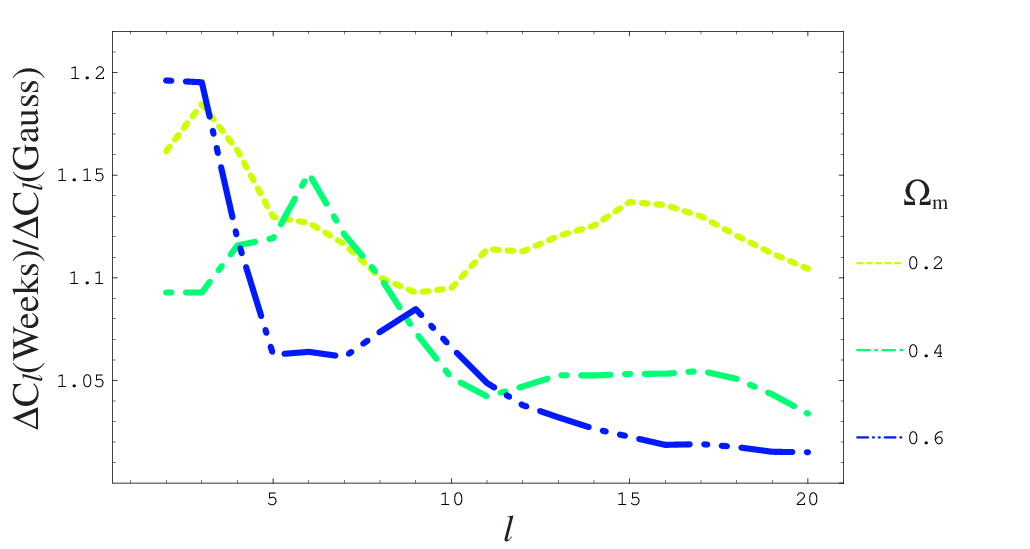,width=15.5cm}}
\caption{Cosmic Variance}
\mycaption{Excess cosmic variances owing to the global inhomogeneity
in the background geometry.
The ratios of the standard 
deviation in the angular power spectrum $\Delta C_l$  
for the Weeks models relative to that for the Gaussian
model $\Delta C_l(Gauss)$ are plotted. In order to compute 
$\Delta C_l$ Monte-Carlo simulations based on 200 realizations
of the initial perturbation $\Phi_\nu(0)$ and 200 realizations of 
the position of the observer have been carried out.}
\label{fig:CosmicVarW}
\end{figure} 
\\
\indent
Roughly speaking the cosmic variance consists of the ``initial'' variance
owing to the uncertainty in the initial conditions  
and the ``geometric'' variance owing to the uncertainty
in the location and the orientation of the observer.
The ``initial'' variance is equal to the ``standard'' 
cosmic variance $\Delta C_l/C_l\!=\!\sqrt{2/2l+1}$ while
the additional ``geometric'' variance is smaller than 
the ``initial'' variance. The effect of the ``geometric''
variance is prominent for low density models since 
the relative size of the spatial hypersurface to the 
present observable region becomes small (figure \ref{fig:CosmicVarW}).
Enhancement in the large-angle initial power  (``blue-shifted
power'') also leads to a large ``geometric'' variance since
the effective degree of freedom of the relevant fluctuations 
is decreased. 
\subsection{Topological measurements}
Topological measures:total area of the excursion regions, total length 
and the genus of the isotemperature contours 
have been used for testing Gaussianity 
of the temperature fluctuations in the COBE DMR
data\cite{Colley,Kogut}. In integral geometry,
they are known as the Minkowski functionals
which characterize the morphological property of the
fluctuations that are additive and invariant
under translations and rotation.
Let us first summarize the known results for 
Gaussian fields (see \cite{Gott,Adler}).
\\
\indent
The genus $G$ of the excursion set for
a random temperature field on a connected and simply-connected 2-surface 
can be loosely defined as 
\BEA
G&=&\textrm{number of isolated high-temperature connected regions}
\\
\nonumber
&-& \textrm{number of isolated low-temperature connected regions}. 
\label{eq:G}
\EEA
For instance, for a certain threshold, a hot spot will contribute $+1$
and a cold spot will contribute $-1$ to the genus. If a hot spot
contains a cold spot, the total contribution to the genus is zero.
\\
\indent
The genus can be represented as the integration of the local properties
of the field which can be easily simulated by a computer.
From the Gauss-Bonnet theorem, the genus of a closed curve $C$
being the boundary of a simply-connected region $\Omega_C$ 
which consists of $N$ arcs with exterior angles 
$\alpha_1,\alpha_2,...\alpha_N$ can be written in terms of
the geodesic curvature $\kappa_g$ and the Gaussian curvature $K$ as 
\BE
G=\f{1}{2 \pi}\Biggl[ \int_C \kappa_g ds+\sum_{i=1}^N \alpha_i+ \int_{\Omega_C}
K dA \Biggr ]. \label{eq:GB}
\EE
For a random field on the 2-dimensional Euclidean space $E^2$
where the N arcs are all geodesic segments (straight line segments),
$K$ and $\kappa_g$ vanish. Therefore, the genus is written as 
\BE
G_{E^2}=\f{1}{2 \pi}\sum_{i=1}^N \alpha_i \Bigr. \label{eq:F}
\EE
\\
The above formula is applicable to the locally flat spaces
such as $E^1\times S^1$ and $T^2$ which have $E^2$ as the 
universal covering space since 
$K$ and $\kappa_g$ also vanish in these spaces. In these 
multiply connected spaces, the naive 
definition Eq.(\ref{eq:G}) is not correct for
excursion regions surrounded by a loop which cannot be contracted 
to a point. 
\\
\indent 
In order to compute the genus for a random field on a sphere $S^2$ 
with radius equal to 1, it is convenient to use 
a map $\psi$:$S^2-\{p_1\}-\{p_2\}\!\rightarrow\! S^1\times (0,\pi)$
defined as 
\BE
\psi:(\sin\theta \cos\phi,\sin\theta \sin\phi,\cos\theta)
\rightarrow (\phi,\theta),~~0\le\phi<2\pi,0<\theta<\pi,
\EE
where $p_1$ and $p_2$ denote the north pole and the south pole, respectively. 
Because $S^1\times (0,\pi)$ can be considered as locally flat spaces
($\phi,\theta$) with metric $ds^2=d\theta^2+d\phi^2$ which have 
boundaries $\theta\!=\!0,\pi$,
the genus for excursion regions that 
do not contain the poles surrounded by straight segments
in the locally flat ($\phi,\theta$) space is given
by Eq.(\ref{eq:F}). It should be noted that the straight segments do not 
necessarily correspond to the geodesic segments in $S^2$. 
If a pole is inside an excursion region 
and the pole temperature is above the threshold then the genus is 
increased by one. If the pole temperature is below the threshold, it
does not need any correction. Thus the genus for the excursions is 
\BE
G_{S^2}=\f{1}{2 \pi}\sum_i \alpha_i+N_p, \label{eq:Gs}
\EE
where $\alpha_i$ is the exterior angles at the intersection of two 
straight segments in the ($\phi,\theta$) space and $N_p$ is the number
of poles above the threshold. 
\\
\indent
Now consider an isotropic and homogeneous 
Gaussian random temperature field on a sphere $S^2$
with radius 1. Let $(x,y)$ be the local Cartesian coordinates on $S^2$ and 
let the temperature correlation function be
$C(r)\!= \langle (\Delta T/T)_0(\Delta T/T)_r\! \rangle \!$ with $r^2\!=\!x^2+y^2$
and $C_0\!=\!C(0)\equiv \sigma^2$, where $\sigma$ is the standard
deviation and $C_2\!=\!-(d^2C/dr^2)_{r=0}$. Then the expectation
value of the genus for a threshold $\Delta T /T=\nu\sigma$ is given 
as \cite{Adler}
\BE
\angle G_{S^2} \rangle =\sqrt{\f{2}{\pi}}\f{C_2}{C_0}\nu\textrm{e}^{-\nu^2/2}+\textrm{erfc}
\Biggl(\f{\nu}{\sqrt{2}}\Biggr ), \label{eq:aveGs}
\EE 
where erfc($x$) is the complementary error function. The first term in 
Eq.(\ref{eq:aveGs}) is equal to the averaged contribution for the excursions
which do not contain the poles while the second term in Eq.(\ref{eq:aveGs})
is the expectation value of $N_p$.  
\\
\indent
The mean contour length per unit area for an isotropic homogeneous 
Gaussian random field is \cite{Gott,Adler}
\BE
\langle s \rangle =\f{1}{2}\Biggl ( \f{C_2}{C_0}\Biggr)^{\f{1}{2}}\textrm{e}^{-\nu^2/2},
\label{eq:LG}
\EE
and the mean fractional area of excursion regions for the field is
the cumulative probability of a threshold level,
\BE
\langle a \rangle =\f{1}{2}\textrm{erfc}\Biggl(\f{\nu}{\sqrt{2}}\Biggr),
\EE
which gives the second term in Eq.(\ref{eq:aveGs}).
\\
\indent
\begin{figure}[tpb]
\centerline{\psfig{figure=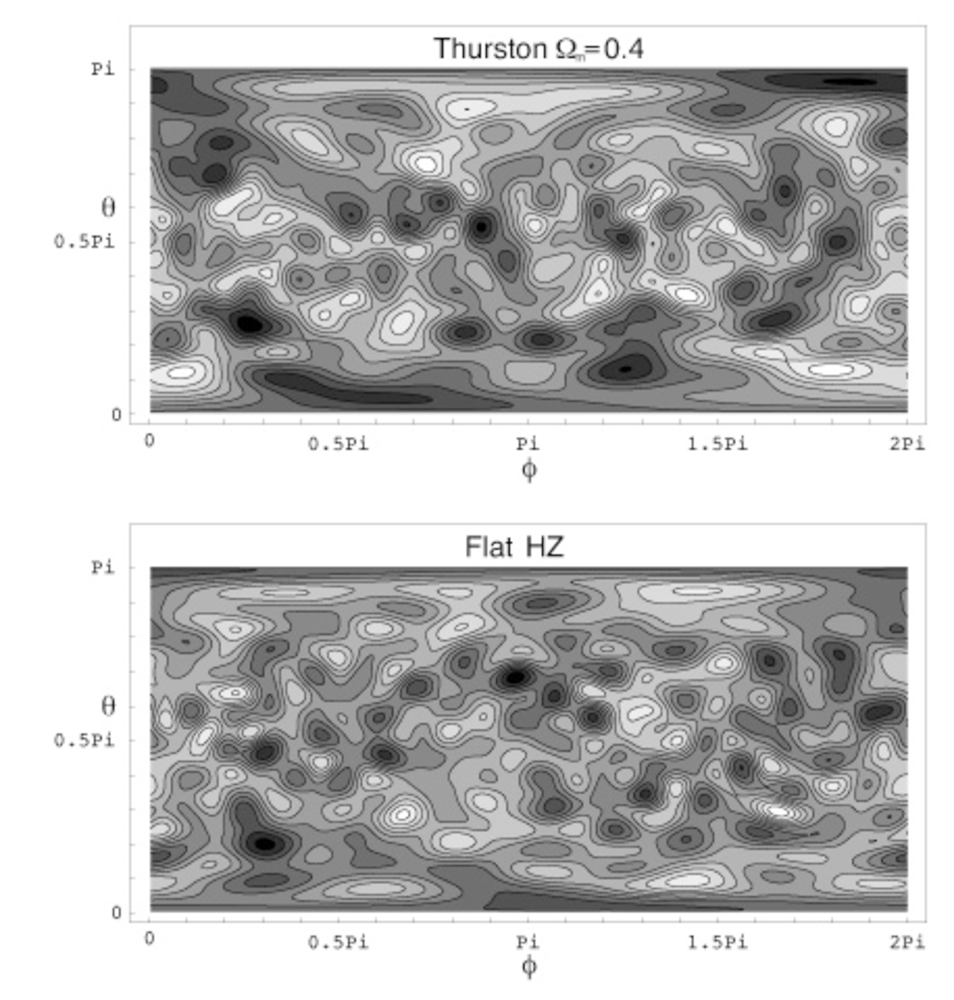,width=16cm}}
\caption{Simulated CMB Sky Map}
\mycaption{Contour maps of the 
CMB (not smoothed by the DMR beam) for the Thurston model 
$\Omega_m\!=\!0.4$ and 
a flat (Einstein-de-Sitter) 
Harrison-Zel'dovich model $C_l \propto
1/(l(l+1))$ in which all multipoles $l\!>\!20$
are removed.}
\label{fig:Sky}
\end{figure}
The CMB anisotropy maps for the Weeks and the Thurston 
(adiabatic) models are produced by using eigenmodes $k<13$ and angular 
components $2\!\leq\!l\!\leq\!20$ for $\Omega_0\!=\!0.2$ and $0.4$. The
contribution of higher modes are approximately 7 percent and 10
percent for $\Omega_0\!=\!0.2$ and $0.4$, respectively. The initial
power spectrum is assumed to be the extended Harrison-Zel'dovich 
one. The beam-smoothing effect is not included. 
For comparison, sky maps for the Einstein-de-Sitter model with
the Harrison-Zel'dovich spectrum $C_l\propto1/(l(l+1))$ are also
simulated. 
\\
\indent
In order to compute the genus and the contour length for each model,
10000 CMB sky maps on a 400$\times$200 grid in the $(\phi,\theta)$
space are produced. 
The contours are approximated by oriented straight
segments. The genus comes from the sum of the exterior angles at the
vertices of the contours and the number of poles at which the
temperature is above the threshold.
The total contour length is approximated by 
the sum of all the straight segments.  
Typical realizations of the sky map are shown in figure \ref{fig:Sky}.
\\
\indent
One can see in figure \ref{fig:LGW02} that the mean 
genera and the mean total
contours for the two CH models 
are well approximated by
the theoretical values for the Gaussian models. This is a natural
result since the distribution of the expansion coefficients $b_{lm}$ 
is very similar to
the Gaussian distribution in the modest threshold levels.
On the other hand, at high and low threshold levels,
the variances of the total contour lengths
and the genera are much larger than that for the Gaussian models,
which can be attributed to the positive skewness in the distribution
function of $b_{lm}$.  
The excess variances for 
the Weeks model with $\Omega_0\!=\!0.2,0.4$ compared with the Gaussian flat
Harrison-Zel'dovich model are observed at the absolute 
threshold level approximately $|\nu|\!>\!1.4$ for genus and
$|\nu|\!>\!0.7$ for total contour length, respectively. The reason why
the non-Gaussian signature appeared in the variance of the 
topological measures rather than the means is explained as follows:
Suppose that the temperature map consists of hexagonal pixels whose
distance between the center of any two adjacent pixel is $\Delta$. 
From the additivity of Minkowski functionals, one can show that 
they can be decomposed into the components
for each pixel and the intersections of adjacent pixels
which allows the explicit calculation of means and variances of them
\cite{Winitzki}. For instance, the mean of the total
length per unit area for the map is given by
\BE
\langle s \rangle =\f{4}{\Delta}(P_1(\nu)-P_2(\nu;\Delta))
\label{eq:L}
\EE
where
\BE
P_n(\nu,r_{ij})\equiv \int_\nu^\infty dx_1 \dots 
\int_\nu^\infty dx_n~ p_n(x_1, \dots, x_n;r_{ij})
\EE
that describes the probability of $\Delta T/T>\nu$ at points 
$(x1, \dots, x_n)$ with distance $r_{ij}$ for each pair 
$(x_i,x_j)$ and $p_n(x1, \dots, x_n;r_{ij})$ is the 
$n$-point distribution function in real space. For ``smooth'' random fields,
the limit of (\ref{eq:L}) as $\Delta \rightarrow 0$ gives the   
mean value. In fact, for isotropic random Gaussian field,
one can show that the limit of (\ref{eq:L}) gives (\ref{eq:LG}). 
The variance of $s$ is written as
\BEA
\textrm{Var}(s)&=&\lim_{\Delta \rightarrow 0}
\Bigl (\f{4}{\Delta N} \Bigr )^2 \sum_{s_1,s_2}[
P_2(\nu;r_{12})-2P_3(\nu;\Delta,r_{12})+P_4(\nu;\Delta,r_{12})]
\nonumber
\\
&&-\Bigl (\f{4}{\Delta} \Bigr )^2 [P_1(\nu)-P_2(\nu;\Delta)]^2, 
\EEA
where $N$ is the total number of pixels and $s_i$ denotes 
a pixel and $r_{12} \equiv |s_2-s_1|$. $P_3(\nu;\Delta,r_{12})$
represents the distribution of $\Delta T/T\!>\!\nu$
for 3 points configured as equilateral triangles 
with sides $r_{12}$ and $\Delta$. Similarly, for genus 
one can show that the mean and the variance are written in 
terms of $n$-point distributions 
$n\le 3$ and $n\le 6$, respectively.  
Thus the variance has information of much higher-order correlations.
It is concluded that the non-Gaussian signature is
relevant to the higher-order correlations ($n\!>\!3$) 
in real space. 
\\
\indent
As is well known,  the COBE DMR data excludes
grossly non-Gaussian models \cite{Colley,Kogut}.
However, one should take account of a fact that the signals in the
$10^o$ smoothed COBE DMR 4-year sky maps are comparable to the noises
\cite{Bennett}. This makes it hard to detect the slight non-Gaussian signals
in the background fluctuations.
Because the mean behavior of the topological measures in CH models 
is well approximated by those of Gaussian models, the COBE 
constraints on CH models cannot be so stringent.
In fact, some recent works 
using different statistical tools have shown that the COBE DMR 4-year
sky maps are non-Gaussian \cite{Ferreira,Novikov,Pando} although 
some authors cast doubts upon the cosmological
origin of the observed non-Gaussian signals \cite{Bromley,Banday00}. 
If any non-Gaussian signals in the CMB
on large angular scales are
detected by the future satellite missions, then  
the non-trivial topology of the spatial geometry
surely gives a natural explanation of the origin.
\begin{figure}[b]
\centerline{\psfig{figure=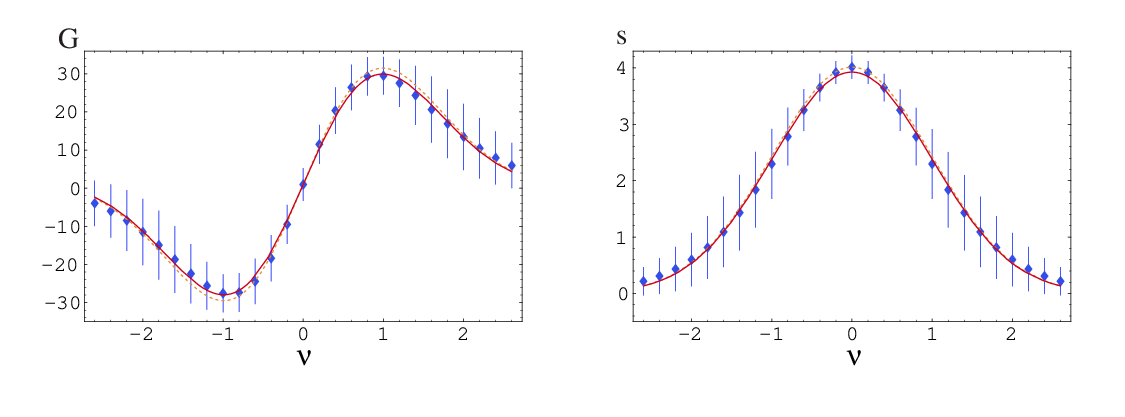,width=16.5cm}}
\caption{Minkowski Functionals}
\mycaption{The means of genus and total length 
of isotemperature contours averaged over 100 realizations of the
initial fluctuations and 100 realizations of the base points
and $\pm 1 \sigma$ run-to-run variations at 27 threshold levels 
for the Weeks model with $\Omega_m\!=\!0.2$.
The dashed curves denote the mean values for a corresponding 
Gaussian model. The solid 
curves denote the mean values for a Gaussian model that are 
best-fitted to the average values.}  
\label{fig:LGW02}
\end{figure}

\begin{figure}
\centerline{\psfig{figure=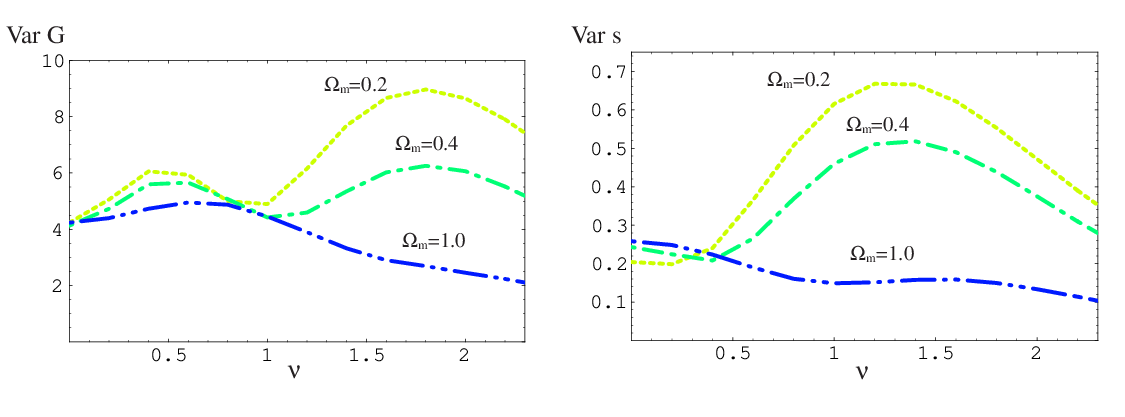,width=16cm}}
\caption{Non-Gaussian Signature in Variance}
\mycaption{The variances 
of genus and total length of isotemperature contours
for the Weeks models (dashed and dashed-dotted curves) for a homogeneous
and isotropic ensemble in comparison 
with those for the infinite flat model $\Omega_m\!=\!1.0$
(dashed-dotted-dotted curve) with Gaussian initial fluctuations. 
At large threshold level $\nu$, the deviation from the Gaussian model
is conspicuous. }  
\label{fig:VarGLWF}
\end{figure}


\chapter{Summary}
\thispagestyle{headings}
In this thesis, we have investigated the CMB anisotropy in 
closed flat and closed hyperbolic models with 
multiply connected spatial hypersurface and obtained 
constraints on these models using the COBE-DMR data and discussed
various observable signatures of non-trivial topology in the CMB
which can be tested by using more accurate data which will be supplied by 
the future satellite missions.
\\
\indent
In chapter 2 mathematical property of closed 3-manifolds 
with constant curvature $K$ of three types, namely flat($K\!=\!0$), 
spherical($K\!>\!0$) and hyperbolic($K\!<\!0$) geometry has been 
reviewed.
\\
\indent
In chapter 3 
we have formulated two types of numerical methods, namely 
the direct boundary element method (DBEM) and the periodic
orbit sum method (POSM) for computing mode functions 
of the Laplace-Beltrami operator for closed hyperbolic(CH) manifolds
and have investigated the statistical property of mode functions
and the relation between the low-lying eigenvalues and various 
diffeomorphism-invariant geometric quantities.
The numerical accuracy of the DBEM is much better than that of
the POSM but it needs mesh generation for each model. On the other
hand the POSM is much suited for analyzing a large samples of
spaces (manifolds and orbifolds) which needs only the fundamental group 
of the space. 
\\
\indent
In section 3.1 we have seen that the low-lying eigenmodes
that are continued onto the universal covering space 
of the two smallest CH manifolds are well described by a set of
random Gaussian numbers. Finding such property for the low-lying 
modes is rather surprising since they are relevant to purely quantum 
nature. The description breaks only for cases where the symmetries
of the mode functions which are equivalent to the symmetries of the
manifold and perhaps of the finite-sheeted cover of the manifold 
are apparent. The origin of the pseudo random Gaussianity of the mode
functions has not yet been well understood. The ``random'' property of a
large sphere in the universal covering space may give a clue
for finding the origin.
We have conjectured that the pseudo
random Gaussian behavior is universal for all CH spaces
which can be used for simulating the CMB anisotropy 
in $(l,m)$ spaces.
\\
\indent
In section 3.2, we have numerically analyzed the length spectra
and the eigenvalue spectra of CH manifolds with small volume.
First, the asymptotic behavior in the classical staircase 
is found to be consistent with the known analytical 
formula which does not depend on the topology
or symmetry of the manifold although the arithmeticity
of the manifolds makes a difference for behavior in the multiplicity
number. Next, we have applied the trace formula to these manifolds
and obtained consistency with those by the DBEM.
No supercurvature mode are found for a total of 308 CH 3-manifolds 
(volume $<3$). We need a further investigation to check
whether small manifolds that are sufficiently 
similar to cusped manifolds can support supercurvature modes. 
Thirdly,  the first eigenvalues are compared to diameter, volume
and the shortest length of the periodic orbits. The 
numerical results imply the existence of much shaper bounds
for the first eigenvalues in terms of diameter.  
Some fitting formulae have been introduced and
their validity has been checked. 
CH 3-manifolds can be roughly divided into two categories:
``slightly anisotropic'' and ``almost anisotropic'' ones.
The former has not any very short periodic orbits while the 
latter has. For example, manifolds which are very similar 
to the original cusped manifold are belonging to the latter
category. It is found that the  
deviation of the spectrum from the Weyl asymptotic formula 
for these manifolds is conspicuous even for manifolds 
with small volume.
Finally, the global ``anisotropy'' in the spatial 
geometry has been measured by
$\zeta$-function and the spectral distance
for 3 examples of CH 3-manifolds.  
It is found that the angular power spectra in the CMB 
are greatly affected by the globally anisotropic structure 
in the spatial geometry.  
\\
\indent
In chapter 4, we have explored the current constraints on the 
closed flat and hyperbolic models using the COBE-DMR data. 
In section 4.1 and 4.2, we have derived the Sachs-Wolfe
formula which describes the CMB anisotropy 
based on the linear perturbation theory in a gauge-invariant way. 
In section 4.3, we have investigated the effect of the 
non-trivial topology on 
the angular power spectra. 
Assuming adiabatic initial perturbation with scale-invariant 
spectrum ($n=1$), a prominent suppression occurs
for a ``standard'' closed flat toroidal model with $\Omega_m=1.0$ since
fluctuations beyond the size of the cube at the last scattering
are strongly suppressed. However, for low matter density models
such a strong suppression does not occur since one cannot 
ignore the contribution from the (late) ISW effect caused 
by the decay of the gravitational potential at the 
$\Lambda$ or curvature dominant epoch. The crucial point is that 
they are produced at \ti{late} time well after the last scattering.
If a fluctuation on scale smaller than the fundamental domain 
is produced at a point sufficiently nearer to us,
then the corresponding angular scale 
becomes large provided that the background geometry 
is flat or hyperbolic. Therefore, the large angular fluctuations 
produced late time ``survive'' even if the size of the cell is finite.
On the other hand, slight suppression in 
large-angle temperature correlations in such models explains rather
naturally the observed anomalously low quadrupole which is
incompatible with the prediction of the ``standard''
FRWL models. 
In section 4.4, we have carried out the Bayesian analyses using the
COBE-DMR 4year data. 
In the case of flat topology, the discrete eigenmodes
have ``regular'' features which leads to significant 
correlations in $a_{lm}$'s which describes the fluctuations in the $(l,m)$
space. Even if marginalized over
the orientation, the correlations do not completely disappear.
The analysis using only the angular power spectrum 
is not enough since the background geometry is globally
inhomogeneous or anisotropic. However, even including the 
non-diagonal elements in the correlations the constraint 
is still less stringent for low matter density models.
This is because the physical size of the relevant fluctuations 
which are produced at late time are much smaller than the 
size of the fundamental domain.
In the case of hyperbolic topology in which
the discrete eigenmodes are ``chaotic''
the correlations in  $a_{lm}$'s for a given $l$ almost disappear
if one takes an average over the position of the observer.
As we have seen, the likelihood of CH models 
is very sensitive on the position of the observer. 
Although, for a large number of choices the 
fits to the COBE data are bad, in some places the fits are
\ti{much better} than those of the FRWL models. Because 
all CH manifolds are globally inhomogeneous\cite{Wolf67},
in order to constrain the models, 
one must compare the expected sky map at \ti{every place and every 
orientation} to the data. Even after marginalized over the 
position and orientation of the observer, the likelihoods are 
still comparable to that of the infinite counterparts. 
It is natural that the previous analyses on CH models \cite{Bond1,Bond2} 
have given stringent constraints since only fluctuations 
at a point where the injectivity radius is maximal with only 24
orientations were compared to the COBE data. 
Nature wouldn't choose such a special point where the 
symmetry of the fluctuations is maximal as our place 
in the universe. 
\\
\indent
In the last chapter, we have investigated the observable
signatures of the spatial non-trivial topology, namely the periodical
structure and the non-Gaussianity in the CMB. 
In section 5.1 we have re-examined the method of finding matched circles. 
We have crudely estimated the detectability
of the periodical patterns in the CMB sky map assuming that 
the signal and noise obeys the Gaussian statistics as in \cite{CSS98a}
but using slightly different statistics. 
The detectability depends
on the signal-to-noise ratio and the angular resolution of the 
sky map as well as the number of
circles and their angular radius. We have also considered the effect from
the ``background noise'' owing to the Doppler 
effect on the scale of acoustic oscillations which might smear the 
perfect identical patterns. 
The chance of detecting the non-trivial 
topology using the COBE data has found to be very low but we can
expect a good chance of the first discovery of the imprint of the
non-trivial topology using CMB maps with much better 
signal-to-noise ratio and the angular resolution which will be 
supplied by the 
future satellite missions, namely  $MAP$ and $Planck$. 
In section 5.2 we examined the non-Gaussianity of the CMB temperature
fluctuations. Inhomogeneous and anisotropic 
Gaussian fluctuations for a particular choice 
of position and orientation are regarded as non-Gaussian fluctuations
for a homogeneous and isotropic
ensemble. For CH models, the distribution of the expansion
coefficients $b_{lm}$ of the 2-dimensional temperature fluctuations  
has positive kurtosis but vanishing skewness provided that 
the initial fluctuations are Gaussian. The non-Gaussianity leads
to a slightly larger cosmic variance where the possibility of
having a rare fluctuation is high. The non-Gaussianity is
much prominent if one considers the statistics of Minkowski
functionals,  namely the total area, the total length and the genus 
of isotemperature contours. The non-Gaussian signature would appear
as large variances of statistics at high and low threshold levels
on large angular fluctuations.
If such a signature should be found in the forthcoming CMB sky maps,
then it would certainly be a strong sign of the non-trivial topology.
Analyzing such statistics for other closed multiply
connected models with low matter density, namely, flat and spherical
models with non-trivial topology will be also interesting issues. 
Even if we failed to detect the identical patterns, 
the imprint of the non-trivial topology -the effect of the
''finiteness''- could be still observable by measuring such statistics.
\\
\indent
The determination of the global
topology of the universe is one of the key issues of the
modern observational cosmology. The observation of the cosmic
microwave background, the distribution of clusters and QSOs
and other astronomical objects in the distant places provides us precise  
information about the global topology of the universe
as well as other cosmological parameters, such as matter 
contents, primordial spectrum, 
Hubble parameters and so on. In the first decade of the 21st century, 
we will be able to answer the fundamental and the simple question 
whether the universe is finite or ``sufficiently big''.
In the former case, it surely gives a great leap in our 
comprehension of the universe.
Even in the latter case, our
quest for searching the global topology will continue until
we obtain a well established unified theory which integrates 
the microscopic and the macroscopic worlds. 
\\
\\
\\
\centerline{\bf Acknowledgments}
\indent
First I would like to thank my advisor Kenji Tomita for his helpful
suggestions, support and continuous encouragement and  
Jeffrey R. Weeks for devoting much time and effort to 
share with me his expertise in topology and geometry of 
3-manifold.  I would also like to thank Naoshi Sugiyama, Takeshi Chiba, 
A.J. Banday for sharing their knowledge on the CMB and Ralph Aurich, 
Neil J. Cornish, Boudewijin F. Roukema and David N. Spergel 
for their helpful comments and suggestions.
Finally I would like to acknowledge stimulating discussions with 
colleagues in YITP and Department of Physics in Kyoto University.


\begin{appendix}
\chapter{Boundary integral equation} 
\thispagestyle{headings}
Here we derive the boundary integral equation (\ref{eq:bem0}) in
section 3.1. For simplicity, we prove the formula in 3-spaces. 
\\
\indent
First, we start with Eq.(\ref{eq:inter}) with dimesnsion $M=3$.
Although the integrand in Eq.(\ref{eq:inter}) is divergent at 
$\x=\y\in \del\Omega$, the integration can be regularized as follows. 
Let us draw a sphere with center $\y\in \del\Omega$ with
small radius $\epsilon$ and let $\Gamma_\epsilon$ be the outer 
spherical boundary and $\alpha$ and $\beta$ be the internal solid angle 
and external solid angle as shown in figure \ref{fig:BEMlimit},
\begin{equation}
u(\y)
+\int_{\del\Omega+\Gamma_\epsilon} G_E(\x,\y) \nabla_i u \,\sqrt{g}\, dS^i
-\int_{\del\Omega+\Gamma_\epsilon} (\nabla_i G_E(\x,\y)) u \,\sqrt{g}\,
dS^i=0.
\label{eq:first}
\end{equation}
The singular terms in Eq.(\ref{eq:first}) can be separated 
from non-singular terms as
\BEA
\lim_{\epsilon \rightarrow 0}\int_{\del \Gamma+\Gamma_\epsilon}
G_E(\x,\y) \nabla_i u \,\sqrt{g}\, dS^i&=&\int_{\del \Gamma}
G_E(\x,\y) \nabla_i u \,\sqrt{g}\, dS^i
\N
\\
\N
&+&\lim_{\epsilon \rightarrow 0}
\int_{\Gamma_\epsilon}
G_E(\x,\y) \nabla_i u \,\sqrt{g}\, dS^i,
\\
\N
\lim_{\epsilon \rightarrow 0}\int_{\del \Gamma+\Gamma_\epsilon}
(\nabla_i G_E(\x,\y)) u \,\sqrt{g}\, dS^i&=&\int_{\del \Gamma}
(\nabla_i G_E(\x,\y)) u \,\sqrt{g}\, dS^i
\N
\\
\N
&+&\lim_{\epsilon \rightarrow 0}
\int_{\Gamma_\epsilon}
(\nabla_i G_E(\x,\y)) u \,\sqrt{g}\, dS^i.
\label{eq:separate}
\EEA
If $\epsilon$ is sufficiently small, the region enclosed by 
$\Gamma_\epsilon$ can be approximated as an Euclidean subspace. 
In this region, the asymptotic form of the free Green's function 
$G_E$ takes the form
\begin{equation}
\lim_{\x \rightarrow \y} G_E(\x,\y)= - \f{\exp(ikd)}{4 \pi d}=-\f{1}{4
\pi d}-\f{ik}{4 \pi}+{\cal{O}}(d),
\end{equation}
where $d$ is the Euclidean distance between $\x$ and $\y$.
Taking the spherical coordinates $(\epsilon,\theta,\phi)$ with center
$\y$, the singular terms in Eq.(\ref{eq:separate}) are estimated as 
\BEA
\lim_{\epsilon \rightarrow 0}
\int_{\Gamma_\epsilon}
G_E(\x,\y) \nabla_i u(\x) \,\sqrt{g}\, dS^i&=&
\lim_{\epsilon \rightarrow 0} - \int_{\beta} \f{1}{4 \pi \epsilon}
\f{\del u(\x)}{\del n} \epsilon^2 d \Omega=0,
\N
\\
\lim_{\epsilon \rightarrow 0}
\int_{\Gamma_\epsilon}
(\nabla_i G_E(\x,\y)) u(\x) \,\sqrt{g}\, dS^i&=&
\lim_{\epsilon \rightarrow 0} \int_{\beta} \f{1}{4 \pi \epsilon^2}
\f{\del \epsilon(\x)}{\del n} u(\x) \epsilon^2 d \Omega
\N
\\
&=&\f{\beta}{4\pi} u(\x),
\label{eq:limit}
\EEA
where $d \Omega$ denotes the infinitesimal solid angle element. 
Taking the limit $\epsilon \rightarrow 0$ in Eq.(\ref{eq:first}), 
we have the boundary integral equation in 3-spaces,
\begin{equation}
\f{1}{4 \pi} \alpha(\y) u(\y)
+\int_{\del\Omega} G_E(\x,\y) \nabla_i u \,\sqrt{g}\, dS^i
-\int_{\del\Omega} (\nabla_i G_E(\x,\y)) u \,\sqrt{g}\, dS^i=0,
\label{eq:bemAPENDIX}
\end{equation}
where $\alpha(\y)$ denotes the internal solid angle at $\y$. If the
boundary is smooth at $\y$, $\alpha(\y)$ is equal to $2 \pi$ which
gives the coefficients $1/2$ in Eq.(\ref{eq:bem0}). Similarly, one can 
prove the formula for $M=2$ and $M>3$.
\BF[t]
\centerline{\psfig{figure=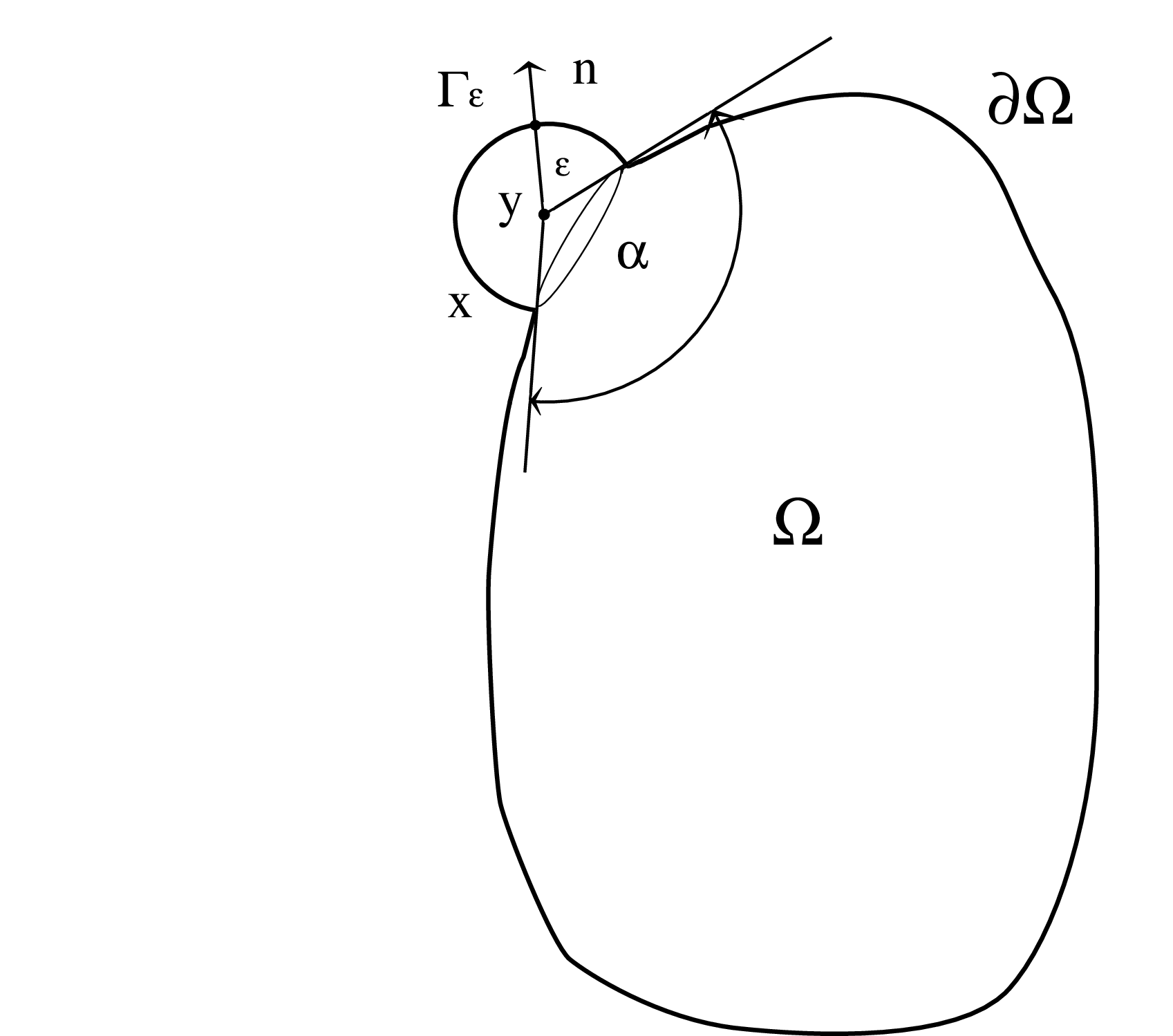,width=10cm}}
\caption{Boundary Integral}
\label{fig:BEMlimit}
\EF
\pagebreak
\chapter{Algorithm of computing length spectra} 
\thispagestyle{headings}
 For a given Dirichlet domain $D$, SnapPea computes
\\
\\
1. Neighboring copies of $D$ (tiles)   
   recursively and stores the corresponding 
   elements $g$ of the discrete isometry group $\Gamma$.
   If it has already been found out, 
   it is discarded. The computation proceeds
   until for all the neighborhoods of $gD$,
   $d(x,hx)> \cosh^{-1} (\cosh R \cosh l/2)$ satisfies
   where $hD'$s are neighborhoods of $gD$($R$ is the spine radius 
   and $h$ is an element of $\Gamma$).
   Since there is no $g$ other than identity 
   where all $h'$s satisfy $d(x,hx)\ge d(x,gx)$,
   this algorithm will not miss any tiles $gD$
   where  $d(x,gx)<2 \cosh^{-1} (\cosh R \cosh l/2)$. 
\\
\\
2. A list of geodesics 
   for all $g's=\{g\}$ where 
   1:the real part of length is not zero and less than $l$;
   2:the distance from $x$ to the geodesic is at most $R$.
\\
\\   
3. The conjugacy class $g'=h g h^{-1}$ or its inverse for 
   each $g$ where $h$ is an element of $\{g\}$. If an identical 
   complex length is found, the complex length which corresponds to
   $g'$ is omitted from the list of geodesics. If the complex
   length of $g$ is conjugate to that of $g^{-1}$, the geodesic
   is topologically a mirrored interval, otherwise it is a circle.
\\
\\
4. Multiplicity of geodesics.
   If a pair of geodesics with two complex lengths being 
   identical within an error range is found, the multiplicity
   number is increased by one. 

\pagebreak
\chapter{Estimate of volume in terms of diameter}
\thispagestyle{headings} 
Suppose a CH manifold $M$
which resembles the original cusped manifold $M_c$ with one cusp.
Let us divide $M_c$ into two parts, the neighborhood
of a cusped point $K_c$, and the complementary part $K_{c0}$.
Similarly, one can divide $M$ into $K$ and $K_{0}$ where
$K_{c0}\approx K_0$ and $K$ corresponds to a ``thin'' part.
Since the neighborhood of a cusp is represented as a ``chimney''(but
having infinite length)
in the upper half space coordinates $(x_1,x_2,x_3)$, 
$K$ can be well approximated by
an elongated box defined by ($-\Delta x/2 \le x_1 \le \Delta x/2,    
-\Delta x/2 \le x_2 \le \Delta x/2, x_{30}\le x_3 \le x_{31}$).
Then the physical length(=diameter) $\tilde d$ of $K$
in the direction $x_3$ is given by
$\tilde{d}=\ln (x_{31}/x_{30})$. On the other hand, the volume 
$\tilde{v}$ of $K$ satisfies
\BE
\tilde{v}=\f{(\Delta x)^2}{2}\biggl( \f{1}{(x_{30})^2}
-\f{1}{(x_{31})^2} \biggr ),
\EE
which gives the ratio of the volume 
$\tilde{v}$ to the volume of $\tilde{v}_\infty$ of $K_c$,
$\tilde{v}/\tilde{v}_\infty=1-(x_{30})^2/(x_{31})^2
=1-\exp(-2 \tilde{d})$. If we approximate the diameter $d$
of $M$ as $d=\tilde{d}+d_0$ where $d_0$ is the diameter of $K_0$,
then we finally have the ratio of the volume $v=v_0+\tilde{v}$ of $M$ to
the volume $v_c=v_0+\tilde{v}_\infty$ of $M_c$,
\BE
\f{v}{v_c}=1-\f{\exp (-2(d-d_o))}{\delta+1},~~~
\delta\equiv v_0/\tilde{v}_\infty
\EE
where $v_0$ denotes the volume of $K_0$.

\BF
\centerline{\psfig{figure=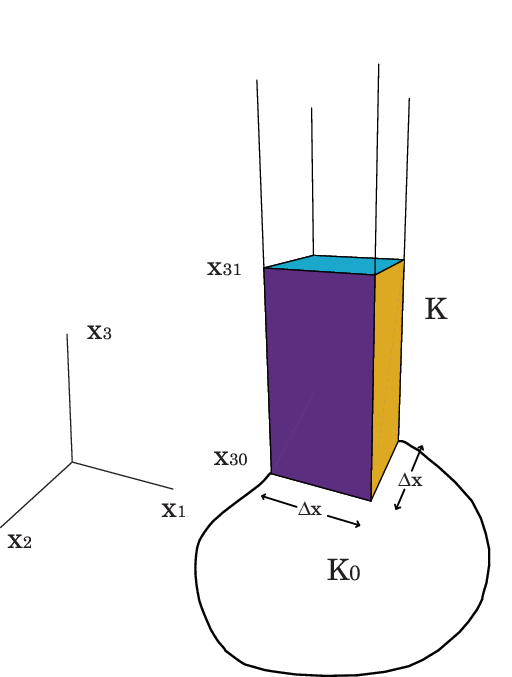,width=10cm}}
\caption{Estimate of Volume of ``thin'' Part}
\label{fig:appendixc}
\EF

\end{appendix}
\end{document}